\documentclass[]{aa}
\usepackage{graphicx}
\usepackage{rotating,subfigure,amssymb,afterpage}
\usepackage{txfonts}
\usepackage{natbib}
\usepackage[pdftitle={X-ray cluster temperature profiles},
pdfauthor={G.W. Pratt et al.},
pdfsubject={}, pdfkeywords={}, bookmarks, bookmarksopen, 
colorlinks=true, linkcolor=blue, citecolor=blue, anchorcolor=blue,
urlcolor=blue]{hyperref} 
\def \xmm {\hbox{\it XMM-Newton }}

\def\kT {{\rm k}T}

\def\rv {R_{200}}

\def \s01 {S_{0.1}}
\def \zs {Z_{\odot}}
\newcommand{\propsim}{\lower 3pt \hbox{$\, \buildrel {\textstyle
      \propto}\over {\textstyle \sim}\,$}}

\begin{document}
     \title{Temperature profiles of a representative sample of nearby
       X-ray galaxy clusters} 

     \author{G. W. Pratt$^1$, H. B\"ohringer$^1$,
       J.H. Croston$^{2,3}$, M. Arnaud$^2$, S. Borgani$^{4}$,
       A. Finoguenov$^{1}$ and R.F. Temple$^5$}
     \offprints{G. W. Pratt, \email{gwp@mpe.mpg.de}}

     \institute{$^1$ MPE Garching, Giessenbachstra{\ss}e, 85748
       Garching, Germany  \\
                $^2$ CEA/Saclay, Service d'Astrophysique,
                L'Orme des Merisiers, B\^{a}t. 709,
                91191 Gif-sur-Yvette Cedex, France \\
                $^3$ School of Physics, Astronomy and Mathematics,
                University of Hertfordshire, College Lane, Hatfield
                AL10 9AB, UK \\
                $^4$ Dipartimento di Astronomia dell'Universit\`a di
                Trieste, via Tiepolo 11, I-34131 Trieste, Italy \\ 
                $^5$ School of Physics and Astronomy, University of
                Birmingham, Edgbaston, Birmingham B15 2TT, UK}
     \date{Received; Accepted}

\abstract
{A study of the structural and scaling properties of the temperature
  distribution of the hot, X-ray emitting intra-cluster medium of
  galaxy clusters, and its dependence on dynamical state, can give
  insights into the physical processes governing the formation and
  evolution of structure.}   
{Accurate temperature measurements are a pre-requisite for a precise
  knowledge of the thermodynamic properties of the intra-cluster
  medium.}
{We analyse the X-ray temperature profiles from XMM-Newton
  observations of 15 nearby ($z<0.2$) clusters, drawn from a
  statistically representative sample. The 
  clusters cover a temperature range from 2.5 keV to 8.5 keV, and
  present a variety of X-ray morphologies. We derive accurate projected
  temperature profiles to $\sim 0.5\,\rv$, and compare structural
  properties (outer slope, presence of cooling core) with a quantitative
  measure of the X-ray morphology as expressed by power ratios. We also
  compare the results to recent cosmological numerical simulations.}
{Once the temperature profiles are scaled by an average cluster
  temperature (excluding the central region) and the estimated virial
  radius, the profiles generally decline in the region $0.1\,\rv
  \lesssim R \lesssim 0.5\,\rv$. The central regions show the largest
  scatter, attributable 
  mostly to the presence of cool core clusters. There is
  good agreement with numerical simulations outside the core
  regions. We find no obvious correlations between power ratio and
  outer profile slope. There may however be a weak trend with the
  existence of a cool core, in the sense that clusters with a central
  temperature decrement appear to be slightly more regular. } 
{The present results lend further evidence to indicate that clusters are
  a regular population, at least outside the core region. } 

%%We present X-ray temperature profiles of 15 nearby ($z <
%%0.2$) clusters, drawn from a statistically representative sample
%%observed with {\it XMM-Newton}. The clusters cover a temperature range
%%from 2.5 keV to 8.5 keV, and present a variety of morphologies. Once
%%the temperature profiles are scaled by an average cluster temperature
%%(excluding the central region) and the estimated virial radius, there
%%is considerable similarity in the region $0.1\,\rv \lesssim R \lesssim
%%0.5\,\rv$, where the profiles generally decline. The central regions
%%show the largest scatter. These results are in good agreement with
%%studies using {\it ASCA}, {\it BeppoSAX} and {\it Chandra}, and with
%%recent numerical simulations. Using power ratios to quantify the
%%dynamical state of the clusters, we find no obvious correlations with
%%outer profile slope. There may however be a weak correlation with the
%%existence of a cool core, in the sense that cooling core clusters
%%appear to be more regular. The present results thus lend more evidence
%%to indicate that clusters are a regular population.


\authorrunning{G.W. Pratt et al.}
\titlerunning{X-ray cluster temperature profiles}
\maketitle
%
%%----------------------------------------------------------------------------

\section{Introduction}

The temperature and density are the key measurable characteristics of
the hot, X-ray emitting intracluster medium (ICM). The determination
of important derived properties such as entropy, pressure, and, under
the assumption of hydrostatic equilibrium, the total mass, is
dependent on accurate estimation of these quantities. Because of
limited photon statistics\footnote{Also the need for an azimuthally
symmetric approximation for purposes of deprojection.} it is usual to
measure the density and temperature in terms of radial
profiles. However, while the density of the ICM is relatively easy to
measure from the surface brightness profile of a given cluster, the
temperature determination requires sufficient photon statistics to
build, and fit, a spectrum. Thus ICM temperature profiles are
typically determined with considerably less spatial resolution than
density profiles. 

The measurement of radial temperature profiles is further complicated
by the density squared ($n_e^2$) dependence of the X-ray emission. The
steep drop of the X-ray surface brightness with distance from the
centre, combined with the background from cosmic, solar and
instrumental sources, makes accurate measurement of the temperature
distribution at large distances from the centre a technically
challenging task.

The earliest temperature profiles were measured with {\it Einstein},
{\it EXOSAT}, {\it Spacelab-2\/} and {\it GINGA\/} only for the
nearest, brightest clusters (e.g.,
\citealt*{fab1,fab2,hughes,eyles,koy}). The low, stable background of
{\it ROSAT} made possible spatially resolved spectroscopy of poor
clusters \citep*[e.g.][]{david}; however, limited spectral resolution
and bandwidth made such measurements difficult for hotter
clusters \citep*[e.g.][]{hbn,bh94,hb95}.
{\it ASCA\/} and {\it BeppoSAX\/} had sufficient high-energy
sensitivity to accurately measure the temperatures of hot
clusters. However both of these satellites suffered from significant
PSF blurring, which, in the case of {\it ASCA},  was exacerbated by a
significant energy dependence.  As a result, at the end of the {\it
ASCA/BeppoSAX\/} era, the exact shape of cluster temperature profiles
was still under vigorous debate
\citep*{mark98,irwin99,whi00,ib00,fin01,dm02}.

{\it Chandra\/} and \xmm do not suffer from major PSF problems. The
on-axis {\it Chandra\/} PSF is negligible, while the \xmm PSF becomes
an issue only for clusters with very centrally peaked core emission;
in addition, neither is energy-dependent. Recent observations of
moderately large samples consisting primarily of nearby cooling core
clusters with \xmm \citep{piff} and {\it Chandra\/} \citep{vikh05}
have largely validated the original {\it ASCA\/} results of Markevitch
et al., which suggested that temperature profiles declined from the
centre to the outer regions. However, other {\it Chandra} and \xmm
observations have found flatter profiles \citep{allen01,kaa,app}. As
of the time of writing, no systematic attempt has been made, with
either \xmm or {\it Chandra}, to look at the temperature profiles of a
representative sample of nearby clusters\footnote{Some work has been
done on medium-distant clusters, see \citep{zhang,kotov}}. Although
other projects on representative samples are in progress
\citep[e.g.,][]{reiprich}, they are not expected to be able to map the
temperature distribution out to large radius.

In this paper we deal with observations of 15 clusters from a
statistically representative sample observed with \xmm. We describe in
detail the data reduction and background subtraction, and compare our
results with previous work and with those from cosmological
hydrodynamical simulations. We also make a preliminary investigation
of correlations with quantitative morphological measures. We present
only projected temperature profiles in this paper -- such profiles are
direct observables and do not depend on complicated PSF and
deprojection algorithms. We will deal with correction of the profiles
in forthcoming papers which make use of observations of the full
sample. All results are given assuming a $\Lambda$CDM cosmology with
$\Omega_m=0.3$ and $\Omega_\Lambda=0.7$ and $H_0 =
70$~km~s$^{-1}$~Mpc$^{-1}$. Unless otherwise stated, errors are given
at the $68$ per cent confidence level.

\section{The sample}
\label{sec:sample}

The \xmm Legacy Project for the study of cluster structure was
initiated to study the structural and scaling properties of a large,
representative sample of clusters. Since full details will appear in a
forthcoming paper, we present here only a short summary of the sample
selection. 

The parent sample is the REFLEX catalogue \citep{reflex}. To ensure
the best quality for potential targets, the REFLEX catalogue was first
screened to include only objects which had (i) a firm detection
threshold of more than 30 source photons in the {\it ROSAT\/} All Sky
Survey and (ii) a low column density ($n_H < 6 \times 10^{20}$
cm$^{-2}$). 

Since the Legacy Project selection was intended to be representative of
an X-ray flux- or $L_X$-limited sample, clusters were chosen purely on
the basis of X-ray luminosity.  Further selection criteria included:
(i) redshift $z < 0.2$ to sample the nearby Universe; (ii) close to
homogeneous coverage of the luminosity space; (iii) a flux limit
corresponding to $\kT > 2$ keV, to sample the mass range from poor
systems to rich clusters; (iv) detectable with \xmm to approximately a
radius of $R_{500}$, with distances selected to optimise $R_{500}$ in
the \xmm field of view.

%%================
%% Figure: Lx-z
%%
\begin{figure}
\begin{centering}
\includegraphics[scale=1.,angle=0,keepaspectratio,width=\columnwidth]{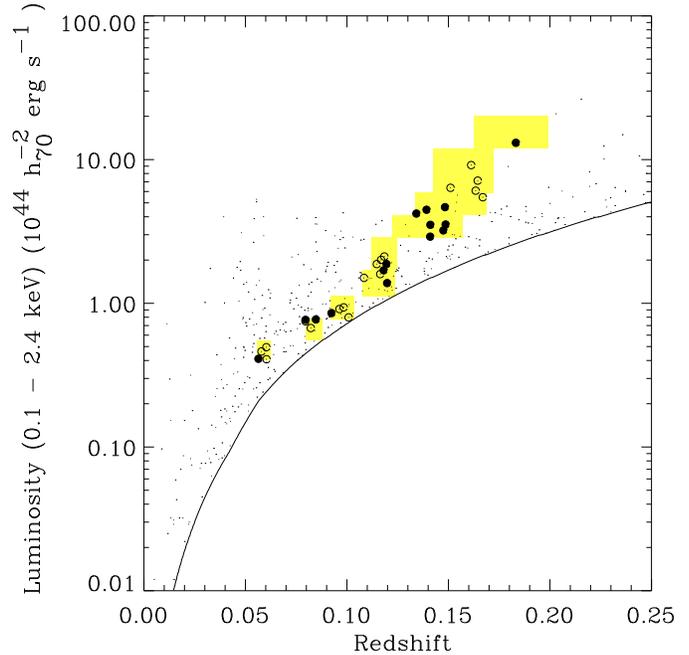}
\caption{{\footnotesize X-ray luminosity - redshift ( $L_X - z$)
    distribution of the REFLEX cluster sample in the redshift range $0
    < z < 0.25$. The redshift/luminosity 
    selection for the Legacy Project sample is indicated by the boxes.
    Filled circles indicate clusters from the
    Legacy Project which are discussed in this paper; dotted circles
    indicate clusters for
    which re-observations are necessary. The solid line is the REFLEX
    flux limit. ({\it This figure is available in colour in the online
    version of the journal.})}}\label{fig:lxz} 
\end{centering}
\end{figure}
%%================

To best assess the scaling relations, the sample should have close to
homogeneous coverage of luminosity space. The luminosity-redshift
space was thus sampled in eight almost-equal luminosity bins (see
Fig.~\ref{fig:lxz})\footnote{One extra bin, containing the most
luminous cluster, uses data from the \xmm archive.}. The lower
redshift boundary of each bin was placed above the flux limit curve or
close to the curve defining the redshift at which $R_{500}$
corresponds to $9\arcmin$ ($10\arcmin$ for the most luminous
clusters). The upper redshift boundaries were defined by the number of
clusters to be included in the bin (4 objects). For the lowest
luminosity bin these criteria were relaxed, and the bin put at lower
redshift, because these clusters are fainter. The distribution of
clusters in luminosity-redshift space is shown together with the
luminosity bins chosen for the Legacy Project sample in
Fig.~\ref{fig:lxz}.

Observations of the full sample of 31 clusters plus 2 archive
observations have now been completed. As detailed below, the quality
of 18 of these observations is not sufficient to derive accurate
radial temperature profiles to relatively large radius. The remaining
15 clusters with good quality data are discussed in this paper. As
shown in Fig.~\ref{fig:lxz}, only one redshift bin is not represented
in this subsample, thus it is representative of the sample as a whole.

\section{XMM data analysis}

Observation data files (ODFs) were retrieved from the XMM archive and
reprocessed with the \xmm Science Analysis System (SAS) v6.1 using the
publicly-available calibration current as of February 2005. The
resulting calibrated EMOS and EPN event files were then used in all
subsequent analysis.

\subsection{ODF preparation}
\label{sec:odfprep}

The data were cleaned for periods of high background due to soft
proton solar flares using a two stage filtering process. A light curve
was first extracted in 100 second bins in the [10-12]/[12-14]
(EMOS/EPN) energy band. A Poisson distribution was fitted to a
histogram of this light curve, and $\pm 3 \sigma$ thresholds
calculated. A Good Time Interval (GTI) file was produced using the
upper threshold, and the event list was filtered accordingly. Since i)
flares often appear to have soft `wings'; ii) the statistics at high
energy are often poor; and iii) softer flares exist, the event list
was then re-filtered in a second pass. In this case light curves were
made in 10 sec bins in the full [0.3-10] band, the smaller bin size
being possible because of the greatly improved statistics. A histogram
was calculated, a Poisson distribution fitted, GTIs generated, and
event lists were filtered as above.

%% =====================
%% Cluster data
%%
\begin{table}
\begin{minipage}{\columnwidth}
\caption{{\footnotesize Basic cluster data.}}\label{tab:obs}
\centering
\begin{tabular}{l l l l l r}
\hline
\hline
\multicolumn{1}{l}{RXCJ} & \multicolumn{1}{l}{$T_X$\footnote{Spectral
    temperature in the radial range 0.1--$0.4\,\rv$, in keV, estimated
    using the $R$--$T$ relation of \citet{app} -- see
    Sect~\ref{sec:sct} for details.}} &
\multicolumn{1}{l}{z} &
\multicolumn{1}{l}{$N_H$\footnote{Column density in units of $10^{20}$
    cm$^{-2}$ (see text for details).}} &
\multicolumn{1}{l}{Exp. \footnote{Cleaned exposure time of EMOS1, EMOS2 and
    EPN in kiloseconds.}} & \multicolumn{1}{l}{Comments} \\ 

%\multicolumn{1}{l}{(RXCJ)} & \multicolumn{1}{l}{} &
%\multicolumn{1}{l}{(keV)} & \multicolumn{1}{l}{} &
%\multicolumn{1}{l}{(ks)} & \multicolumn{1}{l}{}\\ 
\hline
0003+0203 & $3.71\pm0.09$ & 0.085 & 4.7 & 26,26,17 & A2700 \\
0020-2542 & $5.74\pm0.13$ & 0.141 & 2.2 & 16,16,11 & A22 \\
0547-3152 & $6.59\pm0.12$ & 0.148 & 2.1 & 23,24,17 & A3364 \\
0605-3518 & $4.68\pm0.11$ & 0.139 & 4.5 & 22,23,14 & A3378 \\
1044-0704 & $3.56\pm0.05$ & 0.134 & 3.6 & 26,26,18 & A1084 \\
1141-1216 & $3.60\pm0.08$ & 0.120 & 3.2 & 28,28,22 & A1348 \\
1302-0230 & $3.60\pm0.08$ & 0.085 & 1.7 & 25,25,16 & A1663 \\
1311-0120 & $8.45\pm0.12$ & 0.183 & 1.8 & 36,37,29 & A1689 \\
1516+0005 & $4.34\pm0.07$ & 0.120 & 5.4 & 26,27,21 & A2050 \\
1516-0056 & $3.75\pm0.10$ & 0.120 & 5.4 & 29,30,22 & A2051 \\
2023-2056 & $2.83\pm0.08$ & 0.056 & 5.4 & 17,18,10 & S868 \\
2048-1750 & $3.96\pm0.08$ & 0.148 & 4.7 & 25,25,19 & A2328 \\
2129-5048 & $3.84\pm0.10$ & 0.080 & 2.2 & 21,22,11 & A3771 \\
2217-3543 & $4.60\pm0.08$ & 0.148 & 6.6 & 24,24,17 & A3854 \\
2218-3853 & $5.84\pm0.17$ & 0.141 & 5.7 & 21,22,11 & A3856 \\
\hline
\end{tabular}
\end{minipage}
\end{table}
%% =====================

This type of flare filtering is sufficient in the majority of
cases. However, it is not as effective in removing flares in cases of
data sets with softly-varying count rates or gradually increasing or
decreasing count rates. The histograms of such data sets invariably
have a Poisson distribution with a tail, which is not well fitted with
a single component. All light curve histogram fits were thus carefully
examined before further analysis. In problem cases, the data sets were
cleaned by hand (generally by estimating the flare periods by eye, and
excluding them).

Since the object of the present work is to obtain relatively
high-quality temperature profiles, we only use those observations with
a cleaned EPN exposure time greater than 10 ks. Of the 33 clusters in
the Legacy Project sample, 15 meet this criterion at the present
time.\footnote{The remaining poor quality data are being re-observed
under XMM AO4 and AO5.}. These observations are listed in
Table~\ref{tab:obs}. Figure~\ref{fig:lxz} shows the X-ray luminosity -
redshift distribution of the LP clusters.  Filled circles show the
clusters discussed in this paper, while open circles show clusters for
which data are pending. REFLEX clusters appearing in the boxes include
those for which the X-ray criteria (minimum 30 source photons and
column density $n_H < 6 \times 10^{20}$ cm$^{-2}$) were not met. The
current subsample is clearly representative of the whole sample; only
one redshift bin is not represented.

After removal of periods of high soft proton flux, events were
filtered according to {\tt PATTERN} and {\tt FLAG} criteria. For EMOS
event files singles, doubles, triples and quadruples were selected
{\tt (PATTERN < 13)}, while for EPN data sets singles and doubles were
selected {\tt (PATTERN < 5)}. Events not corresponding to these
criteria were removed from the event files before further
processing. In addition, for all cameras events next to CCD edges and
next to bad pixels were excluded {\tt ( FLAG==0)}.

To correct for vignetting, a {\tt WEIGHT} column was added to each
event list using the SAS task {\tt evigweight}. All subsequent science
products were extracted from this column as described in \citet{arnaud01}.

Serendipitous and point sources were detected in a broad band
([0.3-10.0] keV) coadded EPIC image using the SAS wavelet detection
task {\tt ewavdetect}, with a detection threshold set at $5
\sigma$. Detected sources were excluded from the event file for all
subsequent analysis. 

\subsection{Background preparation}

\subsubsection{Flare cleaning}

The basic background files used are those of \citet{rp03}, which have
nominal exposure times of $\sim 1$ Ms/400 ks (EMOS/EPN)\footnote{The
EPN event list is in Extended Full Frame mode and does not necessarily
contain the same fields as the corresponding EMOS event list, hence
the shorter exposure time.}. Close inspection of the high-energy band
light curves showed that considerable periods of high soft proton flux
still existed in the event files. These periods were removed by two
applications of the double pass filtering procedure described in
Sect.~\ref{sec:odfprep} above. The resulting filtered background event
lists have light curve histograms which are adequately described by a
standard Poisson distribution. The loss in exposure time is $\sim
200/100$ ks (EMOS/EPN); the larger relative EPN time loss reflects the
greater EPN sensitivity to flares.

After flare filtering, the same {\tt PATTERN} and {\tt FLAG}
selections as above were applied to the background event files. The
background files were then corrected for vignetting via the addition of
a WEIGHT column to the event lists.

\subsubsection{Exposure correction}

Since the background data sets consist of stacked observations with
sources removed, exposure times can vary by up to a factor of two
across the detector. Using the exposure maps supplied by Andy Read
\footnote{\tt ftp://ftp.sr.bham.ac.uk/pub/xmm/expmap$*$.fits.gz}, a
new exposure map was computed for each event list taking into account
exposure variations due to the point source subtraction. These
exposure maps were renormalised to the new exposure time of the
background files after flare cleaning. An EXPOSURE column, containing
the exposure time at the position of the event, was then added to each
background file. The WEIGHT column was then corrected for exposure
variations by simply dividing by the EXPOSURE column.

\subsection{Background subtraction}

The blank sky backgrounds are recast onto the sky using the aspect
information from the cluster pointing, enabling extraction of source
and background spectra from the same detector regions. This procedure
is necessary because the spatial distribution of the various
instrumental lines is not constant across the field of view. 

\subsubsection{Quiescent background}

The \xmm EMOS and EPN backgrounds are dominated by charged-particle
events above $\sim 2$ keV. The intensity of this component can vary by
typically $\pm 10\%$, and must be accounted for by
renormalisation. The renormalisation factor for each observation was
calculated in the source-free [10-12]/[12-14] kev (EMOS/EPN) energy
band, and the WEIGHT column of each background file was adjusted
accordingly. This renormalisation assumes that the particle induced
background can simply be scaled depending on the count rate. The 
renormalisation factors are listed in Table~\ref{tab:res}.

\subsubsection{Soft diffuse X-ray background}

The observation and blank field event files contain a component due to
the soft diffuse X-ray background, which dominates the flux below
$\sim 1$ keV. This component is variable across the sky, and thus from
pointing to pointing. Correction for this variation is thus needed.

%%================
%% Figure: R0547 residual spectrum
%%
\begin{figure}
\begin{centering}
\includegraphics[scale=0.30,angle=-90,keepaspectratio,width=\columnwidth]{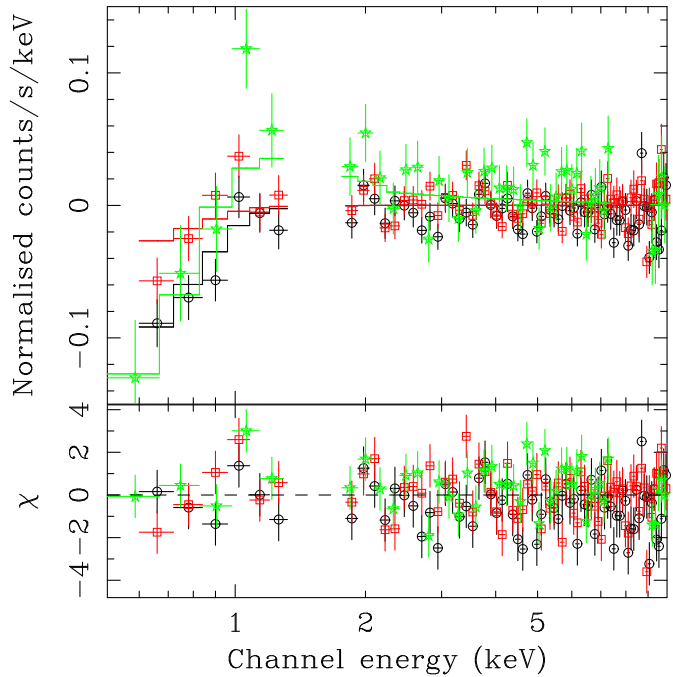}
\caption{{\footnotesize The residual spectrum of R0547, indicating
    oversubtraction of the soft X-ray background (see text for
    details). Black: EMOS1; 
    red: EMOS2; green: EPN. The fit is a Solar abundance MEKAL model
    with $\kT = 0.24$ keV and negative
    normalisation. The EPN
    model has an additional power-law component 
    with positive normalisation. ({\it This figure is available in
      colour in the online 
    version of the journal.})}}\label{fig:res}
\end{centering}
\end{figure}
%%================

As discussed above in Sect~\ref{sec:sample}, the present cluster
sample was explicitly defined so that a certain 
fraction of the detector area is essentially free of cluster emission,
enabling a direct estimation of the local background in the outer
regions of the field of view. For each observation, we extracted
surface brightness profiles in the [0.3-3.0] keV band for each
camera. EMOS and EPN surface brightness profiles were background
subtracted, coadded, and binned to $3\sigma$ significance. The local
background spectrum was built using all events outside the radius at
which the surface brightness profile in no longer significantly
detected. A renormalised 
spectrum from the same region of the blank sky background was then
subtracted, yielding a residual spectrum. We fitted the residual
spectrum in the [0.5-10.0] keV band with an unabsorbed, solar
abundance MEKAL model. Energies with significant instrumental line
emission (1.4-1.6 keV for all instruments; 7.45-9. keV for EPN) were
excluded from the fits. In case of a significant excess of counts in
the $\gtrsim 2$ keV band, presumably due to a remaining component of
soft protons, a power-law was added to the model. The EPIC spectra were
fitted simultaneously, with the temperature of the MEKAL model linked
between instruments. The power-law slope and normalisations of all
components were free to vary in the fitting process. The normalisation
of the MEKAL model was allowed to be negative, to account for
over-subtraction. This purely phenomenological model is capable of
describing a wide variety of residual spectra (Fig.~\ref{fig:res}),
although care must be exercised in deriving the initial model
parameters. This residual model, with all parameters fixed and the
normalisation scaled appropriate to the 
ratio of extraction region areas, was treated as an additional
component in all subsequent annular fits.

%% =====================
%% Cluster data
%%
\begin{table*}
\begin{minipage}{\textwidth}
\caption{{\footnotesize Columns: (1) Cluster name; (2) Radius beyond
    which external region spectra were accumulated (EMOS1,EMOS2,EPN);
    (3-5) Normalisation factor for background rescaling. This factor
    was calculated from the ratio of the count rate in the
    observation and background files in the [10-12]/[12-14] keV
    (EMOS/EPN) energy band; (6) Temperature of {\sc MeKaL} model
    used to describe the residual spectrum; (7-9) {\sc xspec}
    normalisation of the {\sc MeKaL} model used to describe the
    residual spectrum, in units of $10^{-4}$.}}\label{tab:res} 
%%; (8) Was
%%    an additional power-law component needed (Yes/No;
%%    EMOS1,EMOS2,EPN)?; (9) Rough morphological classification:
%%    R=relaxed, D=disturbed.}}\label{tab:res} 
\centering
\begin{tabular}{l l l l l l r r r l l } % ll}
\hline 
\hline
\multicolumn{1}{l}{RXCJ} & \multicolumn{1}{l}{$R_{\rm ext}$} & 
\multicolumn{3}{c}{Norm$_{\rm ext}$} & 
\multicolumn{1}{l}{$\kT_{\rm ext}$} &
\multicolumn{3}{c}{{\sc MeKaL norm}}\\ %% &
%%\multicolumn{1}{l}{PL} & \multicolumn{1}{l}{Morph.} \\ 

\cline{3-5} \cline{7-9}

\multicolumn{1}{l}{ } &
\multicolumn{1}{l}{ ( $^\prime$ ) } &
\multicolumn{1}{l}{EMOS1} &
\multicolumn{1}{l}{EMOS2} &
\multicolumn{1}{l}{EPN} &
\multicolumn{1}{l}{ } &
\multicolumn{1}{l}{EMOS1} &
\multicolumn{1}{l}{EMOS2} & \multicolumn{1}{l}{EPN} \\ %% &
%%\multicolumn{1}{l}{ } & \multicolumn{1}{l}{ } \\

\hline

0003\,+0203 & 11,11,11 & 1.07 & 0.98 & 1.12 & 0.23 & -1.01 & -2.98 &
-0.54 \\ %% & N,Y,Y & D \\

0020\,-2542 & 11,11,12.5 & 1.00 & 0.98 & 1.02 & 0.10 & -14.32 & -20.02 &
-6.00 \\ %%& N,Y,Y & D \\

0547\,-3152 & 11,11,11 & 0.99 & 0.93 & 1.04 & 0.24 & -1.38 & -0.31 &
-0.76 \\ %% & N,N,Y & D \\

0605\,-3518 & 11,11,11 & 1.22 & 1.18 & 1.19 & 0.26 & -1.74 & -1.09 &
-0.42 \\ %% & N,N,Y & R \\

1044\,-0704 & 10,10,10 & 1.14 & 1.11 & 1.16 & 0.26 & -2.16 & -0.67 &
-5.00 \\ %% & N,N,Y & R \\

1141\,-1216 & 11,11,11 & 1.06 & 1.04 & 1.13 & 0.27 & -1.22 & -1.26 &
-0.40 \\ %% & N,N,N & R \\

1302\,-0230 & 11,11,11 & 1.04 & 1.02 & 1.07 & 0.27 & -0.81 & -0.97 &
-0.41 \\ %% & N,Y,Y & R \\

1311\,-0120 & 11,11,11 & 0.97 & 0.93 & 1.05 & 0.19 & 0.95 & 0.95 & 0.95
\\ %% & N,N,Y & R \\

1516\,+0005 & 12.5,12.5,13.5 & 1.16 &  1.14 & 1.29 & 0.24 & 0.48 & 1.27 &
1.01 \\ %% & N,N,Y & D \\

1516\,-0056 & 12.5,12.5,12.5 & 0.98 & 0.96 & 1.07 & 0.25 & 0.75 & 1.04 &
1.67 \\ %% & N,N,N & D \\

2023\,-2056 & 11,11,11 & 0.98 & 1.17 & 1.16 & 0.23 & 1.00 & 1.16 & 2.58 &
\\ %% N,N,Y & D \\

2048\,-1750 & 13.5,13.5,14 & 0.96 & 0.99 & 1.06 & 0.20 & 0.17 & 0.43 & 
0.62 \\ %% & N,N,N & D \\

2129\,-5048 & 11,11,12 & 1.33 & 1.34 & 1.32 & 0.33 & -0.51 & 0.90 & 0.00
\\ %% & Y,Y,Y & D \\

2217\,-3543 & 11,11,12 & 0.93 & 1.14 & 1.16 & 0.65 & -0.75 & -0.45 &
-0.36 \\ %% & N,N,N & D \\

2218\,-3853 & 11.5,11.5,12.5 & 1.24 & 1.25 & 1.25 & 0.26 & -1.32 & 0.30
& 0.05 \\ %% & Y,Y,Y & D\\

\hline
\end{tabular}
\end{minipage}
\end{table*}
%% =====================

\subsection{Spectral fitting}

If the cluster exhibited an obvious bright cooling core region,
spectra were accumulated in annuli centred on this surface brightness
peak. Some clusters have no obvious central peak: in these objects
spectra were centred on the emission centroid evaluated in a 6
arcminute radius. Annular regions for spectral fitting were
then defined i) to have 1500-2500 EMOS1 counts available after
background subtraction, and ii) to have a minimum width of $30\arcsec$
to minimise PSF effects. The spectra were extracted using the {\tt
WEIGHT} column, assuring full vignetting correction. Effective area
and response files corresponding to the on-axis position were
generated using {\tt arfgen} and {\tt rmfgen}, respectively. The
spectra were binned to $3\sigma$ significance after background
subtraction, to allow the use of Gaussian statistics. Spectral fits were
undertaken in the 0.5-10 keV energy range, excluding the 1.4-1.6 keV
band (due to the Al line in all three detectors), and, in the EPN, the
7.45-9.0 keV band (due to the strong Cu line complex).

Spectra were fitted with absorbed MEKAL models with abundances from
the data of \citet{agr}. The residual model
described above, with all parameters fixed and the normalisation
scaled appropriate to the ratio of extraction region areas, was
treated as an additional component. After first checking whether the
X-ray absorption was in agreement with the HI value, the absorption
was fixed at either the HI value or the best-fitting X-ray value. The
EPIC spectra were fitted simultaneously, with temperatures and
metallicities tied and the EPN spectral normalisation as an additional
free parameter. Annuli with abundance uncertainties $\delta Z / Z >
0.3$ were frozen at the average value of the two preceding fitted
annuli. This procedure will not affect the temperature estimates in
view of the generally flat abundance profiles in the outer 
cluster regions \citep[e.g.,][]{dm01}. 

%%================
%% Figure: External region of R2048
%%
\begin{figure}
\begin{centering}
\includegraphics[scale=1.,angle=-90,keepaspectratio,width=\columnwidth]{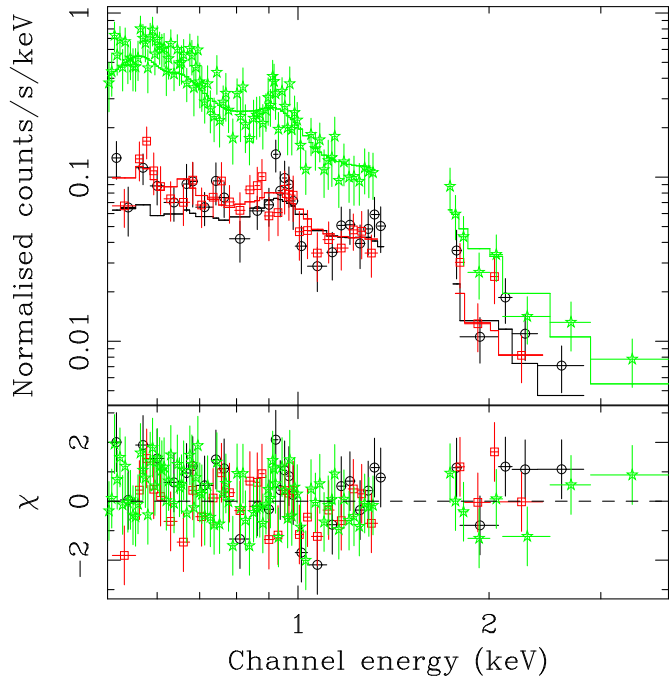}
\caption{{\footnotesize Observed (background subtracted) spectrum of
    the outermost annulus of 
    RXC\,J2048\,-1750 ($6\farcm75 < R < 8\farcm9$) The solid line is
    the best fitting model spectrum consisting of a cluster component
    plus a Galactic component (see text for details). ({\it This
      figure is available in colour in the online 
    version of the journal.})}}\label{fig:ext}
\end{centering}
\end{figure}
%%================

To take into account systematic uncertainties, the spectra were initially
fitted with nominal background normalisation. They were then re-fitted
with the background normalisation fixed at $\pm 10\%$ of nominal. The
changes in the best fitting cluster temperature were treated as the
corresponding systematic uncertainty; these were added in quadrature
to the statistical uncertainties of each annulus. Since the temperature 
determination is dominated by the exponential cutoff of the
Bremsstrahlung slope at higher
energies, which will depend strongly on the scaling of the particle 
background, we believe that this approach is extremely conservative in
terms of error determination.

Figure~\ref{fig:ext} shows the observed background subtracted spectrum
of the outermost 
annulus of RXC\,J2048\,-1750 ($6\farcm75 < R < 8\farcm9$). The signal
to noise of this spectrum is typical of that in the outer annulus
across the sample. The solid line shows the best fitting model
spectrum consisting of a cluster component plus a Galactic
component. The fit is excellent, with a $\chi^2_\nu = 0.98$ for 157
degrees of freedom.

%%================
%% Figure: Projected temperature profiles in arcmin and in kpc
%%
%%\begin{figure*}
%%\begin{centering}
%%\includegraphics[scale=1.,angle=0,keepaspectratio,width=0.475\textwidth]{tprof_arcmin.eps}
%%\hfill
%%\includegraphics[scale=1.,angle=0,keepaspectratio,width=0.475\textwidth]{tprof_kpc.eps}
%%\hfill
%%\includegraphics[scale=1.,angle=0,keepaspectratio,width=0.33\textwidth]{tprof_scaled.eps}
%%\caption{{\footnotesize Projected temperature profiles for all clusters plotted
%%    as a function of angular distance from the centre (left) and in
%%    physical units ($h_{70}^{-1}$; right).}}\label{fig:tprofs1}
%%\end{centering}
%%\end{figure*}
%%================

\subsection{X-ray images}

We produced images for each cluster to enable readers to judge the
morphology of the 15 objects in the sample. Images of the source and
associated background files were extracted from the WEIGHT column of
the EMOS event files in $3\farcs3$ bins in the [0.5-2.0] keV band. (We
do not use the EPN for image generation due to severe problems with
artifacts caused by the large gaps between CCD chips in this
detector.) EMOS1 and EMOS2 images were exposure corrected and
background subtracted separately, after which they were coadded.
The total EMOS image was then binned to $5\sigma$ significance using
the weighted Voronoi tesselation method of \citet{ds}.

\section{Results}

Cluster images and projected temperature and abundance profiles are
described in detail in Appendix~\ref{apx:tprofs}. 
%%In this Section, we discuss the
%%ensemble properties of the temperature profiles.
%%In Fig~\ref{fig:tprofs1}, we show the projected temperature profiles of the
%%sample plotted in units of arcminutes and $h_{70}^{-1}$ kpc,
%%respectively. 
It is clear that there is a general trend for the cluster temperature
profiles to decline with distance from the centre. For a better
understanding of how similar, or otherwise, the profiles are, it is
instructive to look at the scaled temperature profiles.

\subsection{Scaled temperature profiles}
\label{sec:sct}

%%================
%% Figure: temperature profiles
%%
\begin{figure*}
\begin{centering}
\hfill
\includegraphics[scale=1.,angle=0,keepaspectratio,width=0.98\columnwidth]{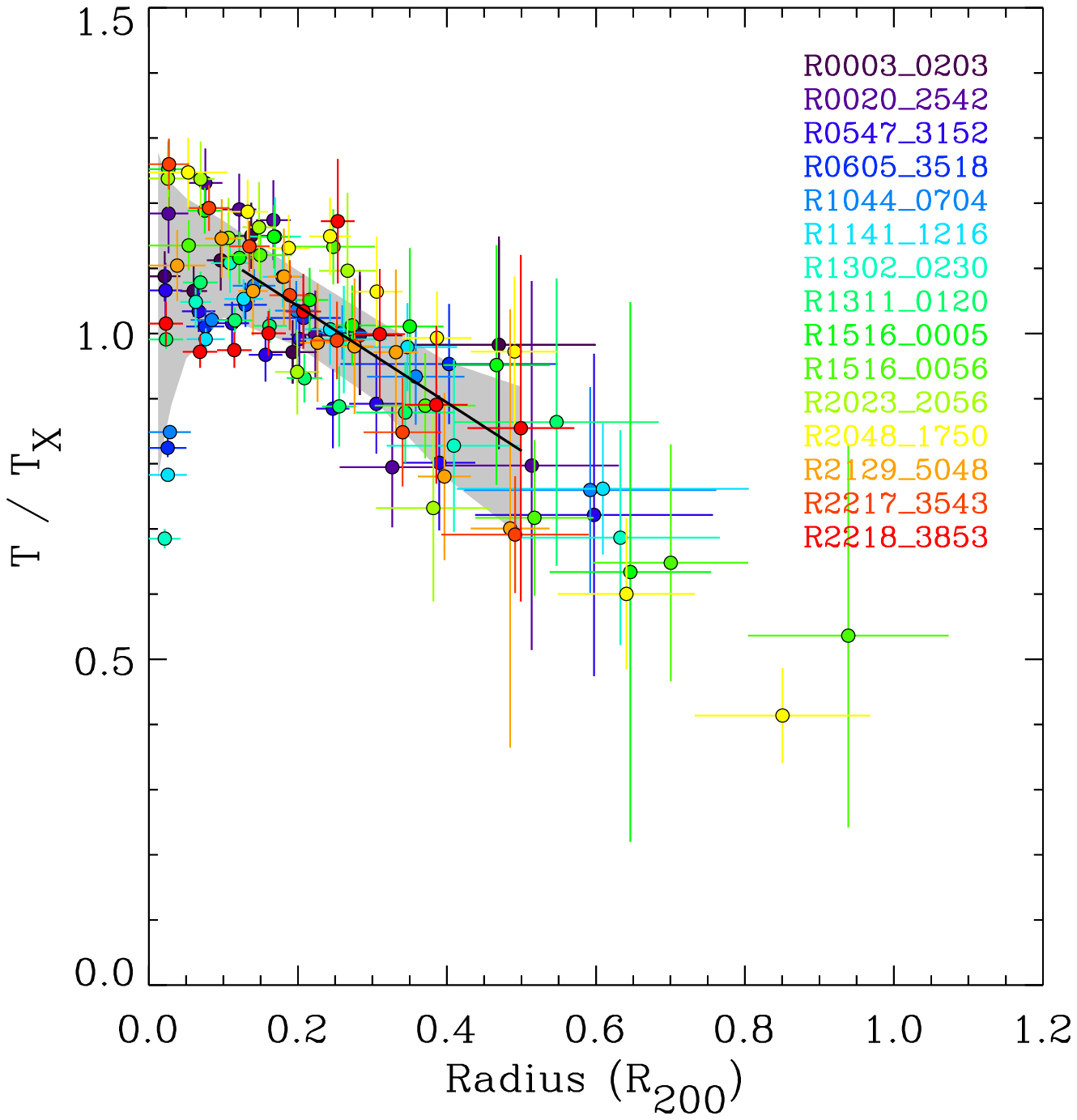}
\hfill
\includegraphics[scale=1.,angle=0,keepaspectratio,width=\columnwidth]{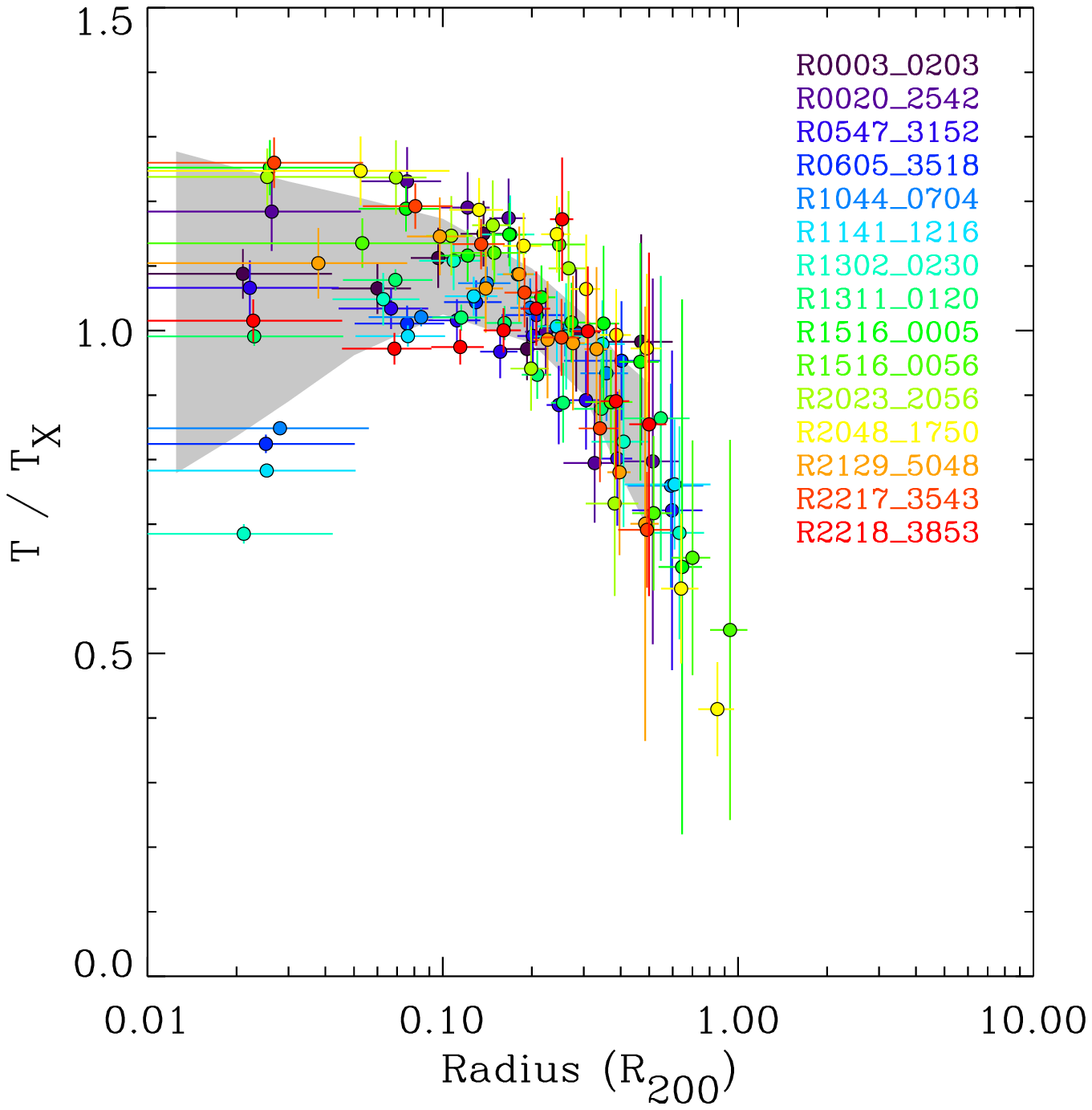}
\caption{{\footnotesize Scaled projected temperature profiles. Left
    panel: linear $x$-axis; right panel: logarithmic $x$-axis. The
    profiles have 
    been normalised to the mean spectroscopic temperature in the
    $0.1\,\rv \leq r \leq 0.4\,\rv$ region, where $\rv$ has been
    determined iteratively using the $R$--$T$ relation of
    \citet{app}. The shaded grey area corresponds to the region
    enclosed by the mean plus/minus the $1\sigma$ standard
    deviation. The solid line in the left-hand panel is the linear fit
    in the radial range $0.125 < \rv < 0.5$ detailed in
    Eqn.~\ref{eqn:tfit}. ({\it This figure is available in colour in the online
    version of the journal.})}}\label{fig:tprofs2} 
\end{centering}
\end{figure*}
%%================

We normalise the radial temperature profile of each cluster by a
global temperature, $T_{X}$, which should be representative of the
`virial' temperature of 
the cluster. Strong cooling core clusters have central temperature
decrements of up to a factor of three which, when combined with the
$n_e^2$ dependence of the X-ray emission, means that average
integrated temperatures of such systems can be biased. However, the
cooling core region rarely extends beyond $\sim0.1\,\rv$. In addition,
our measured temperature profiles do not extend to much further than 1
Mpc even in the best cases, which corresponds to $\sim60$ per cent of
$\rv$ for a 5 keV cluster \citep{app}. We thus chose to use the
overall spectroscopic temperature in the $0.1\,\rv \leq r \leq
0.4\,\rv$ region. We estimated this region in an iterative fashion,
using the $\rv$--$T$ relation of \cite{app} and starting with the mean
temperature from the measured temperature profiles. The measured
values of $T_X$ are given in Table~\ref{tab:obs}.

The resulting scaled temperature profiles are shown in
Figure~\ref{fig:tprofs2}. It is obvious that, despite the large
variety of objects in this sample, from strong cooling core objects to
highly unrelaxed systems, there is some similarity in the
temperature profiles; the profiles generally decline from the centre
to the outer regions. As an initial measure of the scatter in scaled
temperature profiles, we estimated the dispersion at various scaled
radii in the range $0.0125$--$0.5\,\rv$. The shaded region in
Fig.~\ref{fig:tprofs2} shows the region enclosed by the mean
plus/minus the $1\sigma$ standard deviation. Clearly the scatter
increases towards the central regions. The relative dispersion in
scaled profiles remains approximately constant at $\sim 10$ per cent
beyond $0.1\,\rv$. In the core regions, however, this increases to
$\sim 25$ per cent. Since the profiles have not been corrected for PSF
and projection effects, this figure is likely a lower limit.
In fact there is a clear difference between the cool 
core clusters, which have a large temperature drop toward the centre,
and the non-cool core clusters, which generally have 
profiles which increase linearly or flatten toward the centre. 

\begin{figure}
\begin{centering}
\hfill
\includegraphics[scale=1.,angle=0,keepaspectratio,width=\columnwidth]{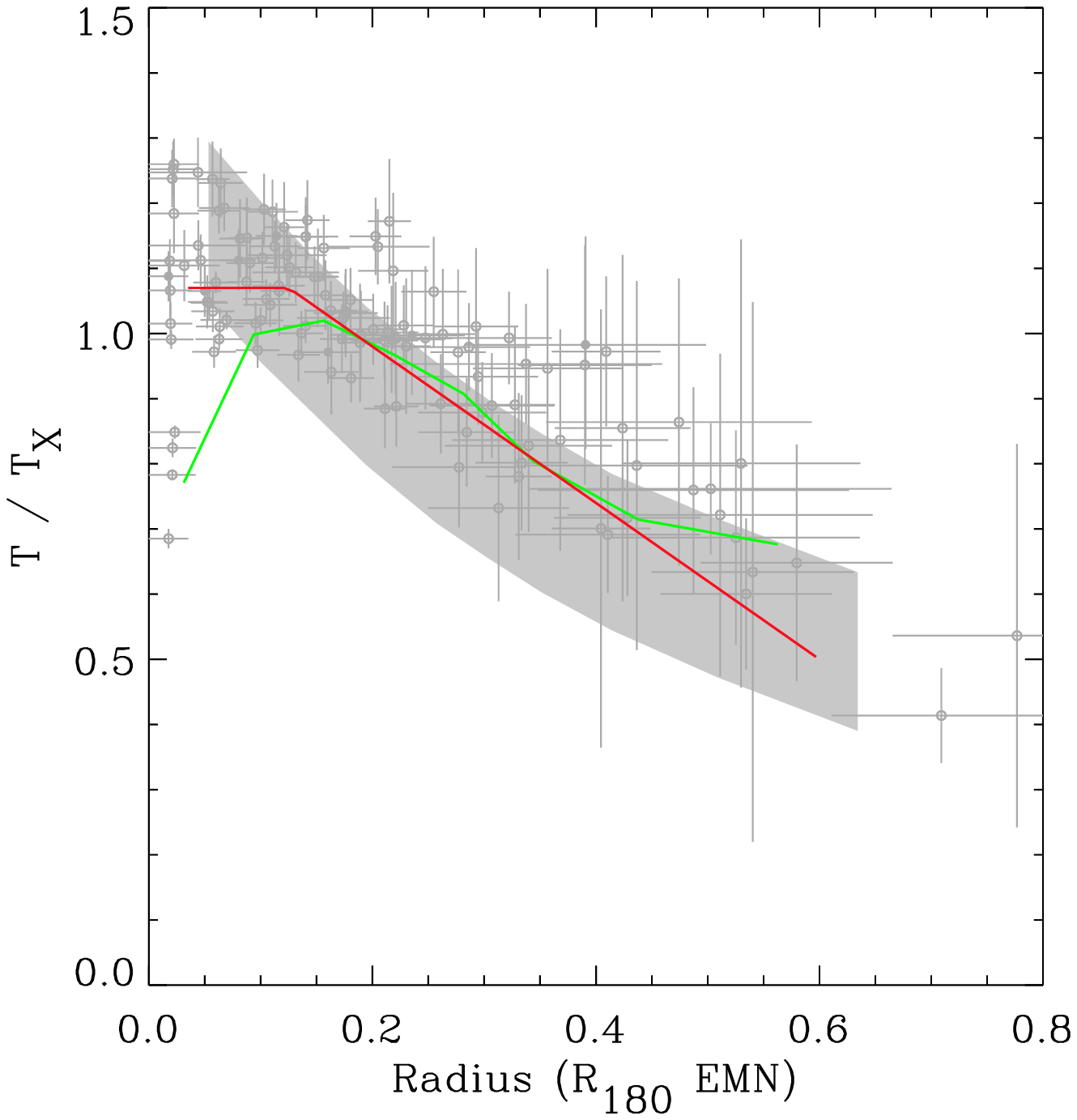}
\caption{{\footnotesize Scaled projected temperature profiles compared
    with the average profiles from {\it ASCA\/} \citep[][grey band]{mark98},
    {\it BeppoSAX\/} observations of cooling core clusters
    \citep[][green line]{dm02}, and {\it Chandra} observations of
  cooling core systems \citep[][red line]{vikh05}. The observed
  profiles have been scaled using $R_{180}$ derived from the
  simulations of \citet{emn}. ({\it This figure is available in colour
    in the online 
    version of the journal.})}}\label{fig:tpcomp} 
\end{centering}
\end{figure}

The largest cluster samples were assembled from {\it ASCA\/} and {\it
BeppoSAX\/} (\citealt{mark98} and \citealt{dm02}, respectively). More
recent investigations with \xmm \citep{piff} and {\it Chandra\/}
\citep{vikh05} have allowed better constraints to be put on the form
of the profiles of cool core clusters out to 0.4-$0.5\,\rv$. In
Fig.~\ref{fig:tpcomp} we show a comparison of our temperature profiles
with those from {\it ASCA}, {\it BeppoSAX} and {\it Chandra}.  We 
use the same average temperature as above to normalise the
temperatures, but as in these previous works, we scale the radial
coordinate to $R_{180}$ using the relation given in \citep{emn}.
There is good agreement between the profiles measured with
different instruments. There is a tendency for our temperature
profiles to scatter around the upper edge of the envelope of the {\it
ASCA\/} results. We note however that the same tendency is seen in the
{\it Chandra\/} observations \citep[see Fig.~16 of][]{vikh05}. 

%%================
%% Figure: temperature profile comparison
%%
%%================

It should be noted that the definitions of the global temperature
and/or the virial radius often differ between samples, making exact
comparison between different results rather difficult. We do not
compare with the \xmm results of \citet{piff} since their results are
quoted for $R_{180}$ measured from the data, rather than derived from
the relation of \citet{emn}. We also note that the normalisation of
the \citet{piff} profiles is $\sim 20$ per cent lower compared to our
results and to those from other satellites. It is possible that this
difference comes from their different definition of global temperature
(\citeauthor{piff} fit the emission-weighted bins outside the cooling
core with a constant temperature). However, the general declining
trend with radius is similar to their results.

Fitting the radial range $0.125 < \rv < 0.5$ with a simple linear
model we find,  

\begin{equation}
T/T_{X} = 1.19 - 0.74 R/\rv \label{eqn:tfit}
\end{equation}

\noindent for a fit with the BCES estimator. A  linear least
squares fit gives identical results. 

%%------------------------------------------

\subsection{Comparison with simulations}

Negative gradients of the temperature profiles on scales
$R>0.1\,R_{200}$ are naturally produced by cosmological hydrodynamical
simulations of galaxy clusters, quite independent of the details of
the physical processes included
\citep[e.g.,][]{emn,lew00,lok02,borg04,kay04}. \citet{mark98} and
\citet{dm02} compared observed temperature profiles from ASCA and
Beppo-SAX data, respectively, to the results from non--radiative
simulations by \citep{emn} and found a reasonable agreement in the
outer cluster regions. \citet{lok02} discussed a universal temperature
profile in their simulated clusters, whose shape agrees well with
observations outside the core regions. However, a number of authors
have shown that including radiative cooling in simulations causes a
substantial steepening of temperature profiles in the central cluster
regions \citep[e.g.,][]{lew00,valda,torna}. The resulting temperature
profiles are at variance with respect to the observed properties of
cool core clusters \citep[e.g.,][]{borg04}. This points towards the
need to introduce a suitable energy feedback scheme to regulate gas
cooling in the central regions \citep[e.g.,][]{kay04}.

%%================
%% Figure: temperature profile comparison with simulations
%%
\begin{figure*}
\begin{centering}
%%\hfill
\includegraphics[scale=1.,angle=0,keepaspectratio,width=\columnwidth]{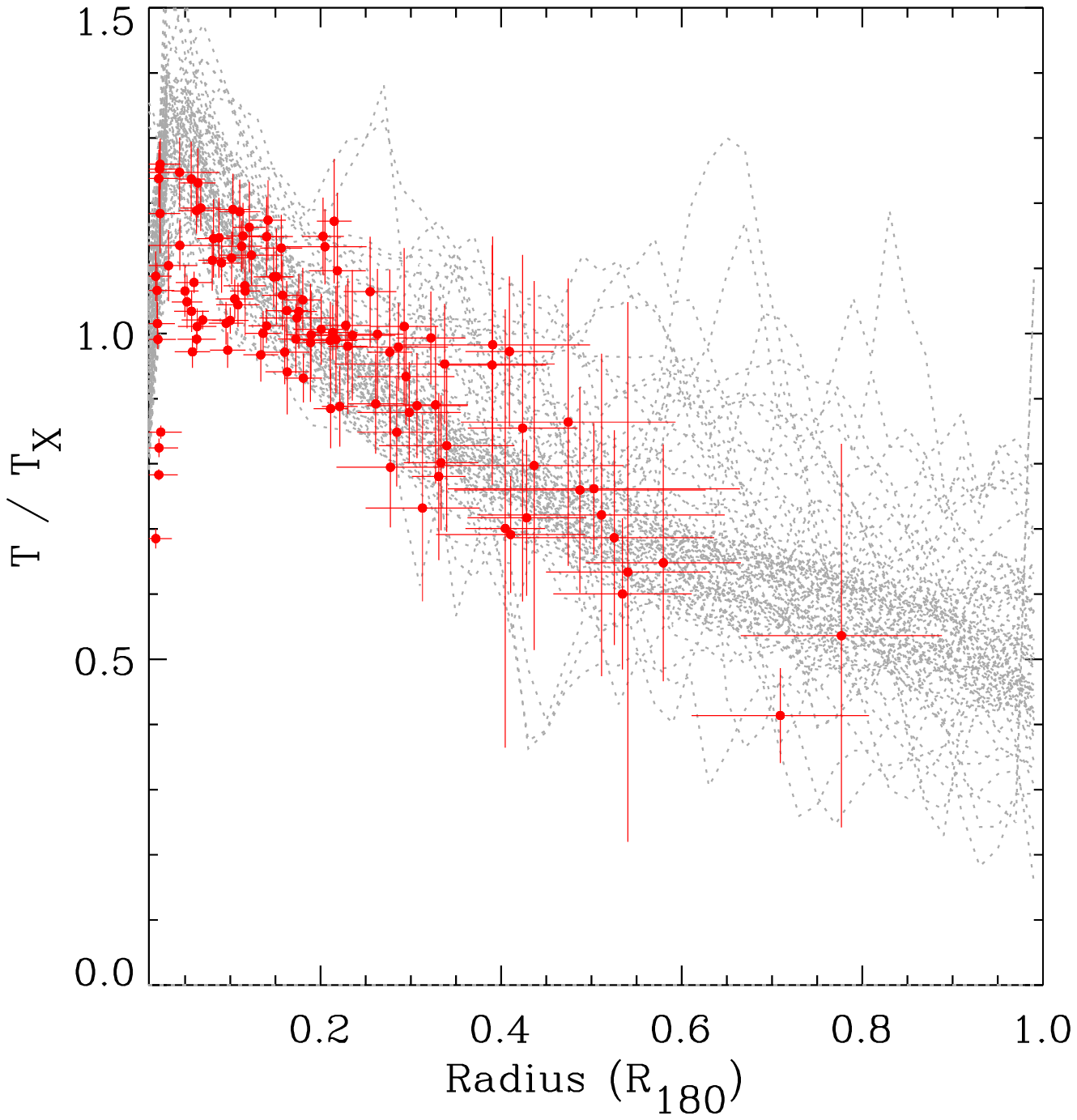}
\hfill
\includegraphics[scale=1.,angle=0,keepaspectratio,width=\columnwidth]{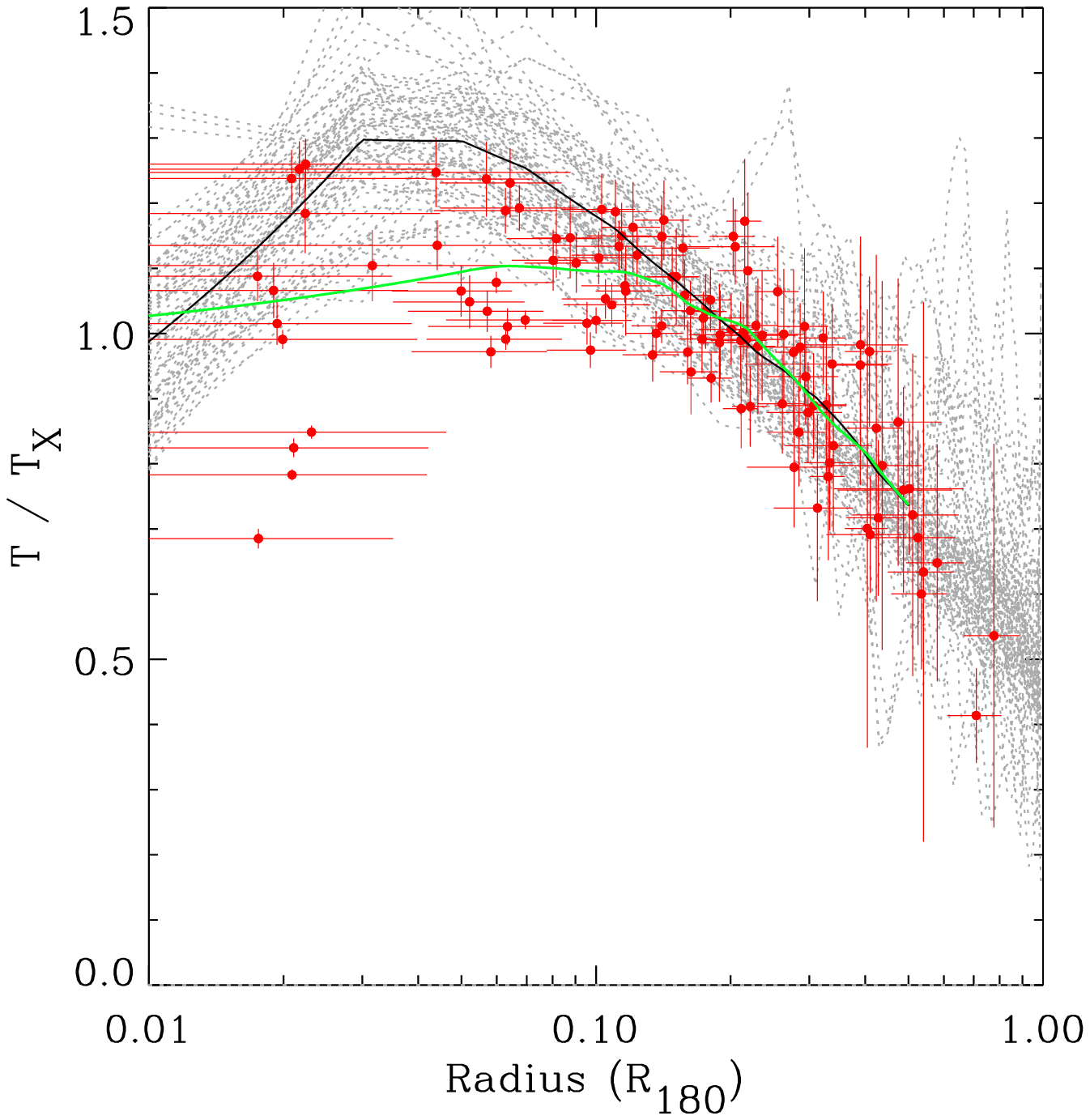}
\caption{{\footnotesize Scaled projected temperature profiles compared
    with the 
    projected profiles of all clusters with $\kT > 2$ keV from the simulations
    of \citet{borg04}. The mean profile of the observed and simulated
    profiles are shown by the black and green lines respectively. The
    observed profiles are scaled by the 
    measured spectral temperature in the $0.1-0.4\,\rv$ region. The
    simulated profiles are scaled using the mean emission weighted
    temperature, with a further adjustment of 8 per cent to take into
    account the difference between the two definitions of global
    temperature used to scale the profiles. See text for 
    details. ({\it This figure is available in
      colour in the online 
    version of the journal.}) }}\label{fig:tscomp}
\end{centering}
\end{figure*}
%%================

In Figure~\ref{fig:tscomp} we show a comparison of the observed
profiles with the projected simulated profiles of
all clusters with $\kT > 2$ keV from \citet{borg04}, in which the SPH
code GADGET-2 \citep{spr05} was used to simulate a concordance
$\Lambda$CDM model ($\Omega_M=0.3, \Omega_\Lambda=0.7, \sigma_8=0.8,
h=0.7$) within a box of $192\,h^{-1}$ Mpc on a side, using $480^3$
dark matter and an equal number of gas particles. The simulation
included radiative cooling, star formation and galactic ejecta powered
by supernova feedback. The observed profiles are scaled using the
spectral temperature $T_X$, measured as described above in
Sect~\ref{sec:sct}. The simulated profiles were scaled using the
emission weighted global temperature, with a further 8 per cent
adjustment so that the normalisation in the
$0.15 < R_{200} < 0.5$ region is the same as that of the observed
profiles (this adjustment is necessary because the emission weighted
global temperature is not the same as the measured spectral global
temperature $T_X$). Three projections are shown for each cluster. The
scatter in the 
simulated profiles is noticeable and comes from colder subclumps
accreting onto the main cluster and shock fronts due to supersonic
accretion. The simulated profiles decline continuously from a peak at
about $0.05\,R_{180}$. The mean observed and simulated profiles are
shown as black and green lines, respectively.

%% =====================
%% PR data
%%
\begin{table*}
\begin{minipage}{\textwidth}
\caption{{\footnotesize Cluster power ratios, calculated in an
    aperture corresponding to $R_{500}$, estimated from the $R$--$T$
    relation of 
    \citet{app}. Columns: (1) Cluster name; (2): $R_{500}$ in
    arcminutes; (3,5,7) power ratios; (4,6,8) $1\sigma$ errors on
    power ratios.}}\label{tab:PR}  
\centering
\begin{tabular}{l l r r r r r r }
\hline 
\hline
\multicolumn{1}{l}{RXCJ} & \multicolumn{1}{l}{$R_{500}$} & 
\multicolumn{1}{l}{$P_2/P_0$} & 
\multicolumn{1}{l}{$\sigma_{P_2/P_0}$} &
\multicolumn{1}{l}{$P_3/P_0$} &
\multicolumn{1}{l}{$\sigma_{P_3/P_0}$} & \multicolumn{1}{l}{$P_4/P_0$}
& \multicolumn{1}{l}{$\sigma_{P_4/P_0}$}\\ 

\hline
0003+0203 &  $9\farcm25$ & $1.06\times10^{-6}$ & $3.67\times10^{-8}$ & $ 8.16\times10^{-8}$ & $1.16\times10^{-8}$ & $  4.01\times10^{-8}$ & $4.66\times10^{-9}$ \\
0020-2542 &  $7\farcm32$ & $9.76\times10^{-7}$ & $3.11\times10^{-8}$ & $-3.48\times10^{-9}$ & $9.68\times10^{-9}$ & $ -1.15\times10^{-9}$ & $4.35\times10^{-9}$ \\
0547-3152 &  $7\farcm47$ & $8.22\times10^{-7}$ & $1.92\times10^{-8}$ & $ 1.04\times10^{-7}$ & $6.10\times10^{-9}$ & $  6.33\times10^{-8}$ & $2.50\times10^{-9}$ \\
0605-3518 &  $6\farcm59$ & $7.10\times10^{-7}$ & $9.90\times10^{-9}$ & $-2.70\times10^{-9}$ & $2.69\times10^{-9}$ & $  2.25\times10^{-9}$ & $1.21\times10^{-9}$ \\
1044-0704 &  $5\farcm91$ & $1.64\times10^{-6}$ & $7.88\times10^{-9}$ & $-1.61\times10^{-9}$ & $2.18\times10^{-9}$ & $  3.27\times10^{-10}$ & $8.32\times10^{-10}$ \\
1141-1216 &  $6\farcm55$ & $4.87\times10^{-7}$ & $1.28\times10^{-8}$ & $ 4.14\times10^{-8}$ & $3.69\times10^{-9}$ & $  2.22\times10^{-9}$ & $1.58\times10^{-9}$ \\
1302-0230 &  $9\farcm10$ & $7.97\times10^{-6}$ & $4.40\times10^{-8}$ & $ 2.76\times10^{-7}$ & $1.31\times10^{-8}$ & $  5.93\times10^{-8}$ & $5.88\times10^{-9}$ \\
1311-0120 &  $7\farcm23$ & $2.54\times10^{-7}$ & $4.35\times10^{-9}$ & $ 1.27\times10^{-8}$ & $1.05\times10^{-9}$ & $  5.36\times10^{-9}$ & $4.90\times10^{-10}$ \\
1516+0005 &  $7\farcm28$ & $2.63\times10^{-7}$ & $2.64\times10^{-8}$ & $ 7.24\times10^{-8}$ & $7.97\times10^{-9}$ & $  9.54\times10^{-8}$ & $3.35\times10^{-9}$ \\
1516-0056 &  $6\farcm69$ & $4.41\times10^{-6}$ & $6.54\times10^{-8}$ & $ 9.63\times10^{-7}$ & $2.20\times10^{-8}$ & $  8.05\times10^{-8}$ & $9.08\times10^{-9}$ \\
2023-2056 & $11\farcm74$ & $3.35\times10^{-7}$ & $1.22\times10^{-7}$ & $-3.43\times10^{-8}$ & $4.09\times10^{-8}$ & $  4.80\times10^{-8}$ & $1.95\times10^{-8}$ \\
2048-1750 &  $6\farcm08$ & $5.89\times10^{-6}$ & $5.53\times10^{-8}$ & $ 2.97\times10^{-7}$ & $1.51\times10^{-8}$ & $  1.46\times10^{6}$ & $6.70\times10^{-9}$ \\
2129-5048 & $10\farcm01$ & $7.18\times10^{-8}$ & $6.68\times10^{-8}$ & $ 3.22\times10^{-7}$ & $1.94\times10^{-8}$ & $ -5.76\times10^{-9}$ & $8.93\times10^{-9}$ \\
2217-3543 &  $6\farcm18$ & $5.33\times10^{-7}$ & $2.21\times10^{-8}$ & $ 3.25\times10^{-8}$ & $6.25\times10^{-9}$ & $  9.83\times10^{-9}$ & $2.61\times10^{-9}$ \\
2218-3853 &  $7\farcm28$ & $5.28\times10^{-6}$ & $1.40\times10^{-8}$ & $ 3.65\times10^{-8}$ & $4.37\times10^{-9}$ & $  1.68\times10^{-8}$ & $2.03\times10^{-9}$ \\

\hline
\end{tabular}
\end{minipage}
\end{table*}
%%================

%%

 There is relatively good agreement in the external regions, with the
 simulated profiles reproducing the observed scatter. In the central
 regions the peak of the simulated temperature profiles lies at $\sim
 0.04\,R_{180}$, in contrast to the observations, which show a less
 pronounced peak at $\sim 0.06\,R_{180}$, a point which is
 particularly evident from the mean profiles. In addition there
 appears to be considerably more dispersion in the observed profiles
 in the central regions. While admittedly our profiles are uncorrected
 for PSF and projection effects, we note that a similar difference in
 peak position (as compared to the simulations) is also evident when
 comparing with the {\it Chandra} results of \citet{vikh05}. Clearly,
 a more rigorous comparison would require estimating temperatures in
 the simulated clusters and rescaling their profiles in exactly the
 same manner as the data. Nevertheless, we believe that the agreement
 between simulated and observed clusters is quite good and lends
 support to the capability of numerical simulations to describe the
 global thermal structure of the ICM.

\subsection{Dependence on cluster morphology/dynamical state}

It is interesting to investigate whether there are correlations
between the form of the temperature profile and the morphology or
dynamical state of the cluster. To this end, we make a preliminary
investigation using the power ratio method of \citet{bt} to
characterise the morpho-dynamical state of the objects in our sample.

%%================
\begin{figure*}
\begin{centering}
\includegraphics[scale=1.,angle=0,keepaspectratio,width=0.32\textwidth]{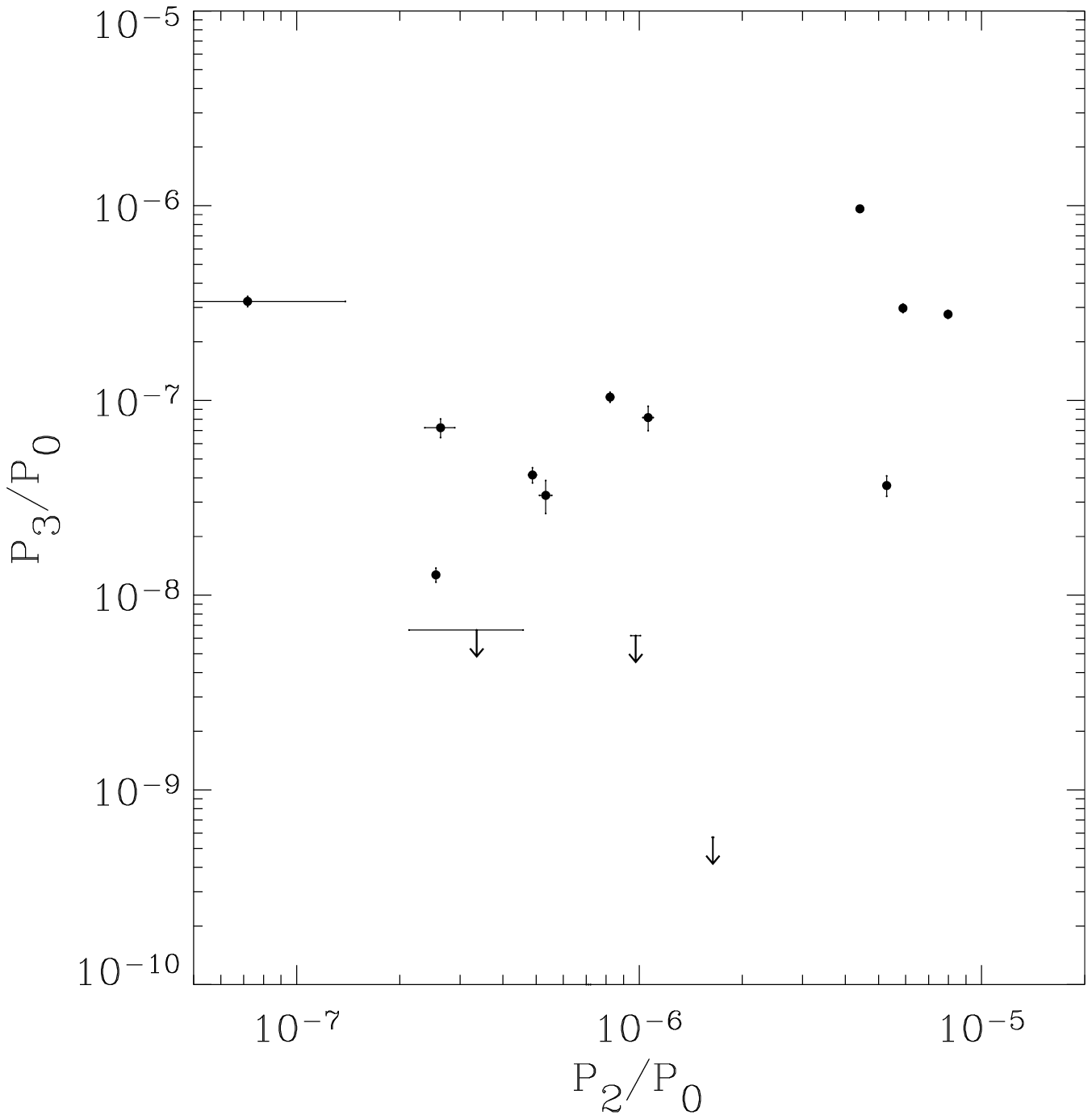}
\hfill
\includegraphics[scale=1.,angle=0,keepaspectratio,width=0.32\textwidth]{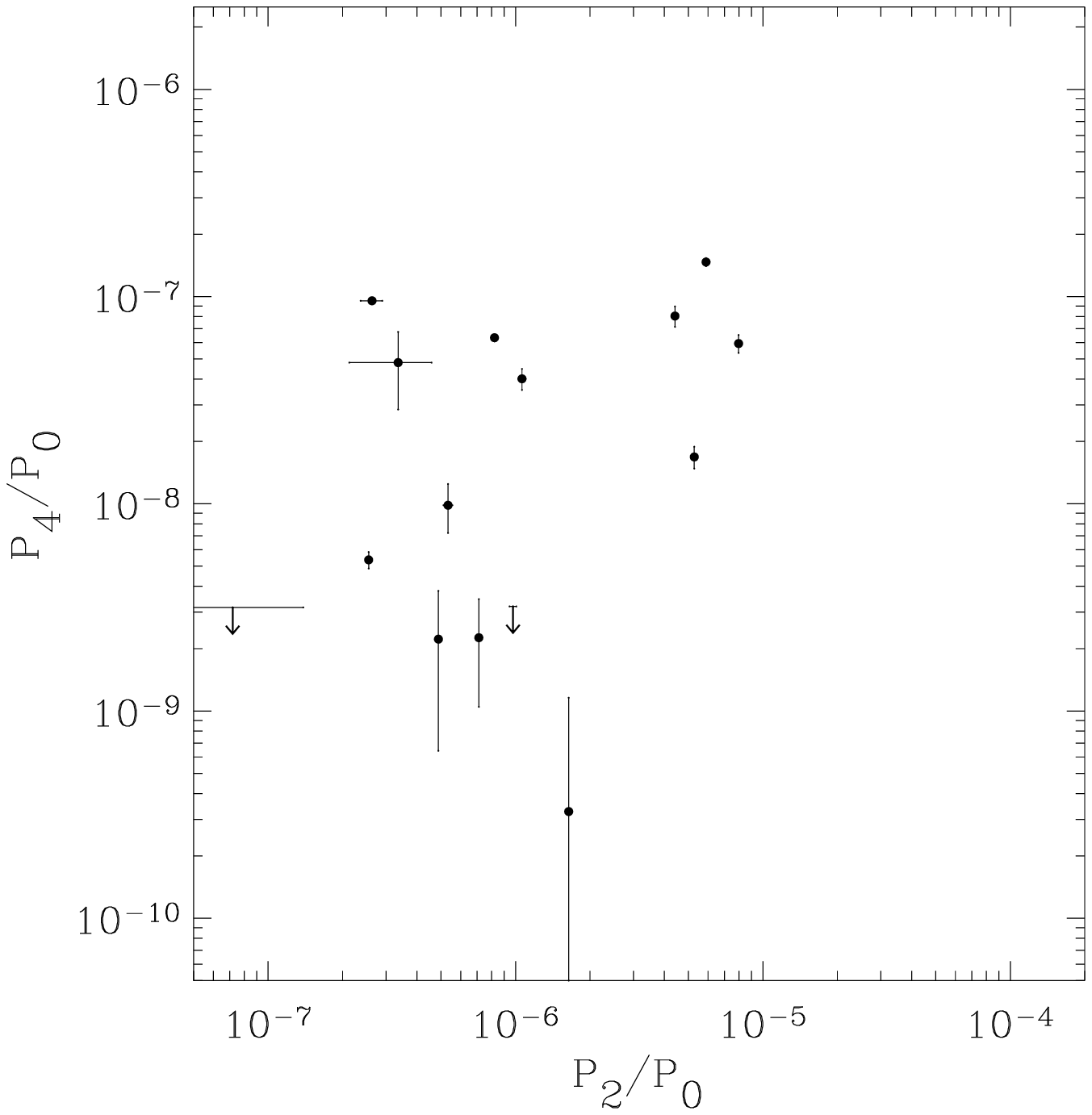}
\hfill
\includegraphics[scale=1.,angle=0,keepaspectratio,width=0.33\textwidth]{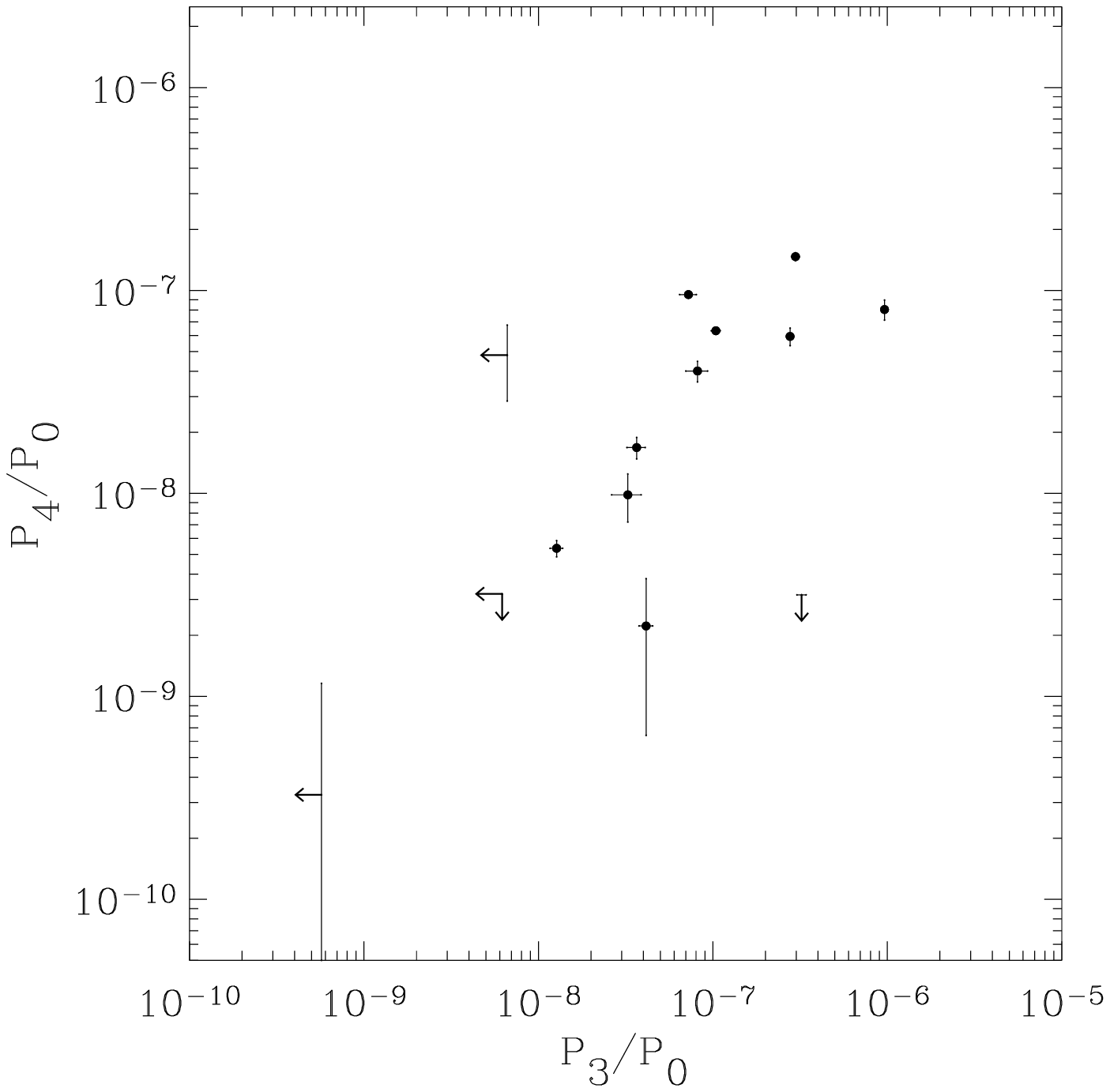}
\caption{{\footnotesize Power ratios. The
      power ratios are computed in an aperture corresponding to
      $R_{500}$ estimated using the $R$--$T$ relation of
      \citet{app}. See text for details.  }}\label{fig:PR}
\end{centering}
\end{figure*}
%%================

%
%%------------------------------------------

\subsubsection{Power ratio calculation}

To obtain a quantitative estimate of substructure in the surface
brightness distribution of the clusters, we applied the analysis
method proposed by \citet{bt}. In this method, the projected mass
distribution is associated with the X-ray surface brightness; a
multipole expansion of the X-ray image then yields a similar expansion
of the gravitational potential. The multipole analysis provides a
measure of the 'power' of each multipole component to the X-ray image
of the cluster in absolute units. To obtain a measure that is
independent of the cluster X-ray luminosity, the 'power' terms are
scaled by the zeroth order (monopole) moment and are consequently
called 'power ratios'. The method was
recently used to compare substructure in nearby and distant cluster
samples observed with {\it Chandra\/} \citep{jelt}. 

The method is applied within an aperture radius as described in
\citet{bt}. We used a minimization of the first (dipole) moment to
obtain an independent centering of the cluster emission within the
aperture. The lowest order power ratios which are of interest for our
study are $P_2/P_0$, $P_3/P_0$, and $P_4/P_0$, which
correspond to the quadruple, the hexapole, and the octopole moments.
Due to the nature of the moment functions, a large weight is given to
the outer parts within the aperture, in particular for the higher
moments. Thus the results depend somewhat on the choice of
aperture. We have explored this effect with a range of aperture radii,
results from which will be described in a future paper. For the
purposes of this initial investigation, we estimate 
the power ratios within $R_{500}$, where this radius is estimated
using the $R$--$T$ relation of \citep{app}. 

%%================
\begin{figure*}
\begin{centering}
\includegraphics[scale=1.,angle=0,keepaspectratio,width=0.32\textwidth]{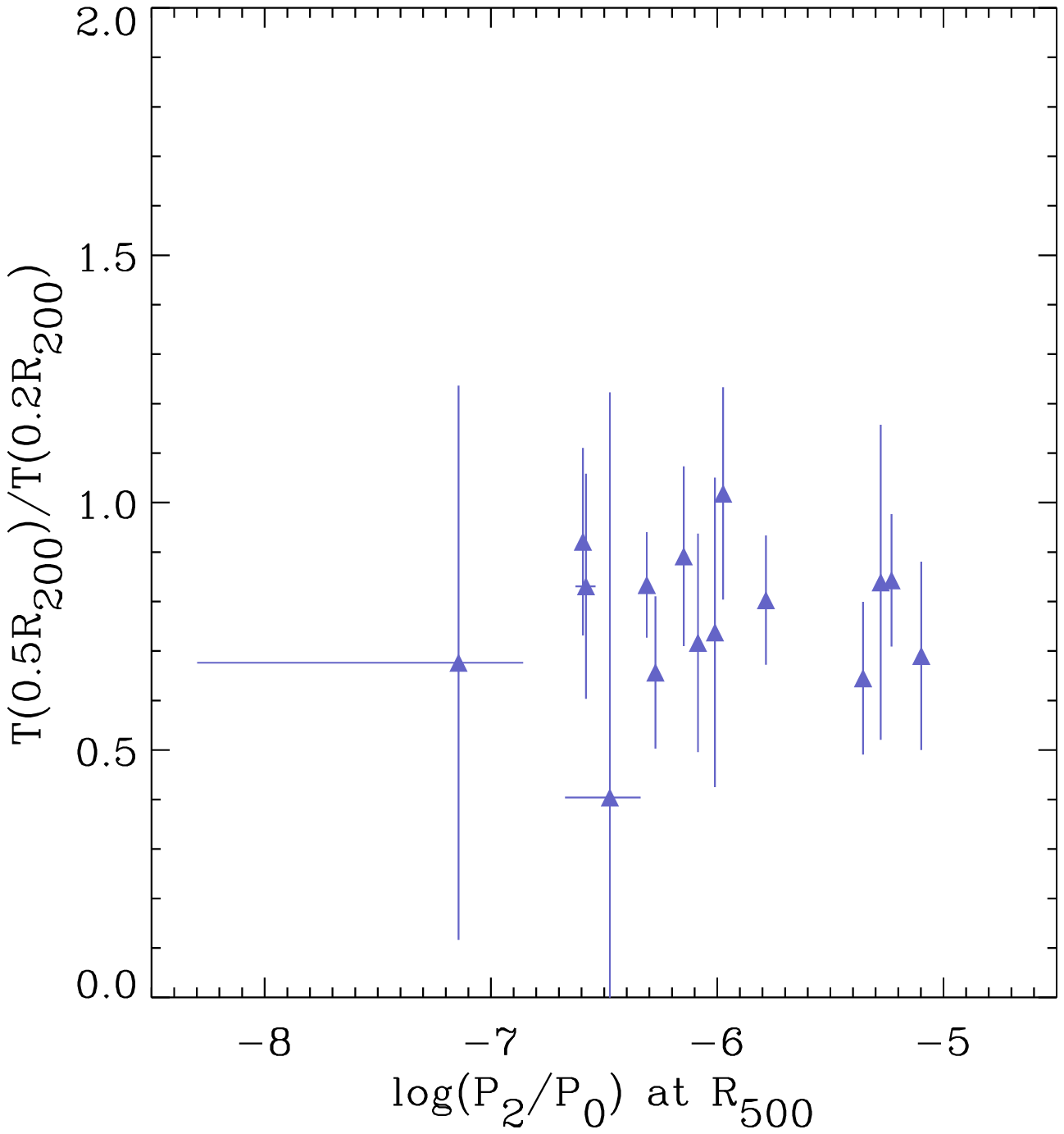}
\hfill
\includegraphics[scale=1.,angle=0,keepaspectratio,width=0.32\textwidth]{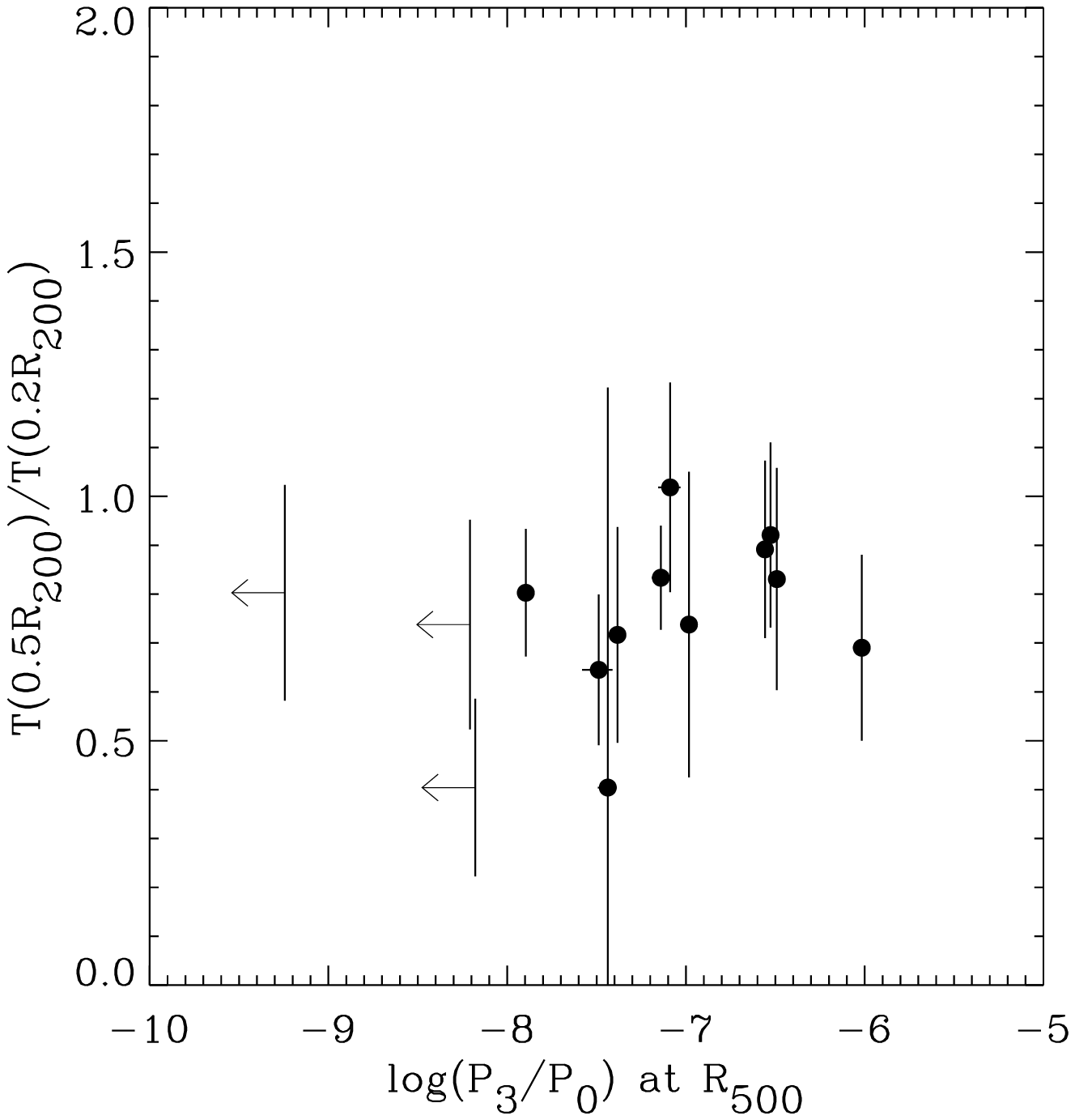}
\hfill
\includegraphics[scale=1.,angle=0,keepaspectratio,width=0.32\textwidth]{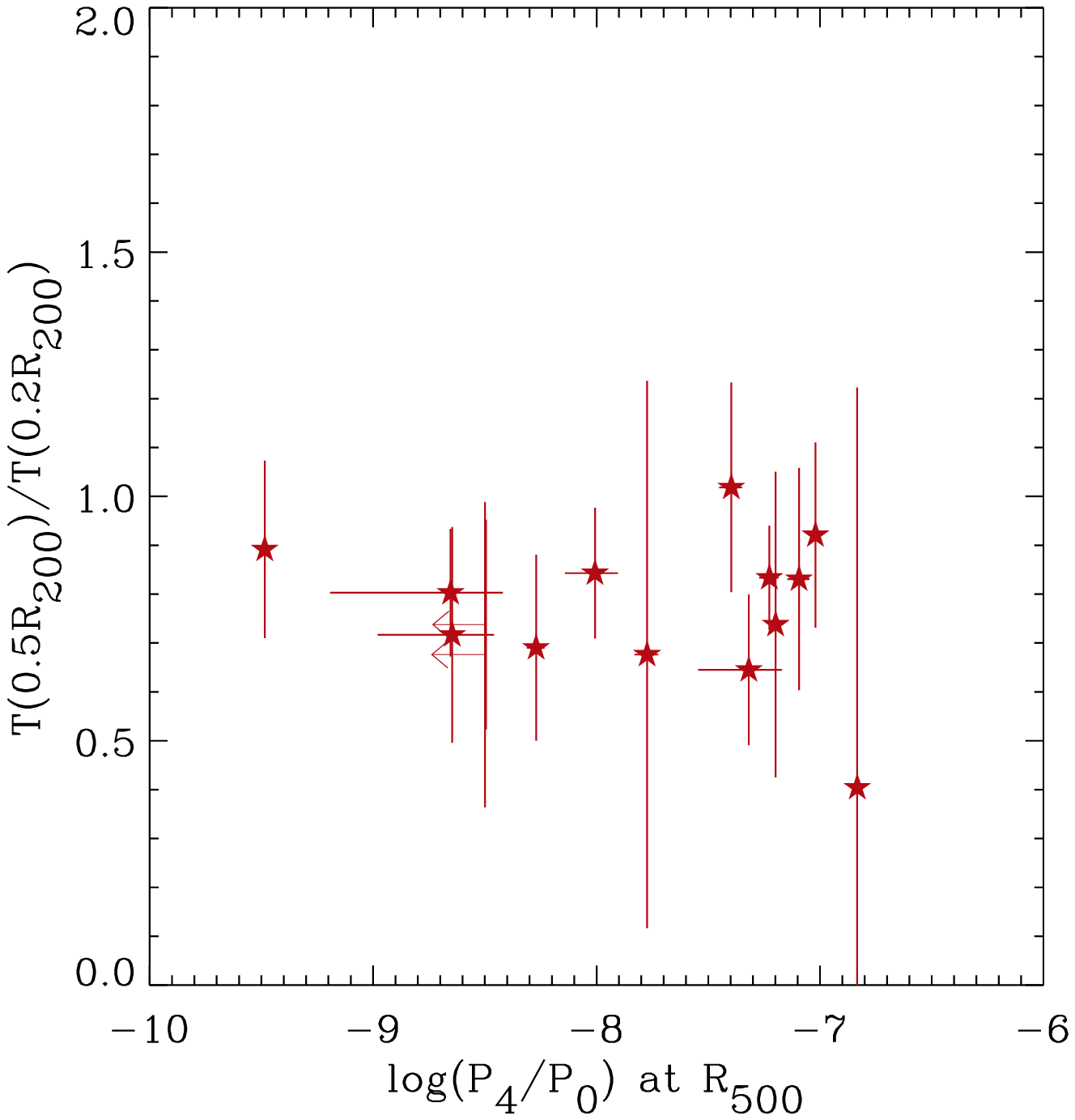}
\caption{{\footnotesize Scatter plots of $T (0.2_{\rv})/T (0.5_{\rv})$
    (a measure of the temperature profile slope) with power
    ratio. There are no obvious correlations.}}\label{fig:PC} 
\end{centering}
\end{figure*}
%%================
%%================
\begin{figure*}
\begin{centering}
\includegraphics[scale=1.,angle=0,keepaspectratio,width=0.32\textwidth]{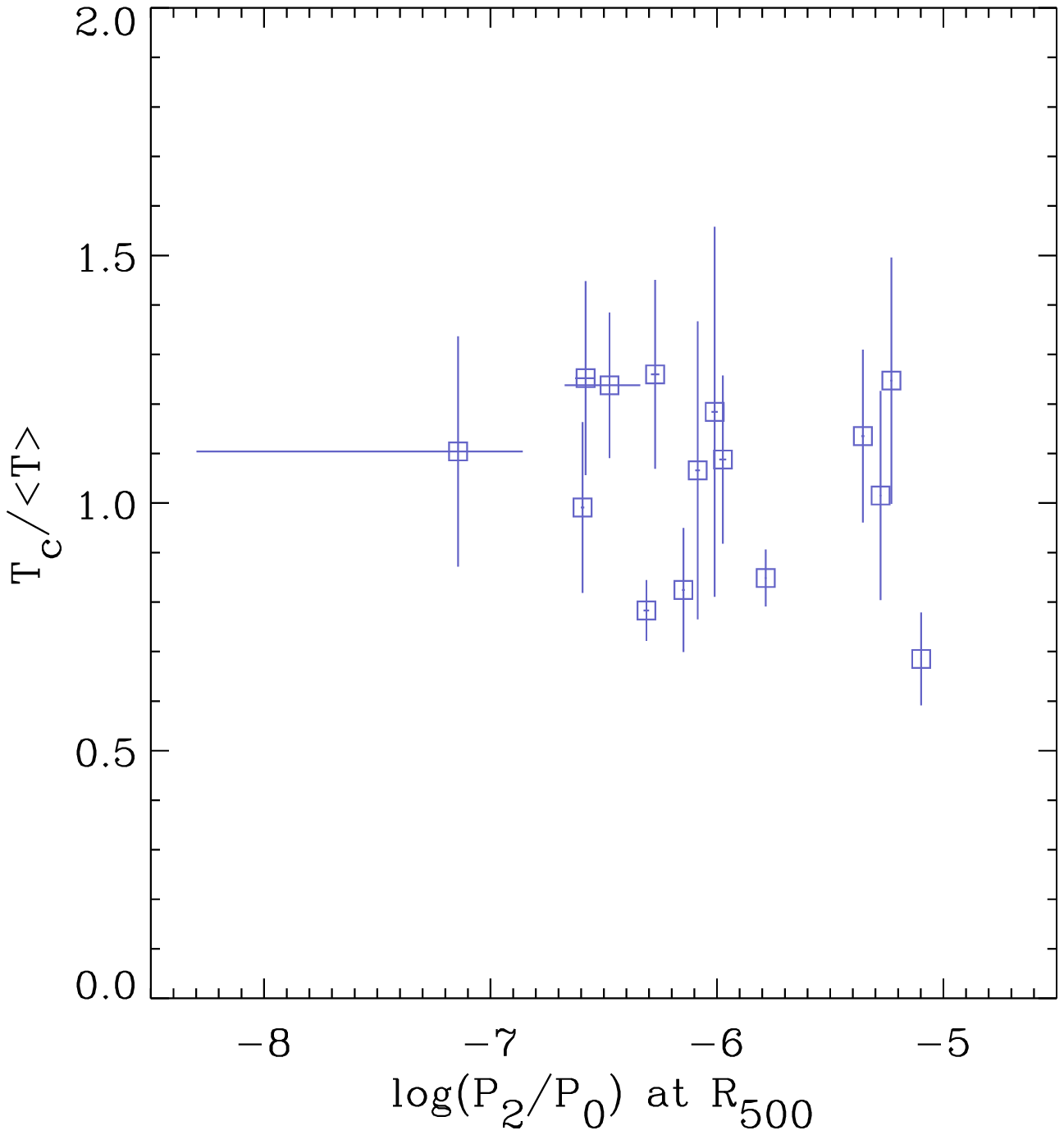}
\hfill
\includegraphics[scale=1.,angle=0,keepaspectratio,width=0.32\textwidth]{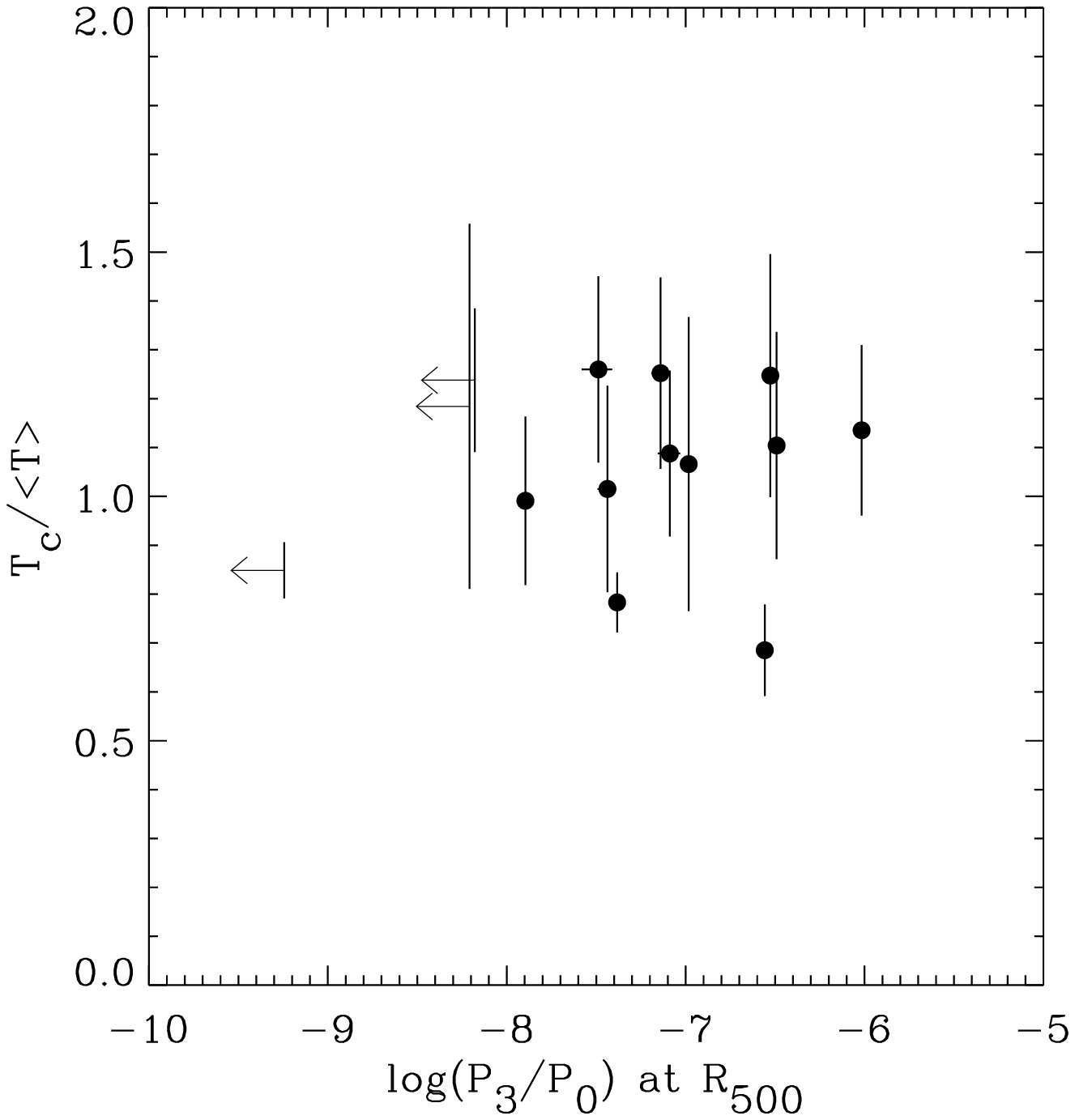}
\hfill
\includegraphics[scale=1.,angle=0,keepaspectratio,width=0.32\textwidth]{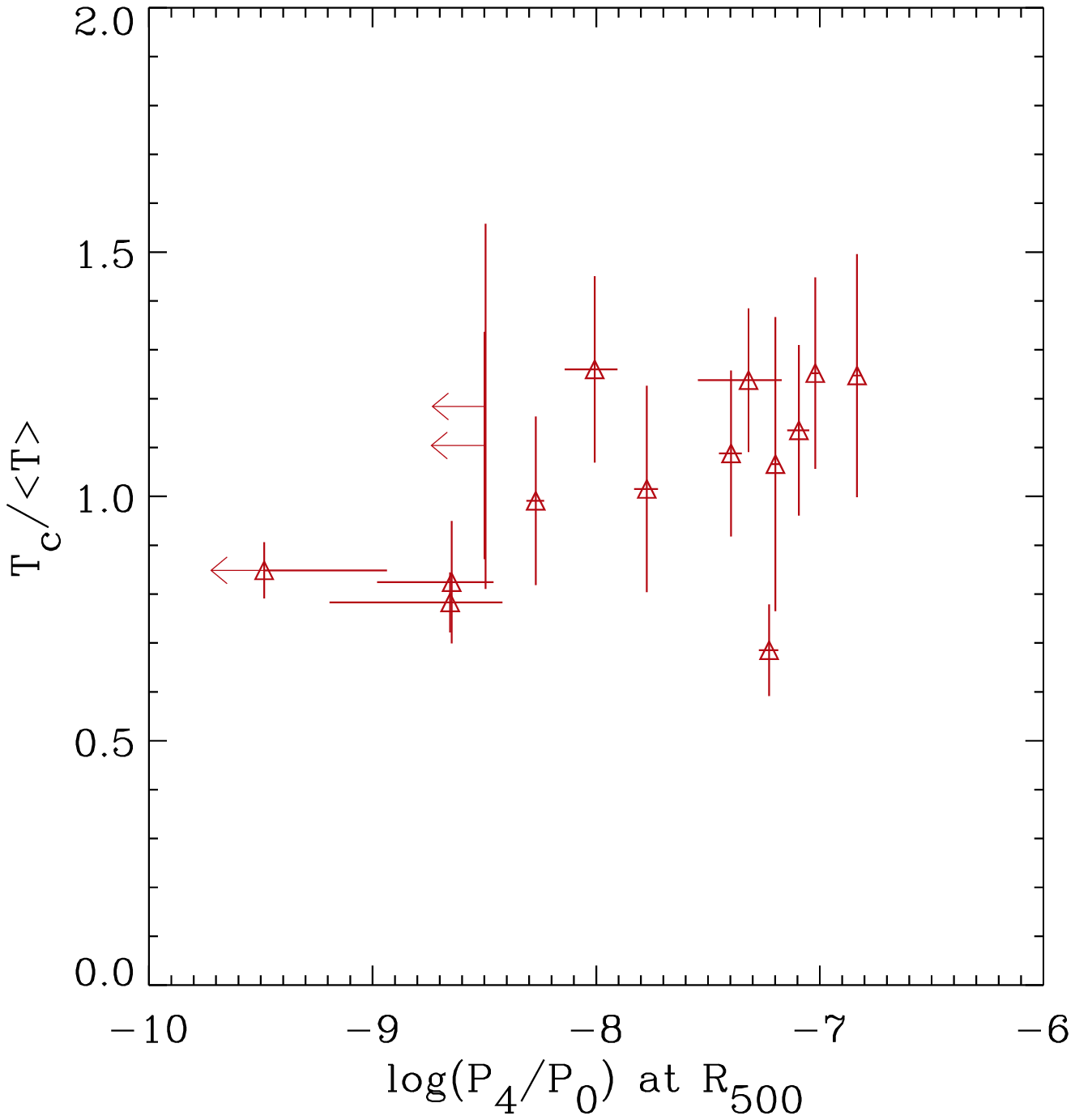}
\caption{{\footnotesize Scatter plots of the (projected) central
    temperature $T_c$ 
    divided by the mean spectroscopic temperature in the $0.1\,\rv
    \leq r \leq 0.4\,\rv$ region (a measure of the central temperature
    dip) with power ratio. }}\label{fig:PC2} 
\end{centering}
\end{figure*}
%%================

Systematic effects and uncertainties for each power ratio were also
taken into account. We created 1000 simulations of each cluster where
the image pixels were azimuthally randomized around the predetermined
cluster centre.  The power ratio signal measured for these simulated
clusters should be solely due to shot noise, providing a measure of
the accuracy for rejection of the hypothesis that the cluster has no
structure. We therefore subtract the
mean of the simulated power ratio signal from the result obtained for
the observed cluster and use the dispersion of the simulated results
as an approximation of the error of the power ratio. We defer a more
precise treatment to a forthcoming paper. The results shown
in Fig.~\ref{fig:PR} demonstrate that we have sufficient photon
statistics to produce robust results with small uncertainties, except for
those cases where the clusters have a highly symmetric appearance. In
general the results for the power ratios correspond very well to the
visual impression of the cluster, where in particular $P_2/P_0$ can
easily be identified with the cluster ellipticity. The parameter
$P_3/P_0$ provides then the best measure for further substructure and
since some ellipticity can be related to the quasi-equilibrium state
of the cluster, the third moment provides often the strongest
indication for a deviation from a relaxed dynamical state.

Power ratio values and $1\sigma$ errors are listed in
Table~\ref{tab:PR}. The power ratios for the clusters in our sample in
general occupy a similar range of values to those derived for a nearby
cluster sample by \citet{jelt}. Plots of $P_2/P_0$ vs $P_3/P_0$,
$P_2/P_0$ vs $P_4/P_0$ and $P_3/P_0$ vs $P_4/P_0$ are shown in
Fig.~\ref{fig:PR}. Of particular note is the strong correlation
in $P_4/P_0$ vs $P_3/P_0$.

\subsubsection{Preliminary comparison with power ratio}

We can check to see if there are any correlations between the power
ratio value and the shape of the temperature profile. We parameterise
the outer temperature slope by measuring the ratio between the
temperature at $0.5\,\rv$ and the temperature at $0.2\,\rv$ ($T
(0.5_{\rv})/T (0.2_{\rv})$. These values are calculated by spline
interpolation and are extrapolated if necessary. In Fig.~\ref{fig:PC}, we
show a scatter plot of the outer slope parameter and each of the power
ratios. We then tested for correlations between $T (0.5_{\rv})/T
(0.2_{\rv})$ and power ratio in the linear-log plane. We calculate the
generalised Kendall's $\tau$ correlation coefficient \citep{ifn},
appropriate for censored data, for each scatter plot. The probability
that a correlation is 
{\it not\/} present is 80, 66 and 88 per cent for $T (0.5_{\rv})/T
(0.2_{\rv})$ vs $P_2/P_0$, $P_3/P_0$ and $P_4/P_0$, respectively.
Given the wide range of morphologies present in the sample, the lack
of significant correlation suggests that, in general, the
morphology/dynamical state  
does not have a significant impact on the {\it outer} slope of the
azimuthally averaged temperature profile, at least at the
currently-available resolution.

Another interesting question is whether there is a correlation between
the presence of a central temperature dip and the power ratio. In
Fig.~\ref{fig:PC2} we show a scatter plot of the ratio of the central
temperature, $T_c$ (the temperature of the first bin in the
temperature profile) to the mean spectroscopic temperature $T_X$, and
each of the power ratios. Calculating the generalised Kendall's $\tau$
correlation coefficient for each scatter plot, we find a probability
that a correlation is {\it not} present of 55, 87 and 9 per cent for
$T_c/T_X$ vs $P_2/P_0$, $P_3/P_0$ and $P_4/P_0$, respectively. Thus
there is evidence for a weak correlation of central temperature drop
with $P_4/P_0$, in the sense that clusters with smaller $P_4/P_0$ are
more likely to have a central temperature drop.
We do not yet have the full sample of clusters from which correlations
can be derived, nor have the temperature profiles been corrected for
PSF and projection effects, which will enhance the observed central
gradients to some degree. 

It should be noted that the spatial resolution
of these and other recent temperature profiles, particularly at large
radius, while much improved over results from previous satellites, is
still the limiting factor when looking for correlations, or comparison
between different cluster subsamples. This will also have a bearing on
comparisons with numerical simulations. 

%%%%%%%%%%%%%%%%%%%%%%%%%%%%%%%%%%%%%%%%%%%%%%%%%%%%%%%%%%%%%%%%%%%%%

\section{Conclusions}

We have used \xmm observations of 15 clusters drawn from a
statistically representative, luminosity-selected 
sample to investigate the behaviour of the temperature profiles. The
clusters range from morphologically relaxed looking objects with
strong central surface brightness peaks (e.g., RXC\,J1044\,-0704), to
diffuse structures with significant amounts of surrounding
substructure (e.g., RXC\,J1516\,+0056), and constitute a
representative subsample. We find that, once scaled
appropriately, the temperature profiles are similar in the
radial range from $0.1\,\rv$ to $0.5\,\rv$, declining steadily from
the central regions to the outer boundary of the measurements with a
relative dispersion of $\sim 10$ per cent out to $0.5\,\rv$. The
region interior to $0.1\,\rv$ is the region of greatest scatter in the
scaled profiles: the relative scatter of $\sim 25$ per cent is likely a
lower limit. A preliminary comparison with numerical simulations
shows relatively good agreement outside $\sim 0.1\,\rv$, with all of
the measured temperature profiles falling within the scatter of the
simulated profiles. 

Calculating power ratios for the sample, we investigate whether there
are correlations between the power ratio measured in an aperture
corresponding to $R_{500}$ and the shape of the
temperature profile. We characterise the temperature profile shape in
two ways: with the ratio $T (0.5\,\rv)/T 
(0.2\,\rv)$, a measure of the outer slope, and with the ratio
$T_c/\langle T \rangle$, a measure of the central temperature drop.
There is no obvious correlation of outer slope with power ratio;
neither is there a correlation of central temperature dip with
$P_2/P_0$ or $P_3/P_0$. There is evidence for a weak correlation of the
central temperature dip with $P_4/P_0$. The analysis thus suggests
that the outer slope of the temperature profile is not particularly
sensitive to the morpho-dynamical state, although there may be some
correlation with the existence of a central
temperature drop. Further investigation with power ratios evaluated in
other apertures, for the entire sample, should be undertaken before
definitive conclusions can be drawm. 

The overall conclusion from this work on a statistically
representative sample indicates that clusters are a relatively regular
population, at least outside the cool core regions, with the caveat
that comparisons between samples or with simulations is limited by the
available temperature profile resolution. The observed similarity in
density \citep{neuarn,croston} and temperature profiles \citep[][this
work]{mark98,dm02,piff,vikh05} indicates both similarity in the
underlying gravitational mass distribution (such as has already been
seen in the X-ray mass profiles of morphologically relaxed clusters,
e.g., \citealt*{point}), and similarity in the entropy of the ICM
(such as has been seen by e.g., \citealt*{psf,pap}). In this case a
single integrated temperature, excluding the core region, should be a
good proxy for the total mass. The observed regularity thus has
important implications for the use of clusters as cosmological probes.

In future papers, we will reinvestigate the trends with the full
sample, make maps of quantities such as temperature, entropy and
pressure, and estimate the mass, baryon fraction and entropy in the
clusters. A more extensive comparison with numerical simulations will
also be undertaken.

\begin{acknowledgements}

We thank E. Belsole, J.P. Henry, K. Pedersen, T.J. Ponman,
T.H. Reiprich, A.K. Romer, P. Schuecker for useful discussions, and
the anonymous referee for comments on the paper. GWP acknowledges partial 
support from a Marie Curie Intra-European Fellowship under the FP6
programme (Contract No. MEIF-CT-2003-500915). The present work is
based on observations obtained with {\it XMM-Newton}, an ESA science
mission with instruments and contributions directly funded by ESA
Member States and the USA (NASA). The \xmm project is supported
in Germany by the Bundesministerium f\"ur Wirtschaft und
Technologie/Deutsches Zentrum f\"ur Luft- und Raumfahrt (BMWI/DLR, FKZ
50 OX 0001), the Max-Planck Society and the Heidenhain-Stiftung.

\end{acknowledgements}

\appendix
\section{Individual cluster profiles}
\label{apx:tprofs}

\subsection{RXC\,J0003\,+0203}

RXC\,J0003\,+0203, also known as Abell 2700, has an average temperature
of $\kT= 3.8$ keV and lies at $z=0.092$. The cluster presents a 
symmetric X-ray image but does not possess a strong central surface
brightness peak. After renormalisation of the quiescent 
background, the spectra extracted in the external region
($r > 11\arcmin$) can be adequately fitted with a {\sc MeKaL} model at
0.23 keV with negative normalisation. An additional power-law is
required for the EMOS2 and EPN spectra.

\begin{figure*}
\begin{centering}
\includegraphics[scale=0.30,angle=0,keepaspectratio,width=0.33\textwidth]{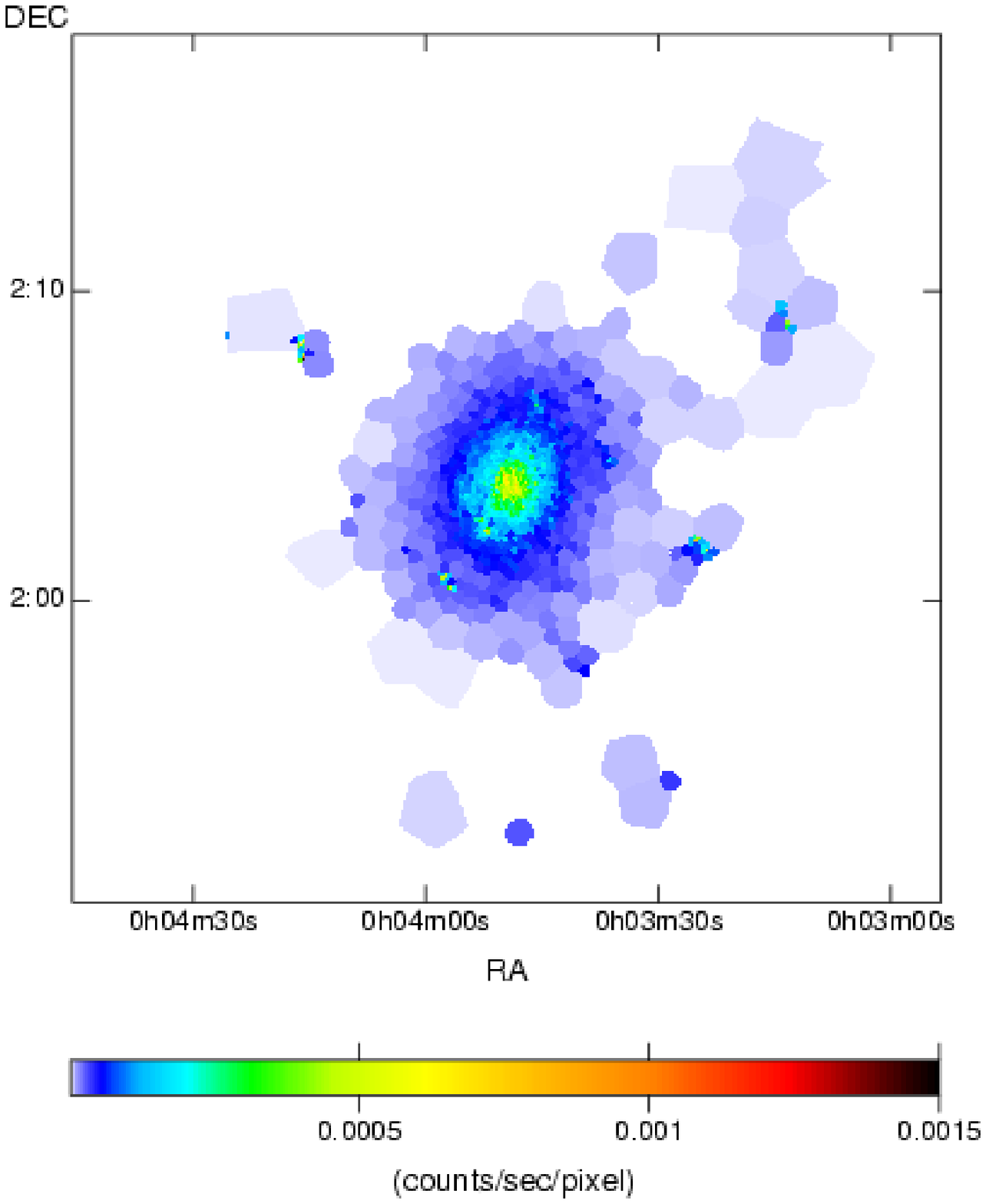}
\hfill
\includegraphics[scale=0.30,angle=0,keepaspectratio,width=0.33\textwidth]{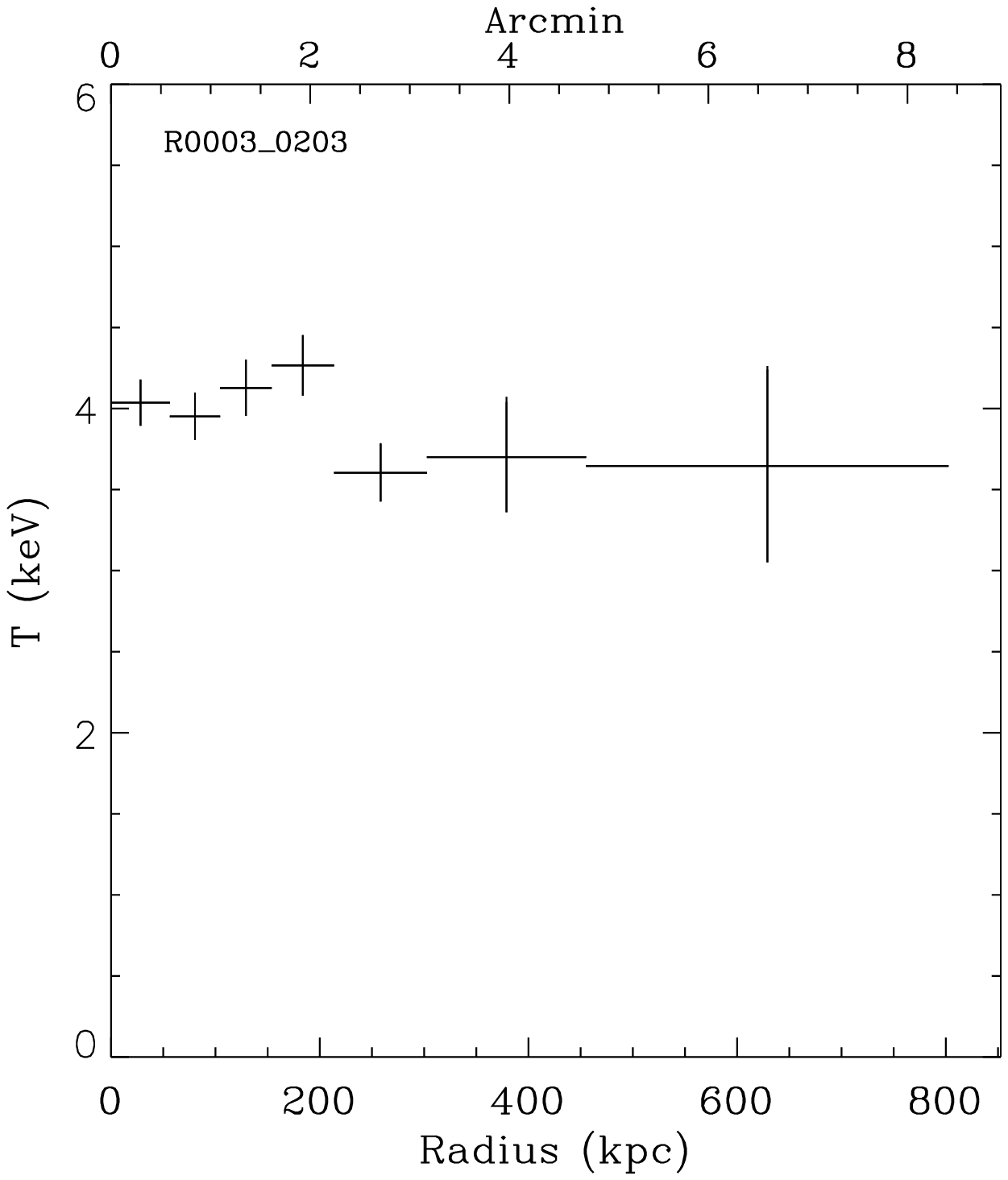}
\hfill
\includegraphics[scale=0.30,angle=0,keepaspectratio,width=0.33\textwidth]{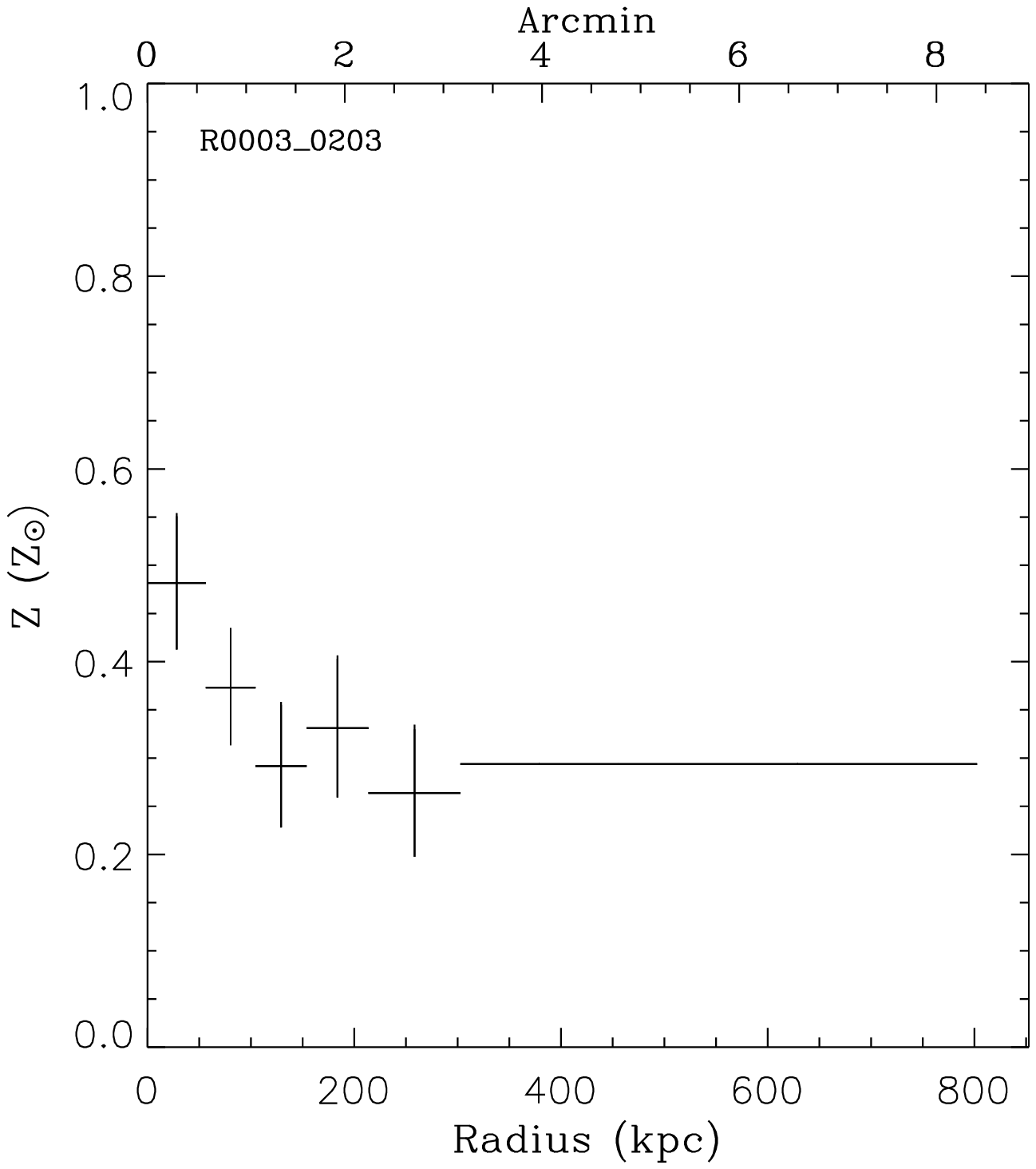}
\caption{{\footnotesize Image (left), temperature (middle) and
    abundance (right) profiles of RXC\,J0003\,+0203. In this and the
    following images, the colour scale is square root with a maximum
    of 0.0015 counts per second (enabling easy comparison between
    images). The angular 
    size of each image has been chosen to approximately match the
    virial radius of the cluster as determined from the average
    temperature $T_X$ (as defined in Sect.~\ref{sec:sct}) and the
    $R$--$T$ relation of \citet{app}. The horizontal line without
    errors in the abundance profile plots denotes the regions where
    the abundance was frozen. ({\it Figures are available in colour
      in the online 
    version of the journal.})}}\label{fig:R0003}
\end{centering}
\end{figure*}

The temperature and metallicity profiles are shown in
Fig.~\ref{fig:R0003}. The temperature profile is consistent with
being isothermal at large radii, although given the $\sim 35\%$
uncertainties in the final bin, a decline cannot be ruled out. The
metallicity profile is highly peaked, declining smoothly from $Z \sim
0.5 \zs$ in the central bin to $Z \sim 0.25 \zs$ at large radii. While
the increase in metallicity towards the centre is reminiscent of a cooling
core \citep{dm02}, the temperature profile does not show a significant
central decline, at least at the resolution of these data. 

%%================
%% Figure: R0003
%%
%%================

\subsection{RXC\,J0020\,-2542}

Moderately luminous, lying at $z=0.141$ and with a temperature of $\kT
= 5.7$ keV, this cluster is also known as
Abell 22. The X-ray image is highly disturbed, with a prominent
surface brightness edge to the N, and an emission extension to the
S. The overall X-ray emission is aligned approximately in the
direction of the line joining the two brightest cluster galaxies. The
annular regions were centred on the surface brightness peak, which
lies on the surface brightness edge and corresponds to the position of
the BCG.  The residual spectrum shows negative residuals and is
adequately described with a combination of a {\sc MeKaL} model at 0.10
keV with negative normalisation, with an additional power-law
component for the EMOS2 and EPN.  

\begin{figure*}
\begin{centering}
\includegraphics[scale=0.30,angle=0,keepaspectratio,width=0.33\textwidth]{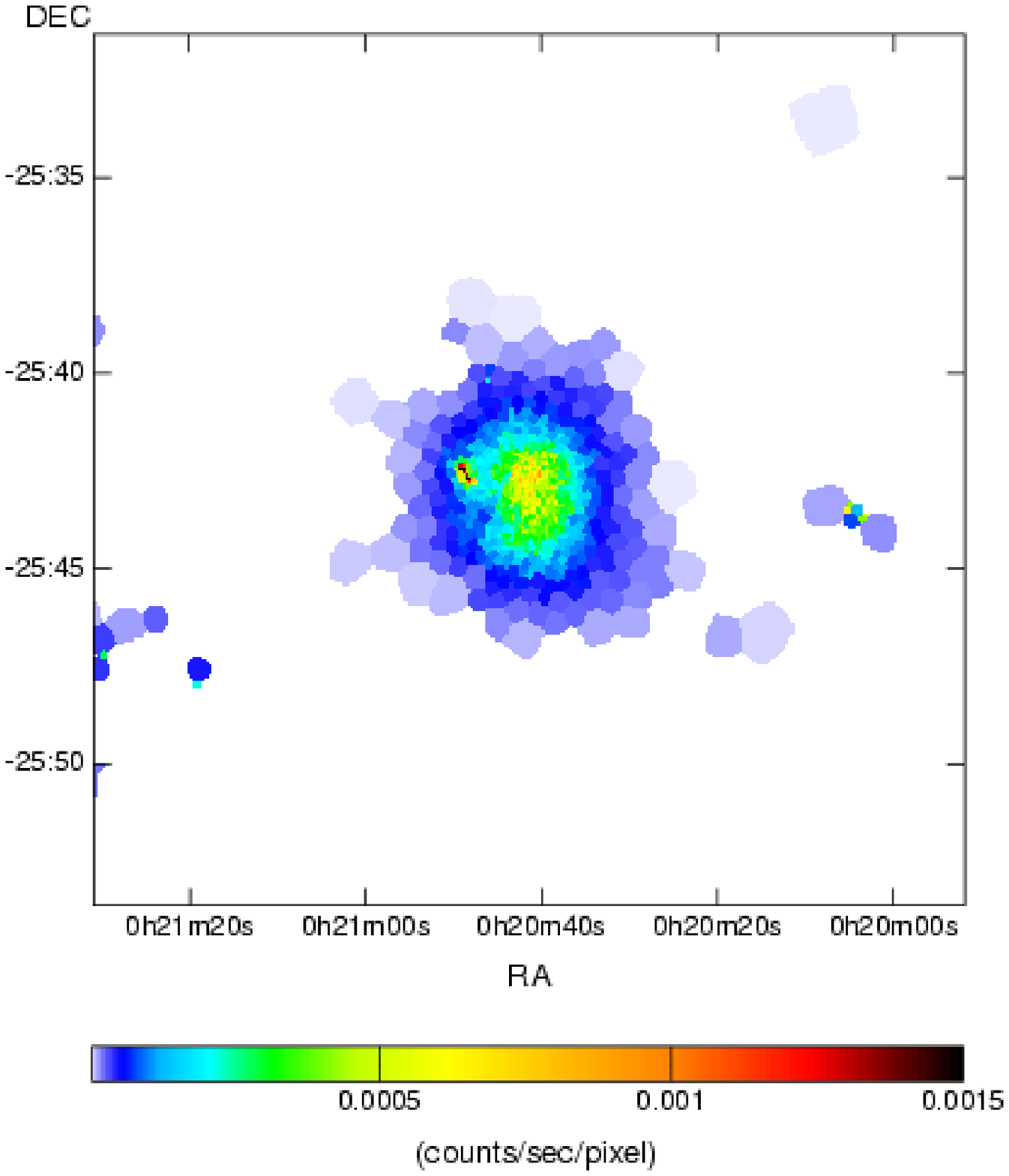}
\hfill
\includegraphics[scale=0.30,angle=0,keepaspectratio,width=0.33\textwidth]{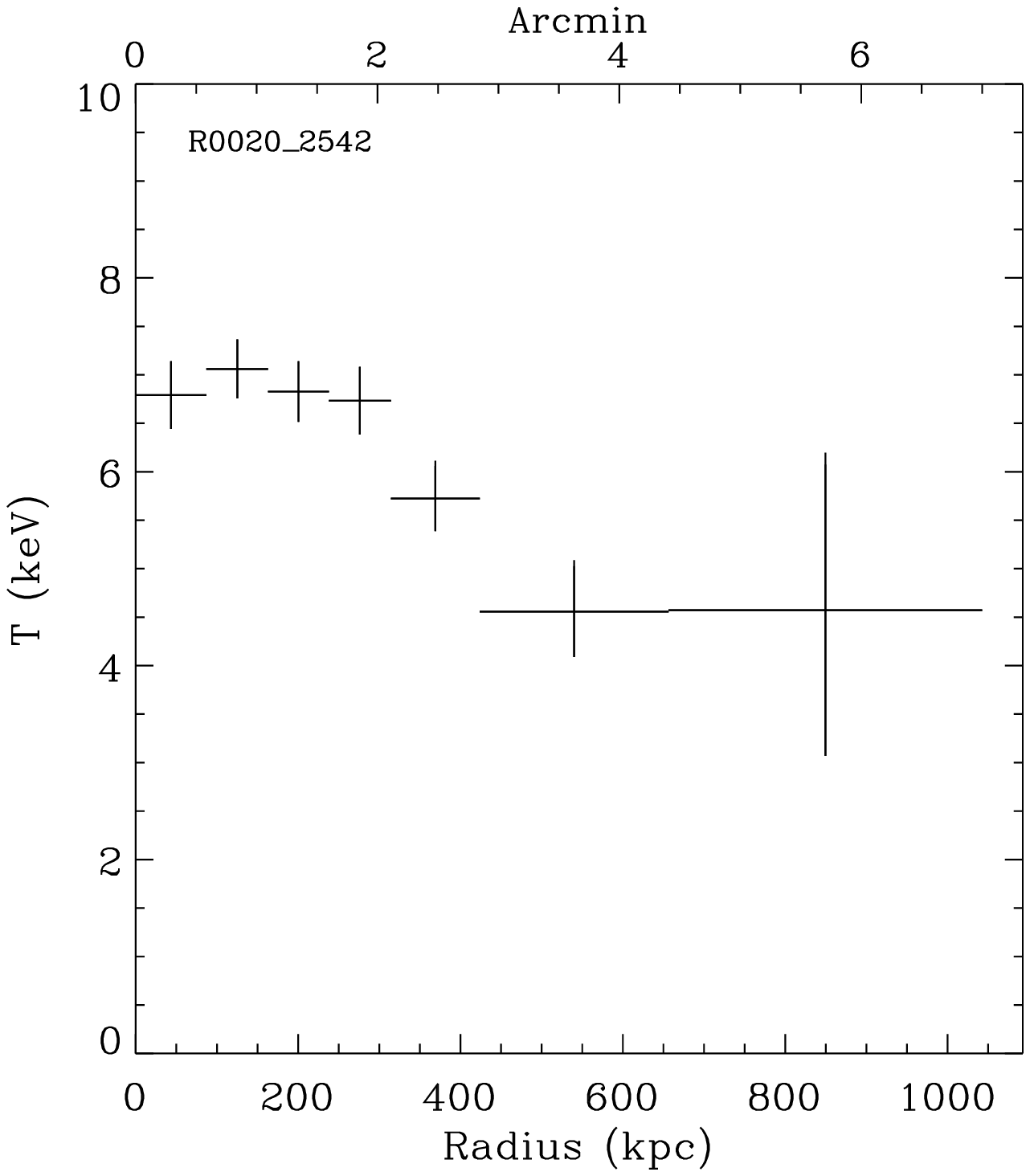}
\hfill
\includegraphics[scale=0.30,angle=0,keepaspectratio,width=0.33\textwidth]{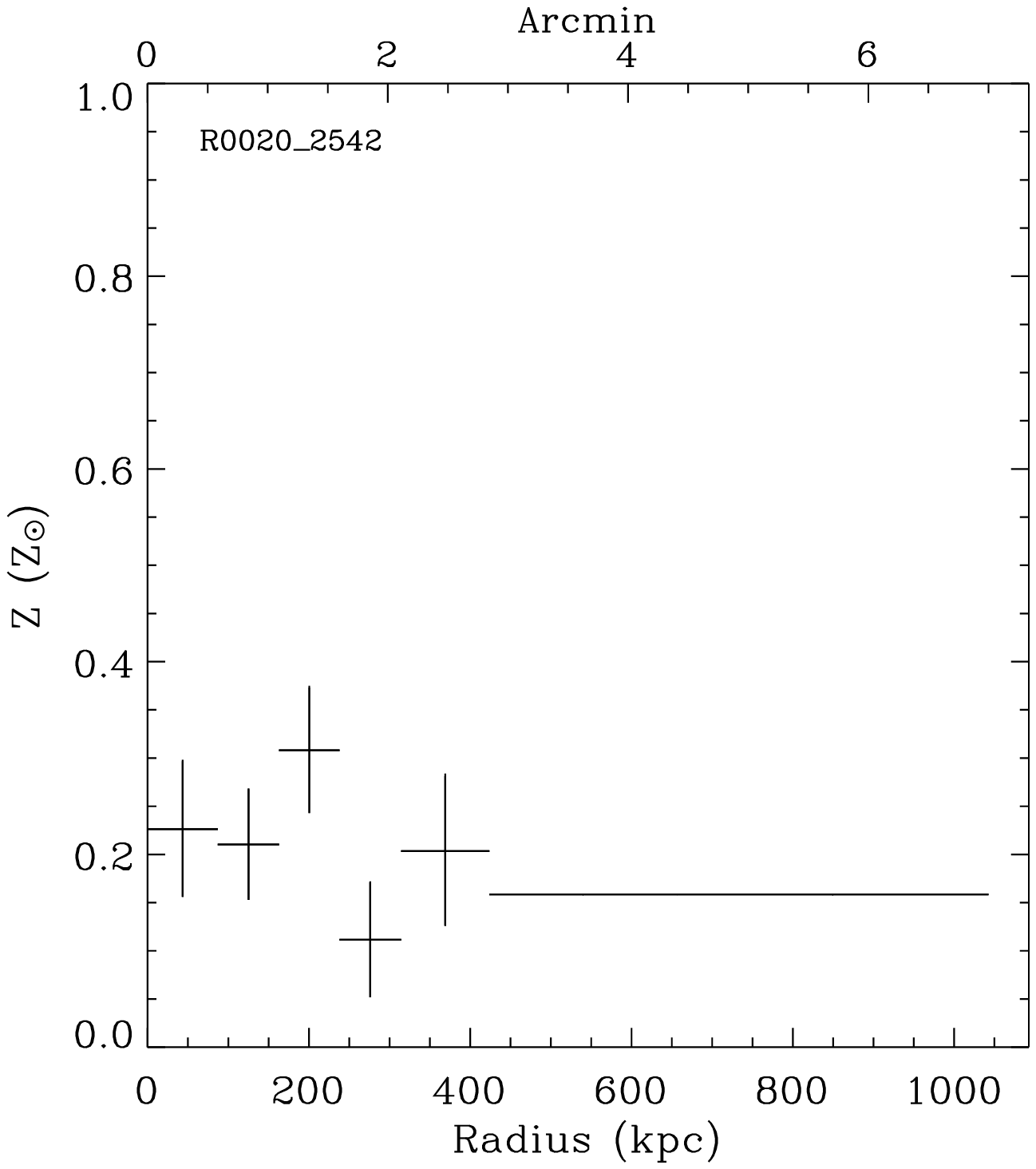}
\caption{{\footnotesize RXC\,J0020\,-2542.}}\label{fig:R0020}
\end{centering}
\end{figure*}

The temperature profile (Fig.~\ref{fig:R0020}) shows a prominent
decline with radius. The abundances are roughly flat but became
unconstrained at only $\sim 400$ kpc from the centre and were frozen
thereafter.  

%%================
%% Figure: R0020
%%
%%================

\subsection{RXC\,J0547\,-3152}

Also known as Abell 3364, this luminous cluster lies at $z=0.148$ and
has an average temperature of $\kT = 6.6$ keV. The 
X-ray image shows a bright, offset core, with obvious surface
brightness edges to the NW and SE. After renormalisation, the 
spectrum of the external region shows negative residuals below
$\sim 1$ keV. These are adequately described with a {\sc MeKaL} model
with negative normalisation and a temperature of 0.24 keV. An
additional power-law 
component is needed to fully describe the EPN spectrum (see
Fig.~\ref{fig:res}).   

The temperature profile declines monotonically, from $\kT \sim 7$
keV to $\kT \sim 4.5$ keV, from the centre to the largest radii at
which we can measure the temperature. The
metallicity profile declines from $Z \sim 0.4Z_{\odot}$ to $Z \sim 0.2
Z_{\odot}$ between $0 < r < 300$ kpc, but then increases once more to
the central value at $r \sim 600$ kpc. The metallicity trends may be
connected to the disturbed nature of the cluster.  

%%================
%% Figure: R0547
%%
\begin{figure*}
\begin{centering}
\includegraphics[scale=0.30,angle=0,keepaspectratio,width=0.33\textwidth]{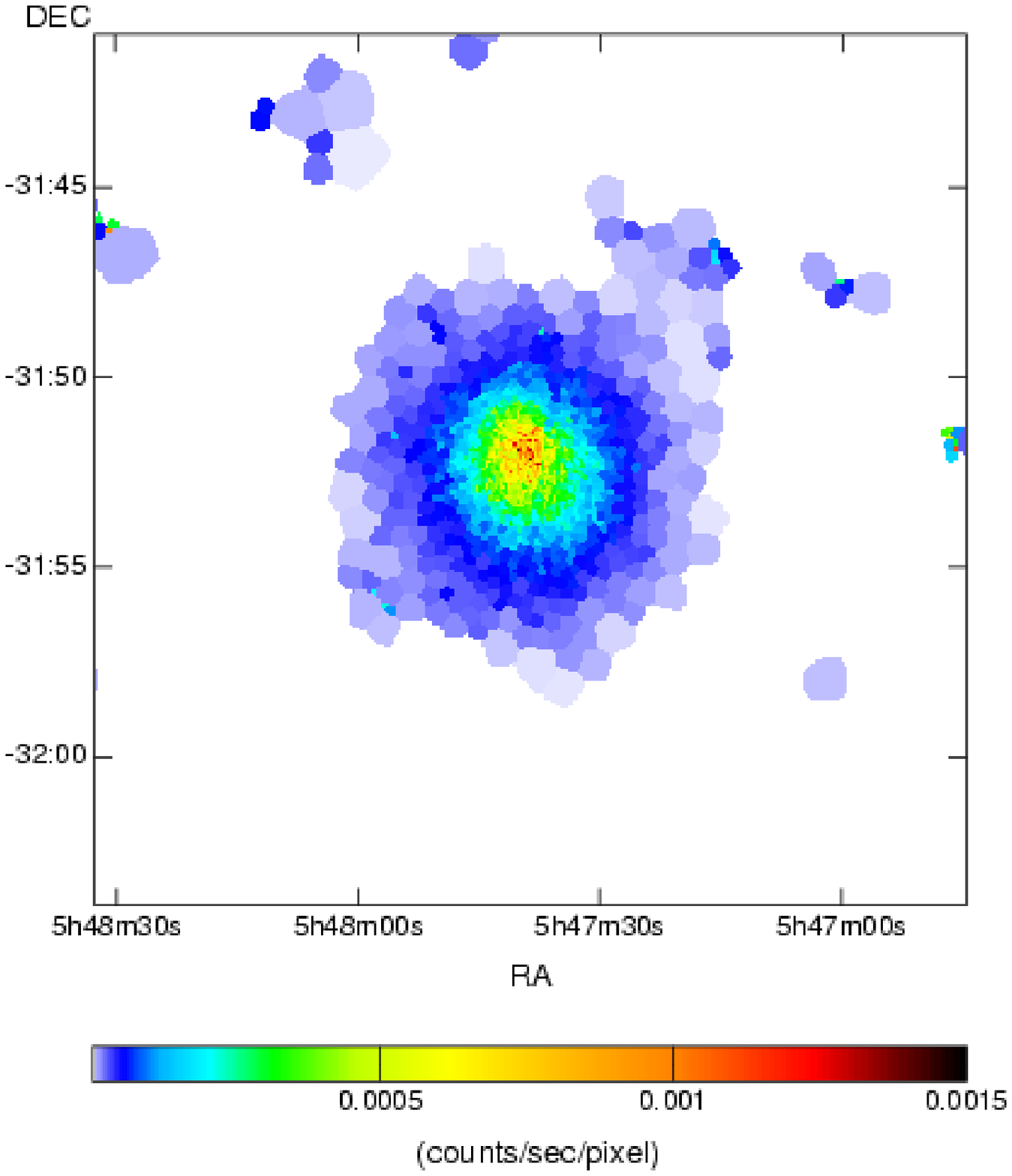}
\hfill
\includegraphics[scale=0.30,angle=0,keepaspectratio,width=0.33\textwidth]{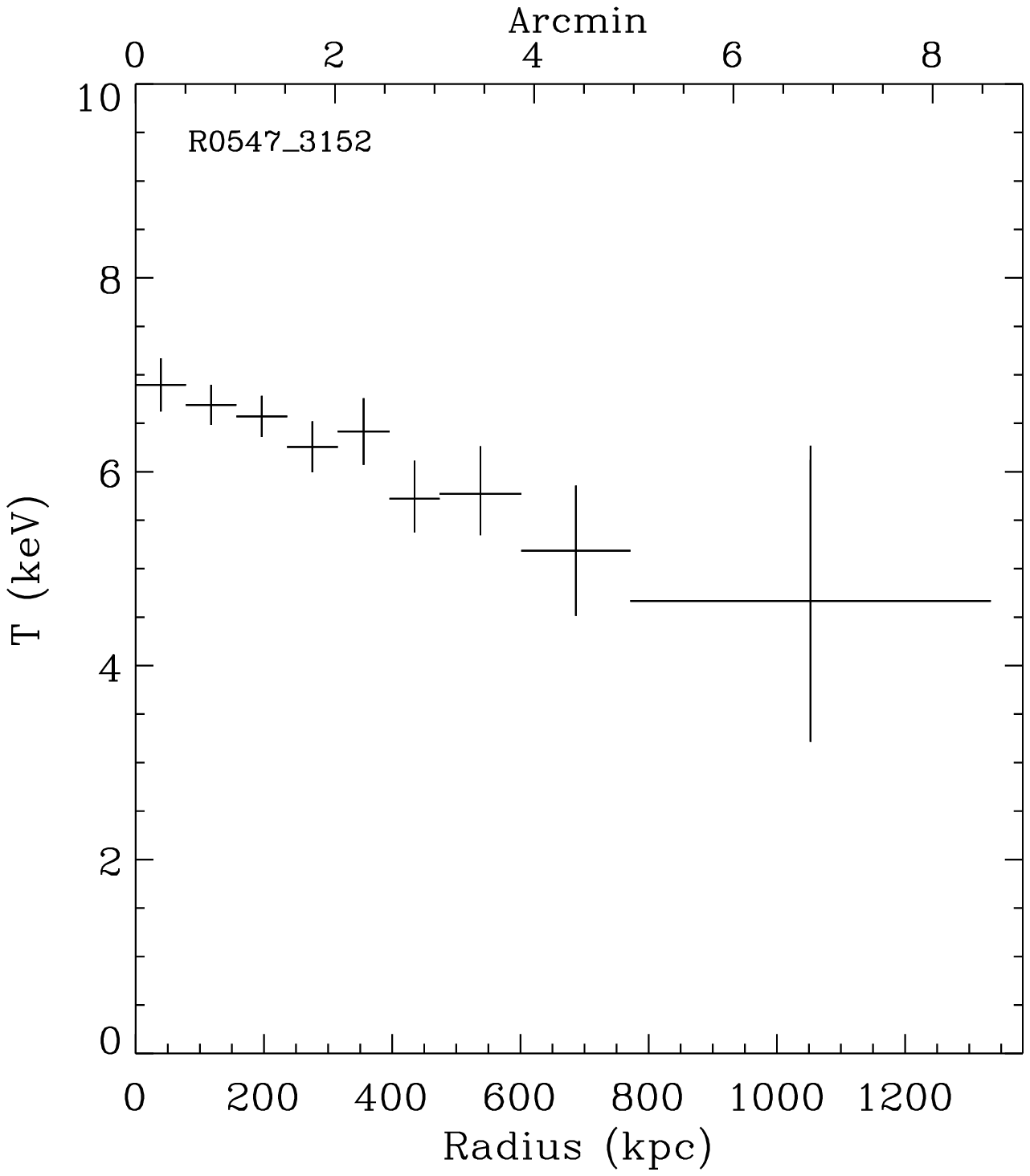}
\hfill
\includegraphics[scale=0.30,angle=0,keepaspectratio,width=0.33\textwidth]{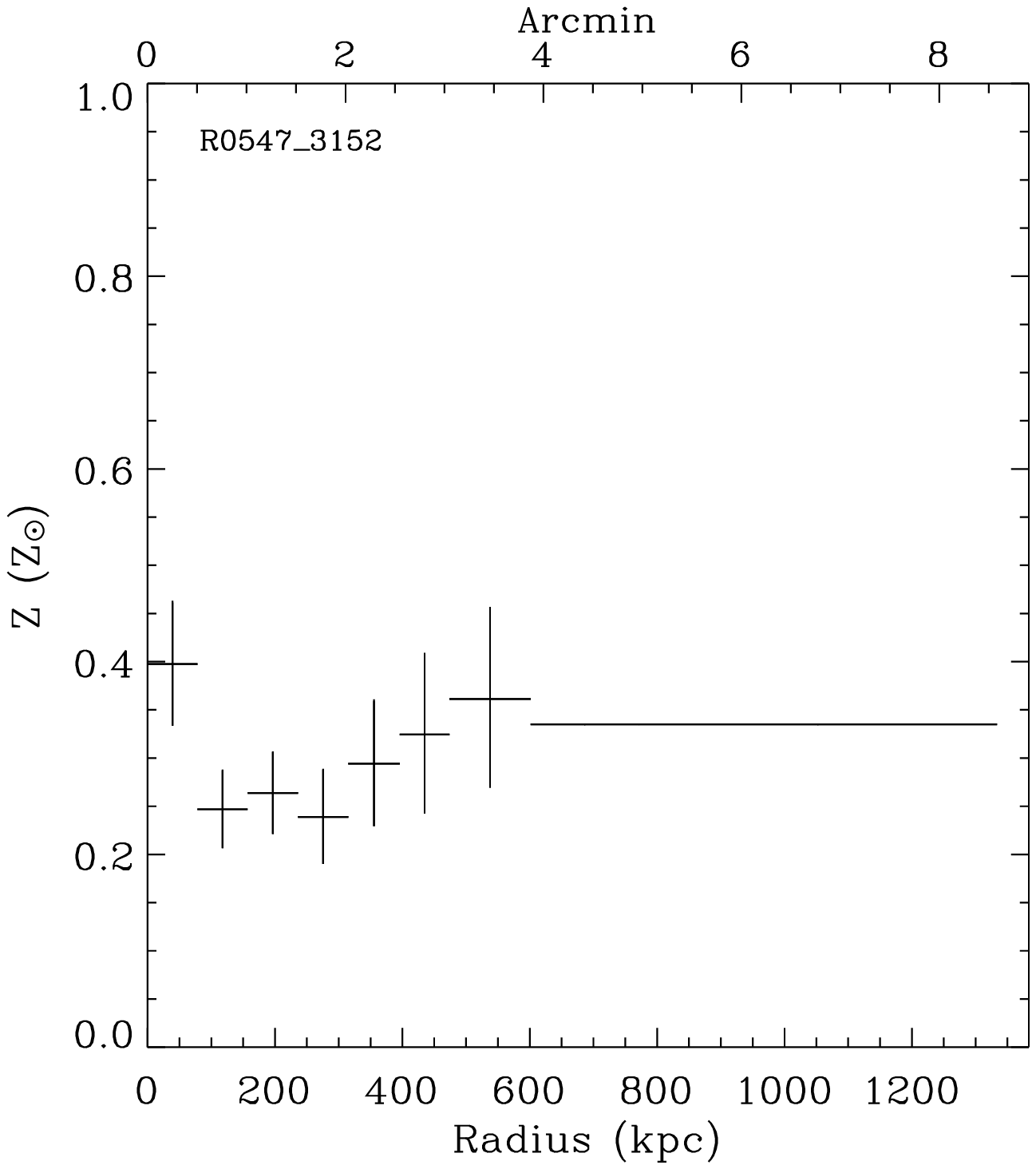}
\caption{{\footnotesize RXC\,J0547\,-3152.}}\label{fig:R0547}
\end{centering}
\end{figure*}
%%================

\subsection{RXC\,J0605\,-3518}

Lying at $z=0.14$, with an average temperature of $\kT =4.7$ keV,
this cluster is also known as Abell 3387. It is highly 
symmetric, presenting a strongly-peaked
central surface brightness and no visible substructure.  The residual
spectrum is well described 
with a {\sc MeKaL} model with negative normalisation and a temperature
of 0.26 keV. An additional power-law component is necessary for a full
description of the EPN data.

\begin{figure*}
\begin{centering}
\includegraphics[scale=0.30,angle=0,keepaspectratio,width=0.33\textwidth]{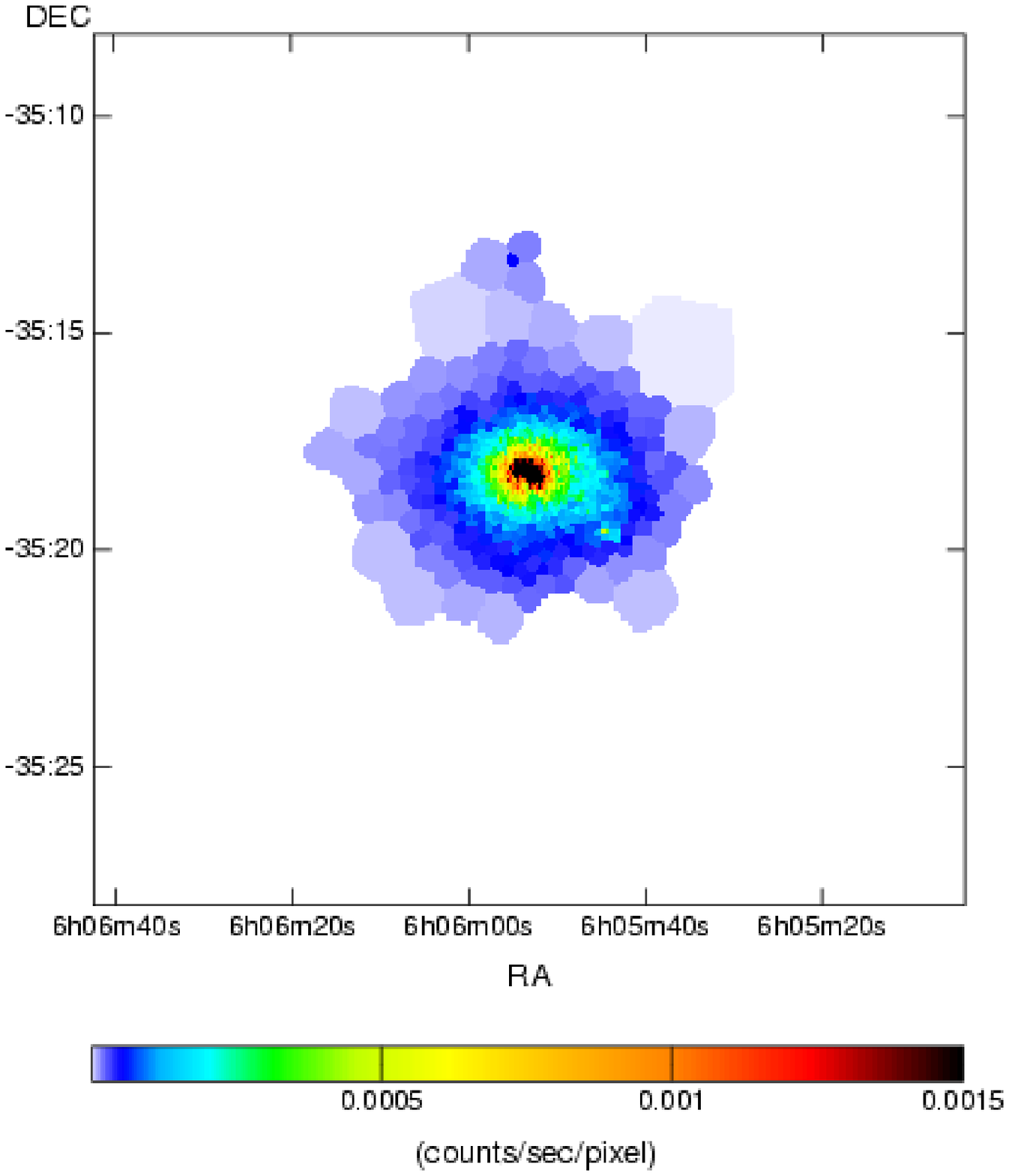}
\hfill
\includegraphics[scale=0.30,angle=0,keepaspectratio,width=0.32\textwidth]{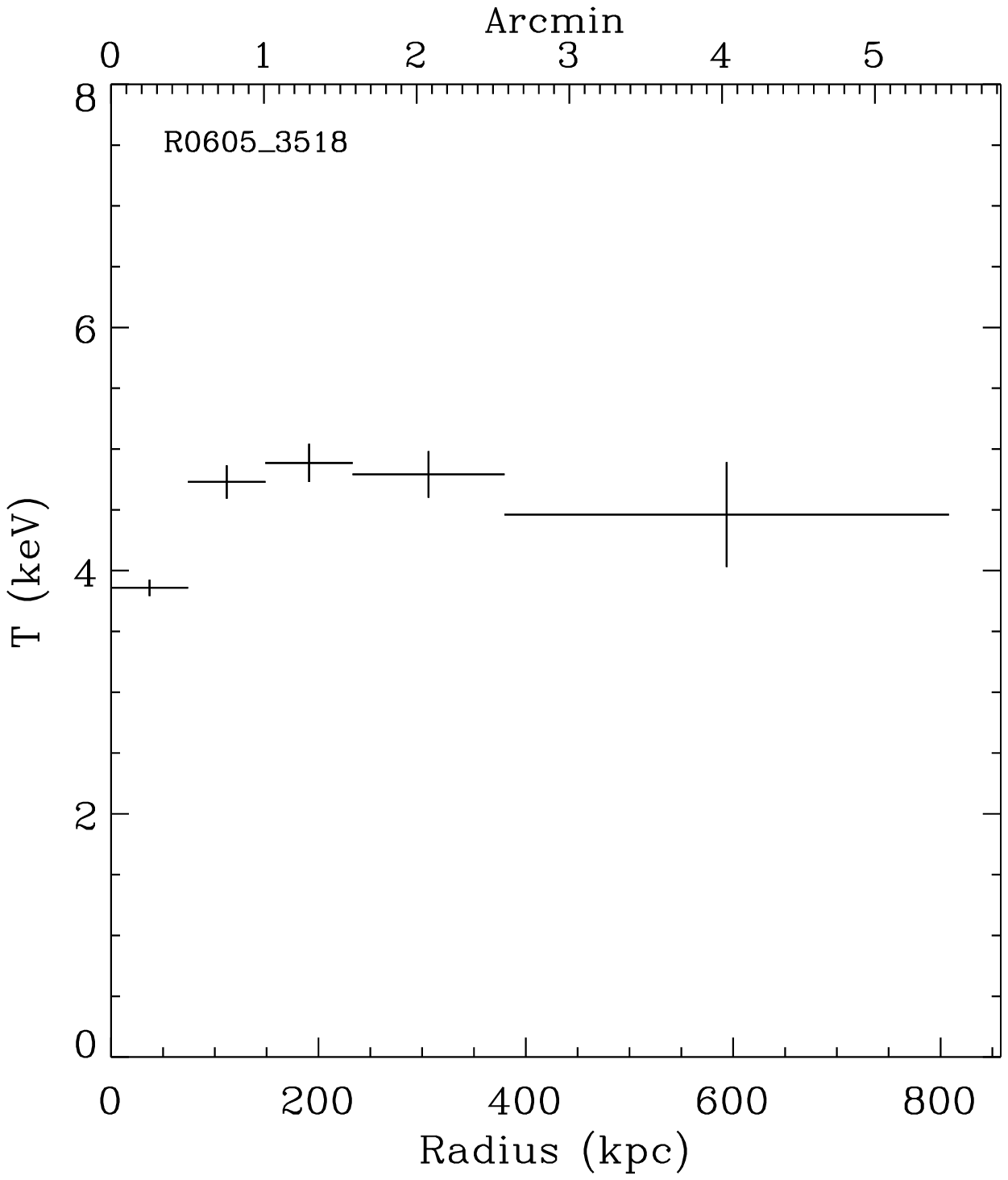}
\hfill
\includegraphics[scale=0.30,angle=0,keepaspectratio,width=0.33\textwidth]{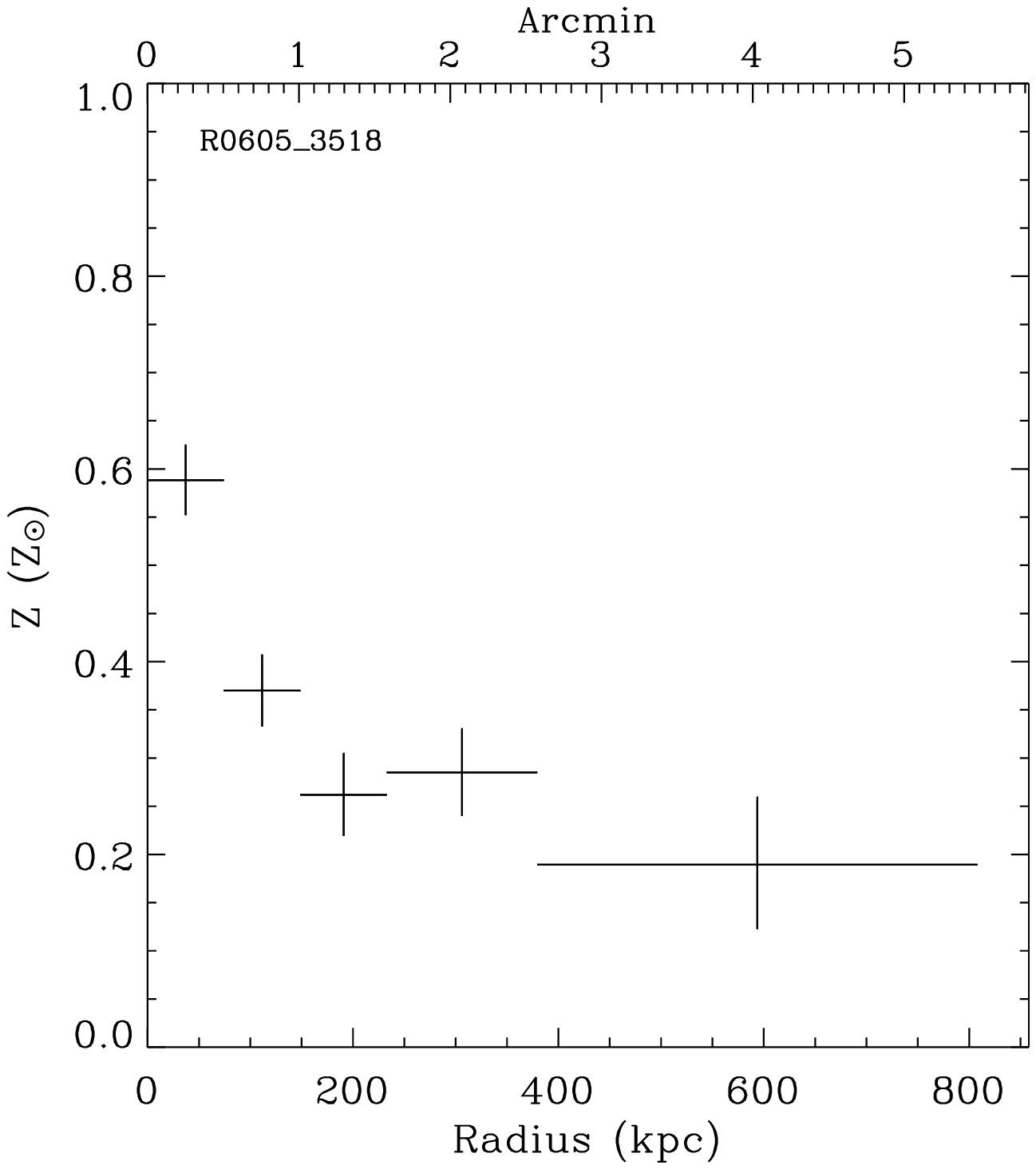}
\caption{{\footnotesize RXC\,J0605\,-3518.}}\label{fig:R0605}
\end{centering}
\end{figure*}

The temperature profile (Fig.~\ref{fig:R0605}) rises from the central
regions to a peak at $R \sim 200$ kpc, after which there is a gentle
decline. The abundance profile declines smoothly from $\sim
0.6\,Z_\odot$ in the central regions to $\sim 0.2\,Z_\odot$ at $R >
500$ kpc. The general behaviour of the temperature and abundance
profiles is very reminiscent of the {\it Chandra\/} temperature
profiles of cool core clusters derived by \citet{vikh05}.

%%================
%% Figure: R0605
%%
%%================

\subsection{RXC\,J1044\,-0704}

RXC\,J1044\,-0704, also known as Abell 1048, lies at $z = 0.13$. It has
an average temperature of $\kT = 3.6$ keV and although slightly
elliptical, is another symmetric cluster with a strongly-peaked
central surface brightness. The residual spectrum is negative in the
0.5-1.0 keV band, indicating oversubtraction of the background in this
band. The residual spectrum can be fitted with a {\sc MeKaL} model at
0.26 keV with negative normalisation. An additional power-law
component improves the fit to the EPN data. 

\begin{figure*}
\begin{centering}
\includegraphics[scale=0.30,angle=0,keepaspectratio,width=0.33\textwidth]{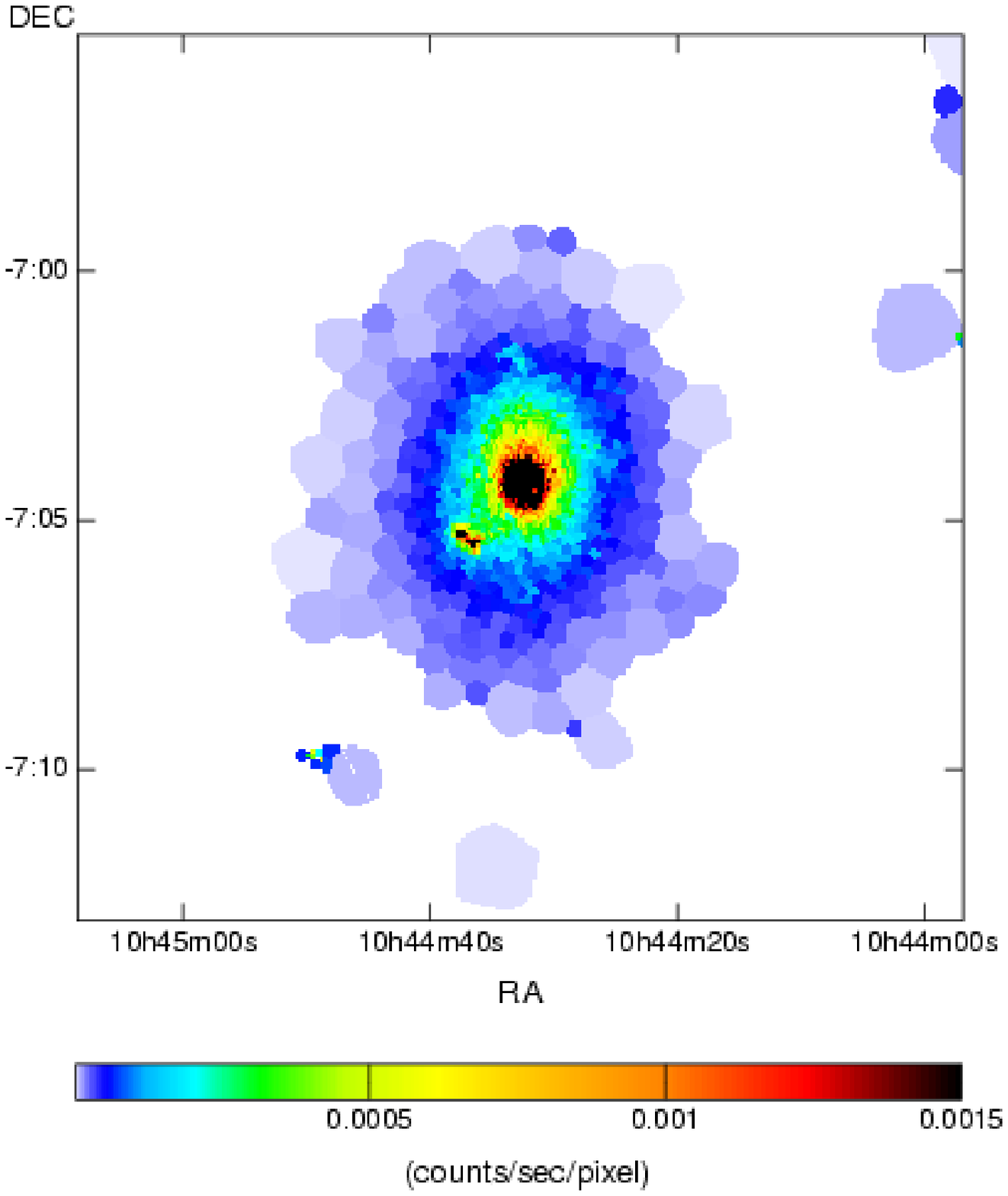}
\hfill
\includegraphics[scale=0.30,angle=0,keepaspectratio,width=0.33\textwidth]{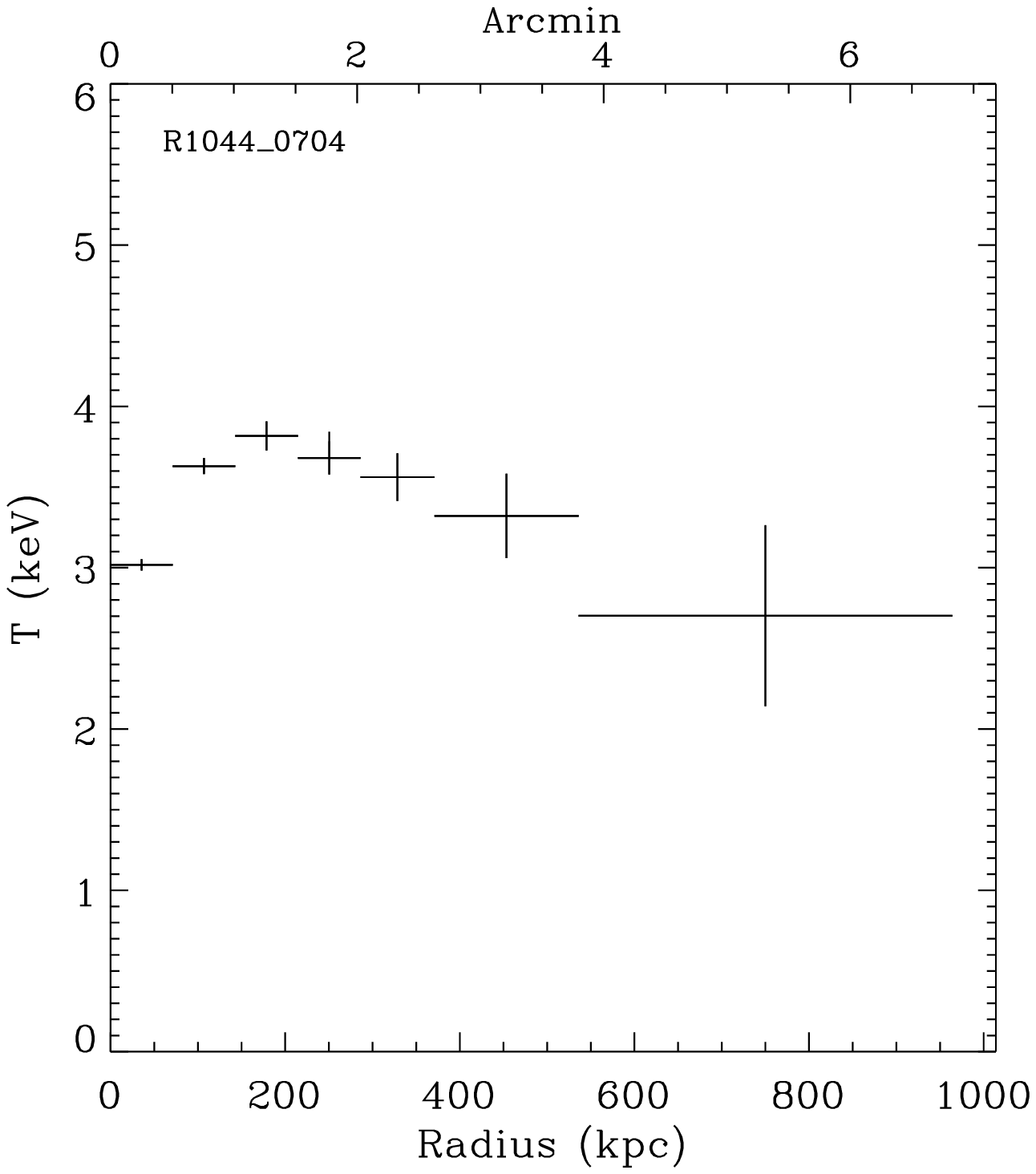}
\hfill
\includegraphics[scale=0.30,angle=0,keepaspectratio,width=0.33\textwidth]{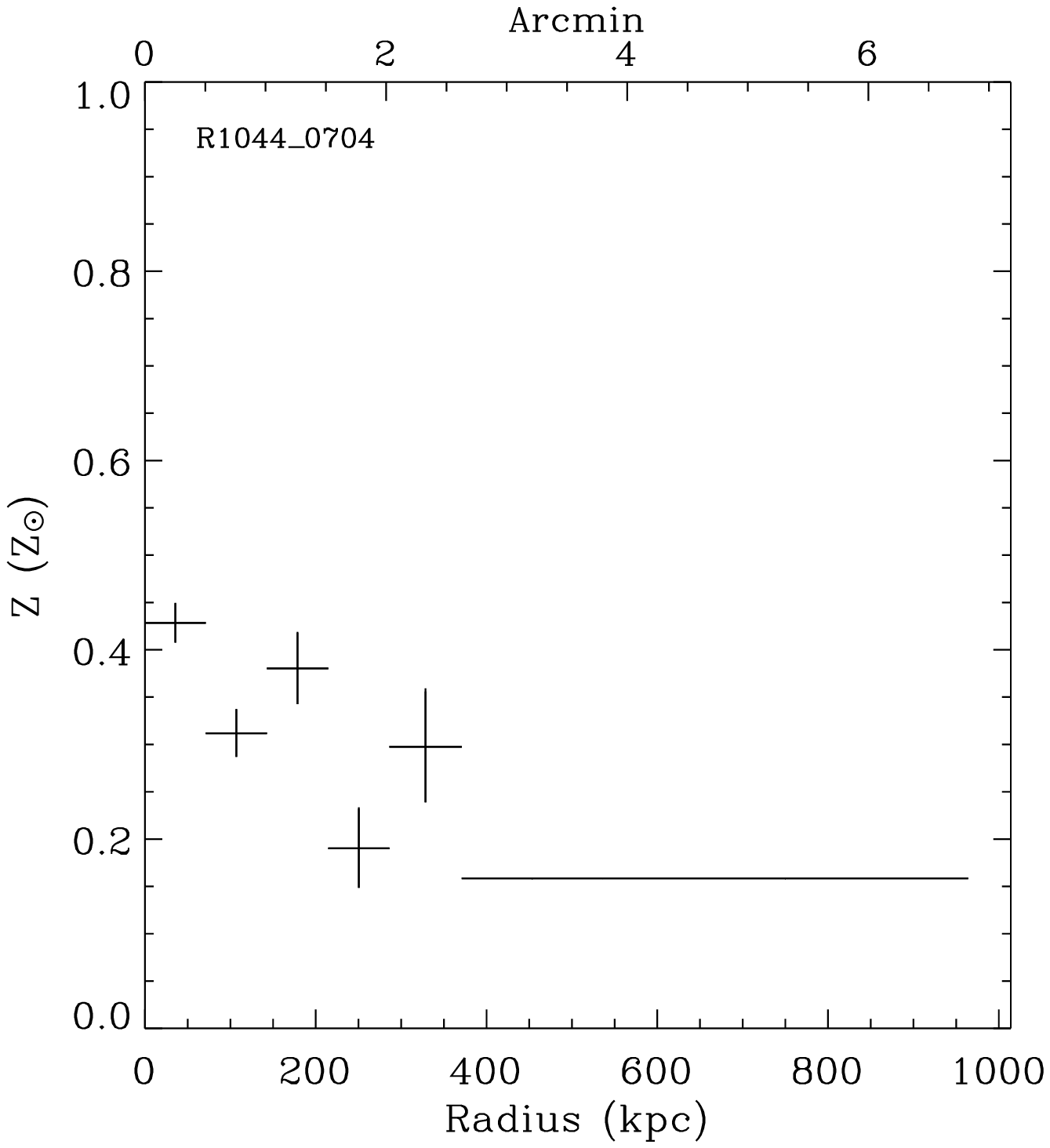}
\caption{{\footnotesize RXC\,J1044\,-0704.}}\label{fig:R1044}
\end{centering}
\end{figure*}

The temperature and abundance profiles (Fig.~\ref{fig:R1044}) are once
again reminiscent of 
those of other cool core clusters. The temperature climbs to a peak at
$R \sim 200$ kpc and declines thereafter, while the abundance profile
declines from the central regions to the outskirts (although in this
case the decline is not smooth).

%%================
%% Figure: R1044
%%
%%================

\subsection{RXC\,J1141\,-1216}

This highly symmetric cluster at $z=0.12$ exhibits strongly peaked
central emission. The cluster is also known as Abell 1348 and has an
average temperature of $\kT = 3.6$ keV. The residual spectrum shows
negative residuals and is well fitted with a simple {\sc MeKaL} model
with negative normalisation and a temperature of 0.27 keV. 

\begin{figure*}
\begin{centering}
\includegraphics[scale=0.30,angle=0,keepaspectratio,width=0.33\textwidth]{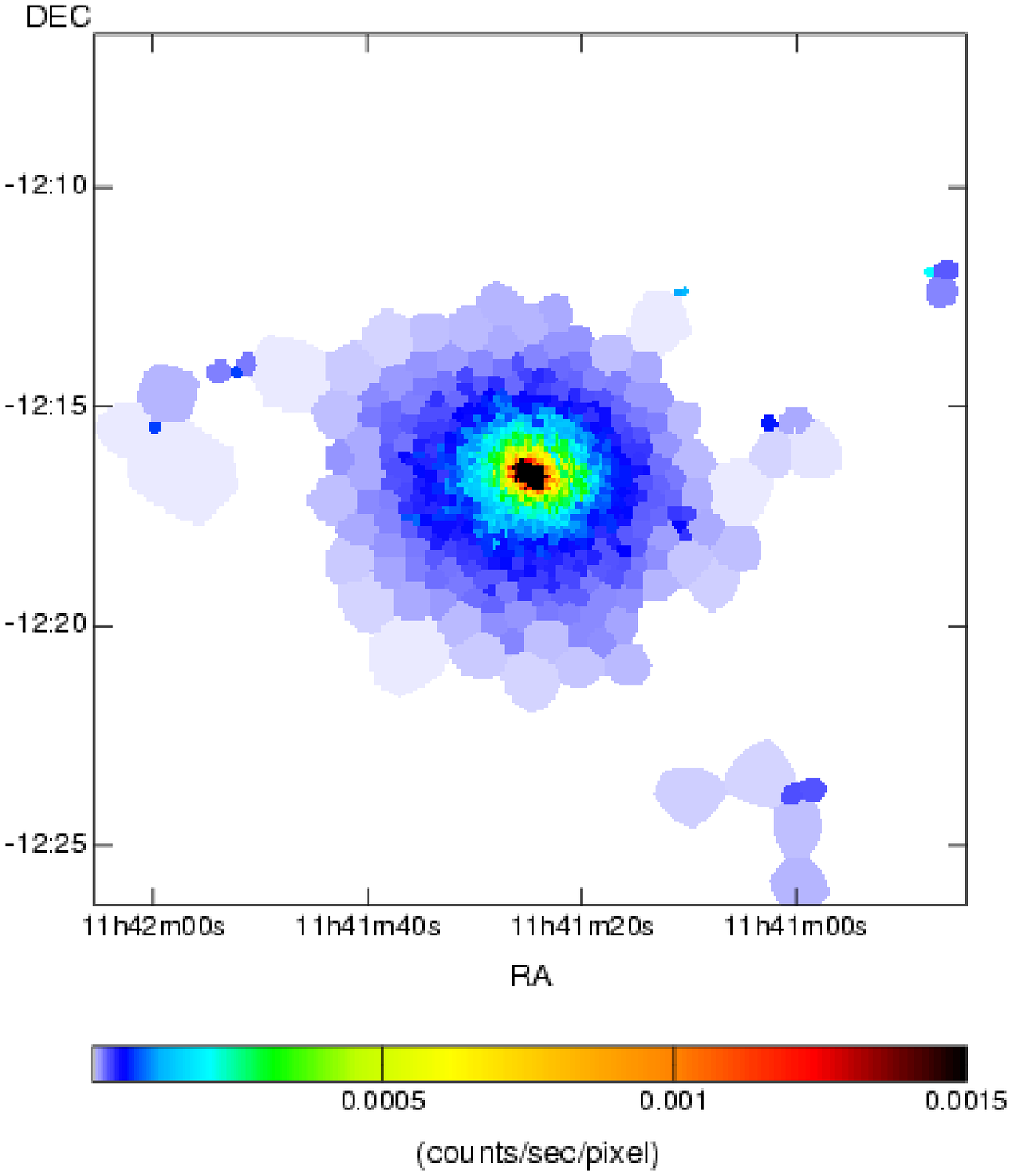}
\hfill
\includegraphics[scale=0.30,angle=0,keepaspectratio,width=0.32\textwidth]{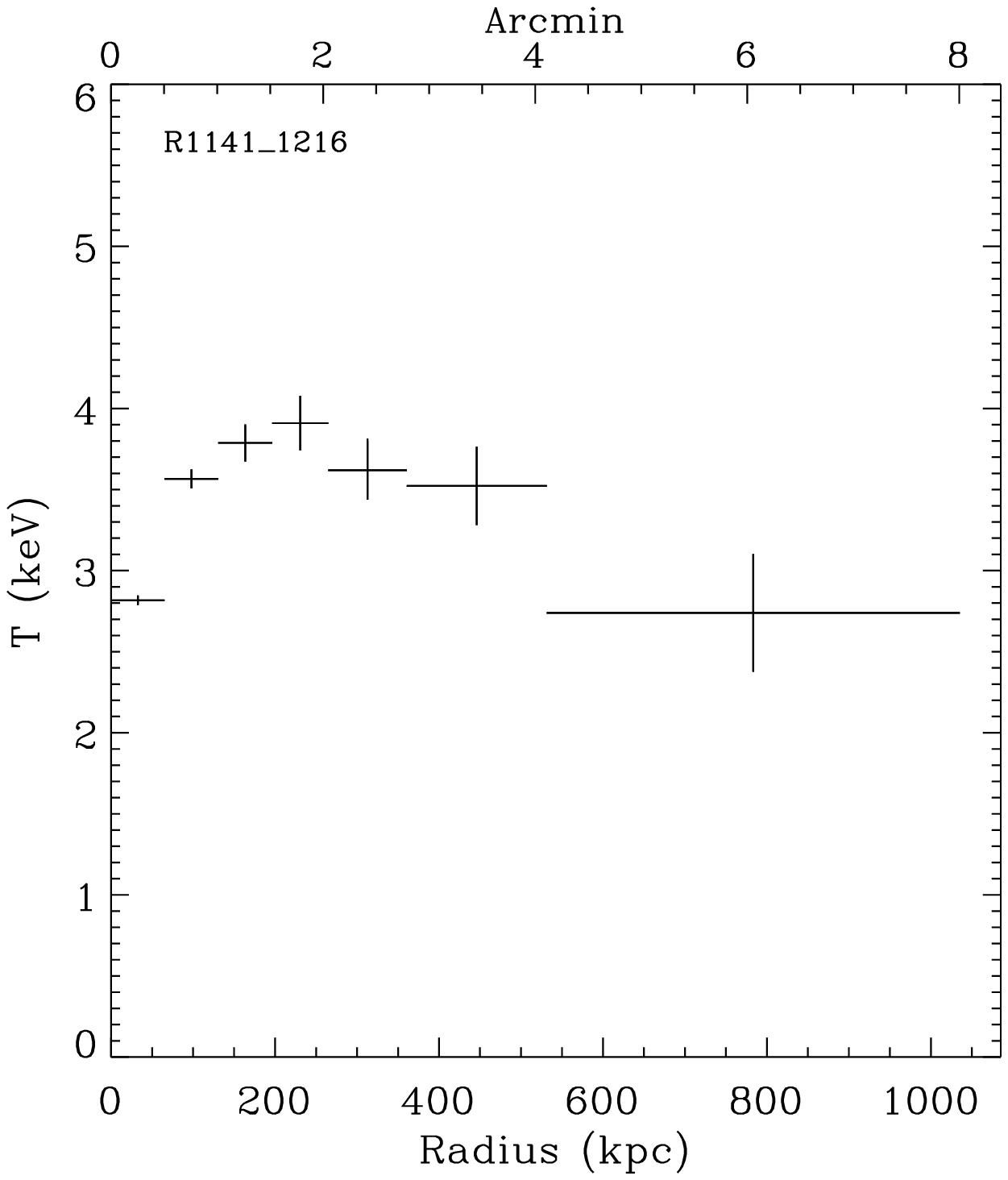}
\hfill
\includegraphics[scale=0.30,angle=0,keepaspectratio,width=0.33\textwidth]{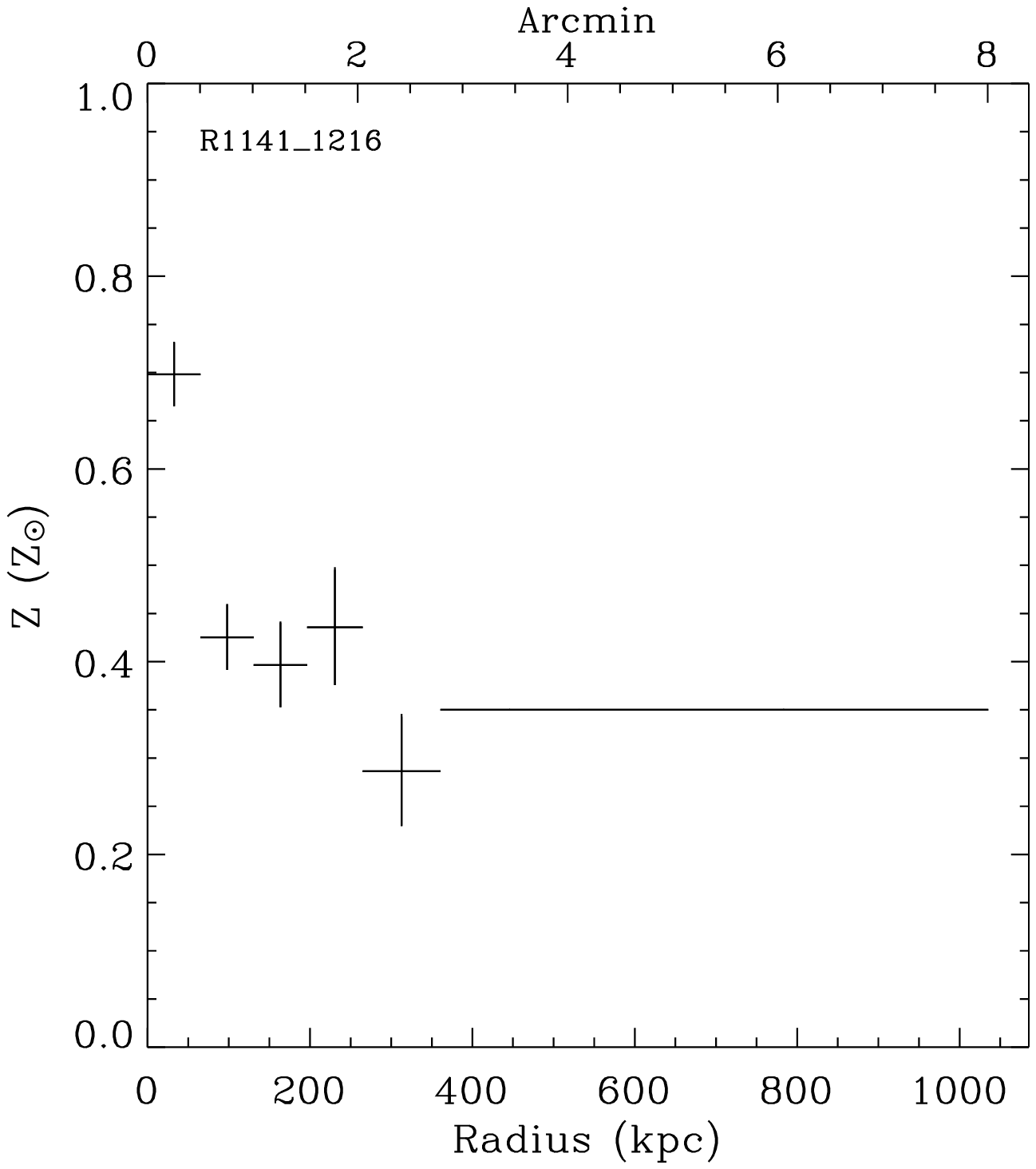}
\caption{{\footnotesize RXC\,J1141\,-1216.}}\label{fig:R1141}
\end{centering}
\end{figure*}

The temperature and abundance profiles shown in Fig.~\ref{fig:R1141}
are very similar to those of the previous two clusters. The central
temperature dip is associated with an abundance enhancement; the
temperature peaks around 200 kpc and declines gently thereafter. The
abundance declines smoothly from the centre to the external regions.

%%================
%% Figure: R1141
%%
%%================

%%\afterpage{\clearpage}

\subsection{RXC\,J1302\,-0230}

Also known as Abell 1663, this is a symmetric looking cluster at $\kT
= 3.6$ keV lying at $z=0.085$. It has a strong 
central emission peak. The residual spectrum of the
external region ($r > 11\arcmin$) can be characterised with a {\sc
  MeKaL} model at 0.27 keV with negative normalisation. An additional
power law component improves the fit for the EMOS2 and EPN
spectra. 

\begin{figure*}
\begin{centering}
\includegraphics[scale=0.30,angle=0,keepaspectratio,width=0.33\textwidth]{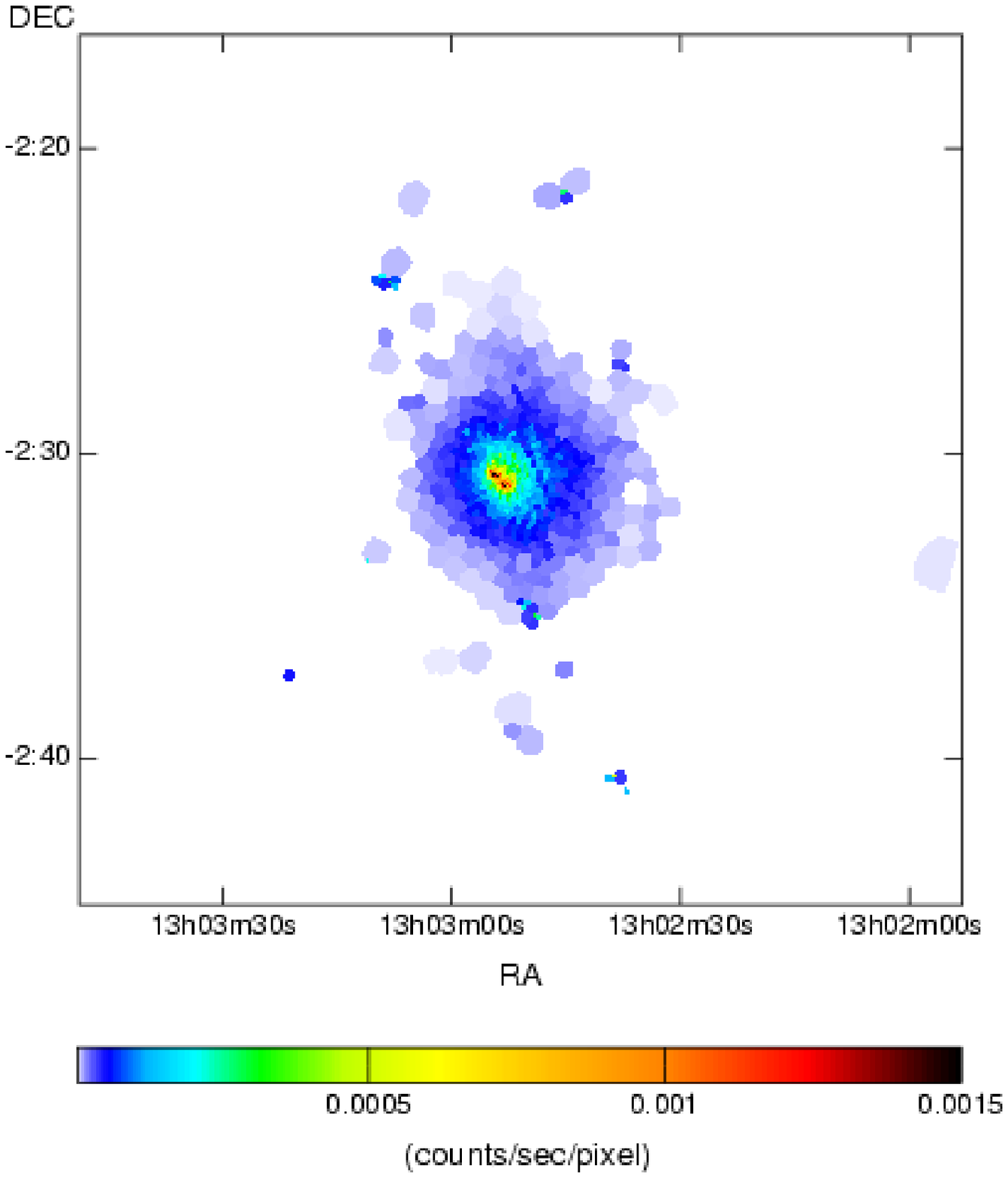}
\hfill
\includegraphics[scale=0.30,angle=0,keepaspectratio,width=0.32\textwidth]{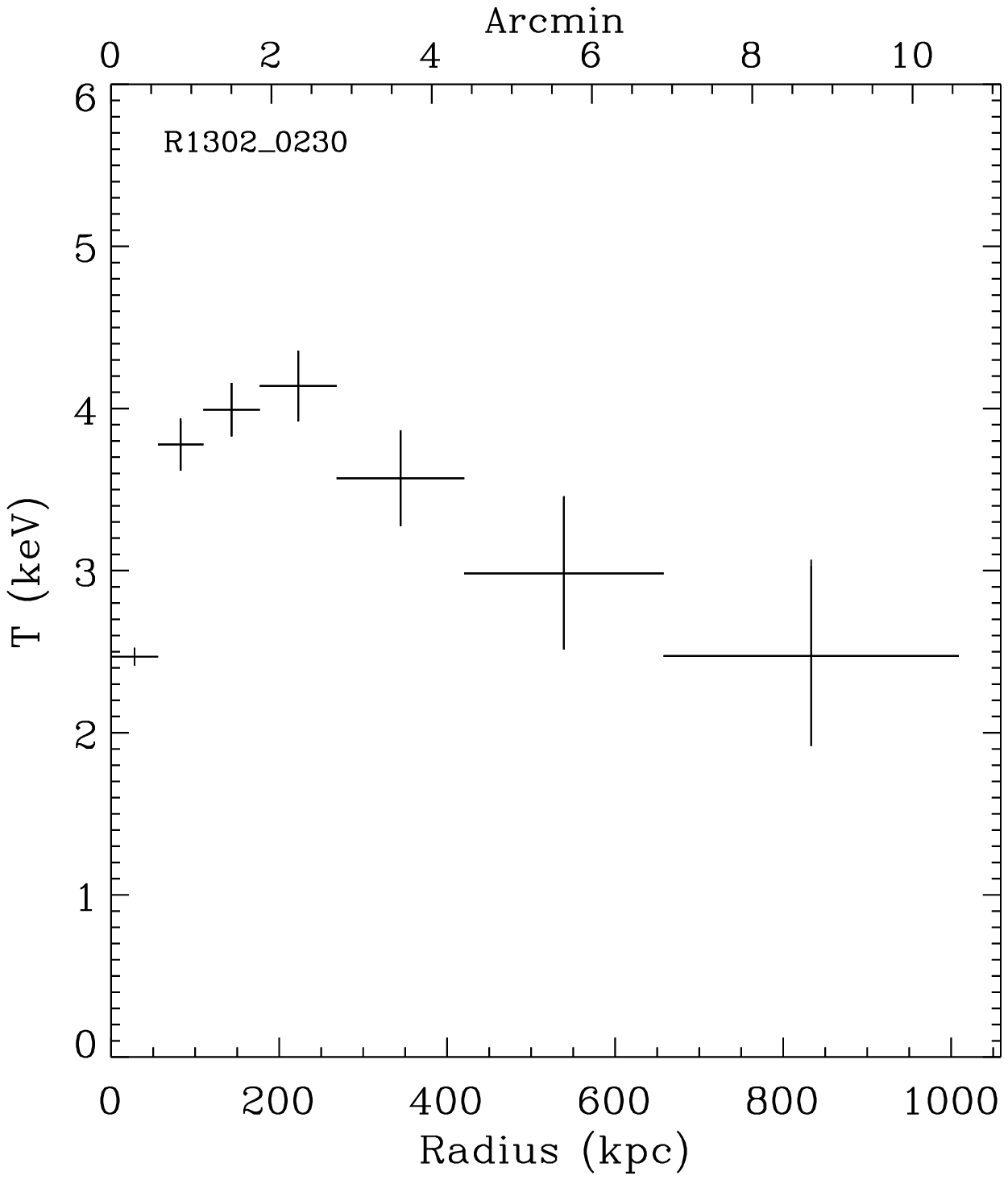}
\hfill
\includegraphics[scale=0.30,angle=0,keepaspectratio,width=0.33\textwidth]{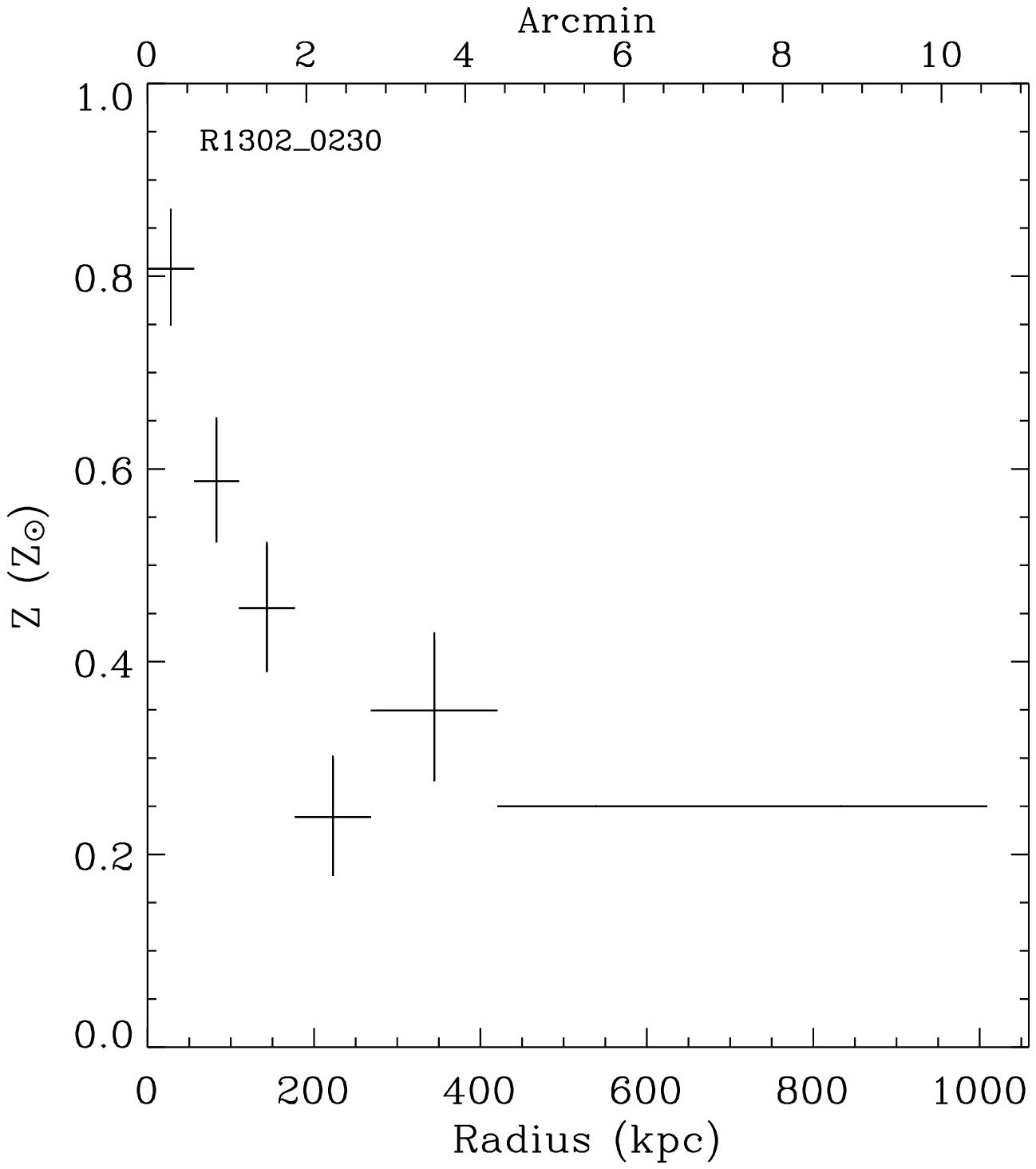}
\caption{{\footnotesize RXC\,J1302\,-0230.}}\label{fig:R1302}
\end{centering}
\end{figure*}

The temperature and abundance profiles (Fig.~\ref{fig:R1302}) are
very characteristic of cooling core clusters. Compared to similar
clusters in this sample, however, RXC J1302 -0230 is characterised by
a particularly steep central temperature drop, and a similarly steep
increase of metallicity towards the central regions.

%%================
%% Figure: R1302
%%
%%================

\subsection{RXC\,J1311\,-0120}

This extremely symmetric cluster lying at $z=0.183$ is the well-known
lensing cluster Abell 1689. This is the most luminous cluster in the
present sample, which is reflected by its particularly high
temperature ($\kT = 8.5$ keV). The residual spectrum of the external
region ($r > 11\arcmin$) shows a positive excess which is well modelled
by a {\sc MeKaL} model at 0.19 keV. An additional power-law component
improves the fit for the EPN detector.

\begin{figure*}
\begin{centering}
\includegraphics[scale=0.30,angle=0,keepaspectratio,width=0.33\textwidth]{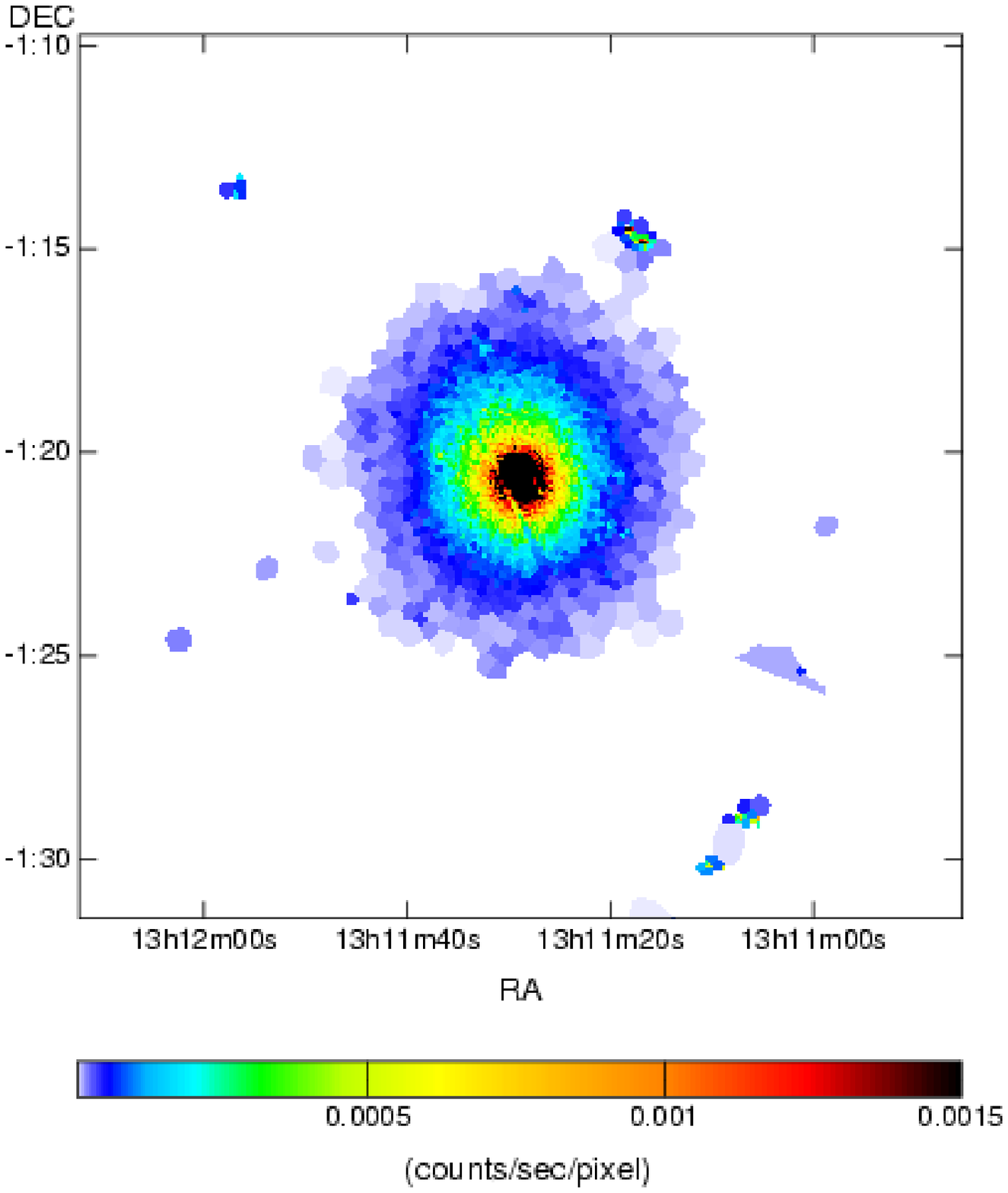}
\hfill
\includegraphics[scale=0.30,angle=0,keepaspectratio,width=0.32\textwidth]{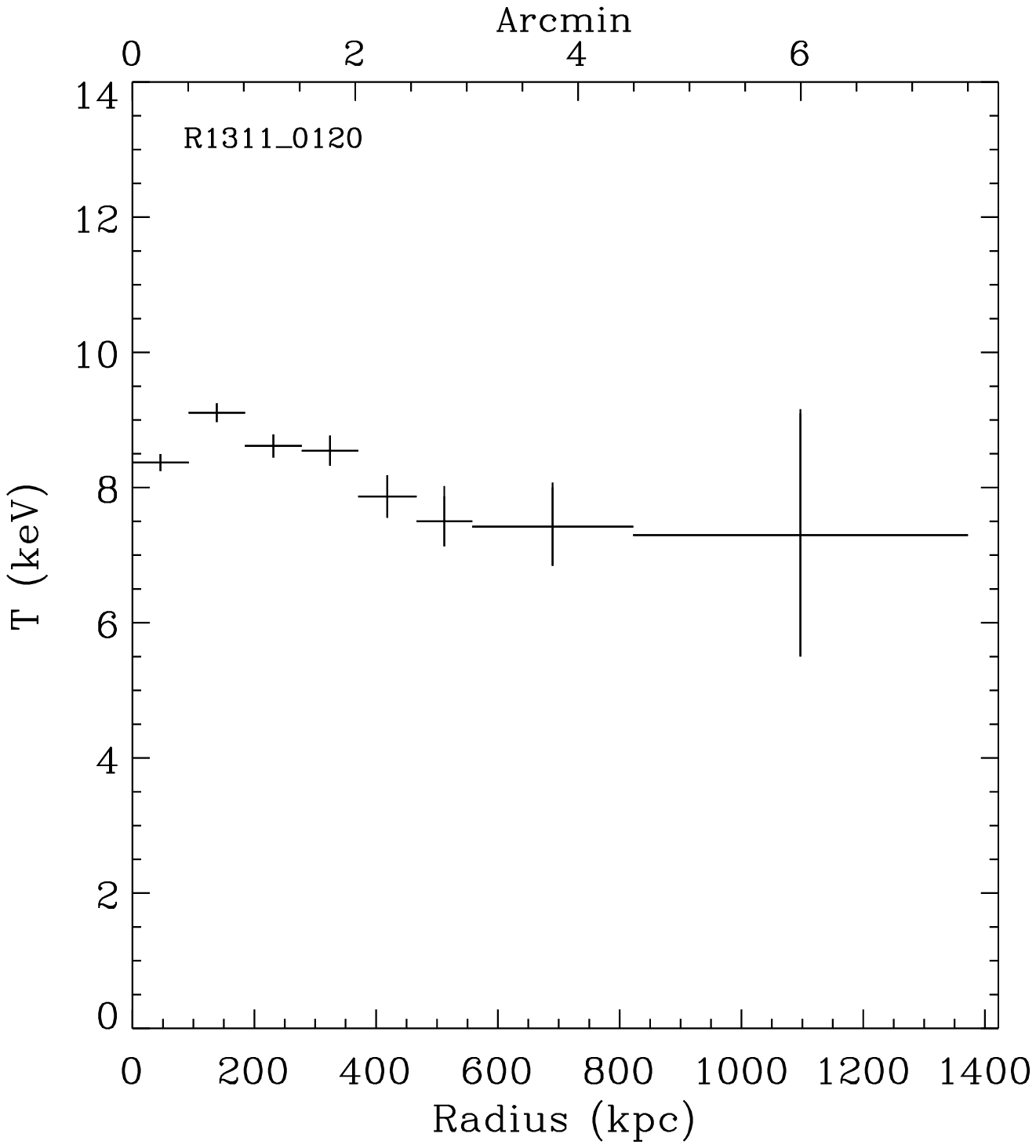}
\hfill
\includegraphics[scale=0.30,angle=0,keepaspectratio,width=0.33\textwidth]{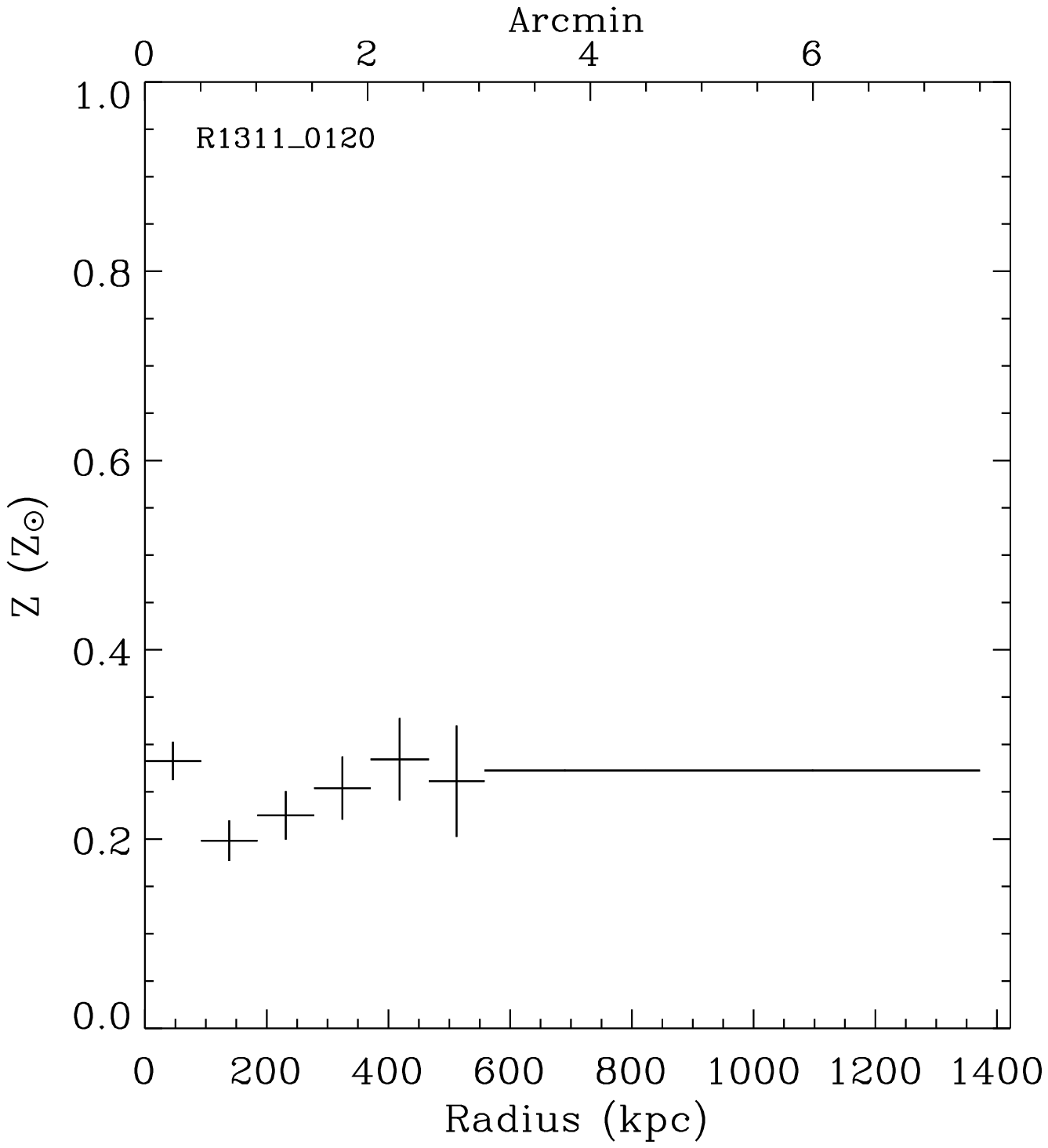}
\caption{{\footnotesize RXC\,J1311\,-0120.}}\label{fig:R1311}
\end{centering}
\end{figure*}

The temperature and abundance profiles (Fig.~\ref{fig:R1311}) are not
characteristic of cooling core clusters, however. While there is a
central temperature drop, it is not nearly as steep as that displayed
by other cool core clusters in this sample. Furthermore, the abundance
profile does not show a central peak. The temperature and abundance
profiles we have derived are in good agreement with those derived
(from the same \xmm data) by \citet{am}. \citet{gir} use galaxy
velocity date to describe this cluster 
as a line of sight merger. This may explain why the clusters is quite
symmetric but does not appear to possess a strong cooling core.

%%================
%% Figure: R1311
%%
%%================

\subsection{RXC\,J1516\,+0005}

Also known as Abell 2050, this moderate temperature ($\kT =
4.6$ keV) symmetric looking cluster lying at
$z=0.120$ does not, however, exhibit peaked central emission. The
external residual spectrum, accumulated from events from beyond
$12.5\arcmin$ from the cluster centre exhibits an excess of counts at
$E < 1$ keV and is adequately fitted with a {\sc MeKaL} model at 0.25
keV. An additional power law component improves the EPN fit.

\begin{figure*}
\begin{centering}
\includegraphics[scale=0.30,angle=0,keepaspectratio,width=0.33\textwidth]{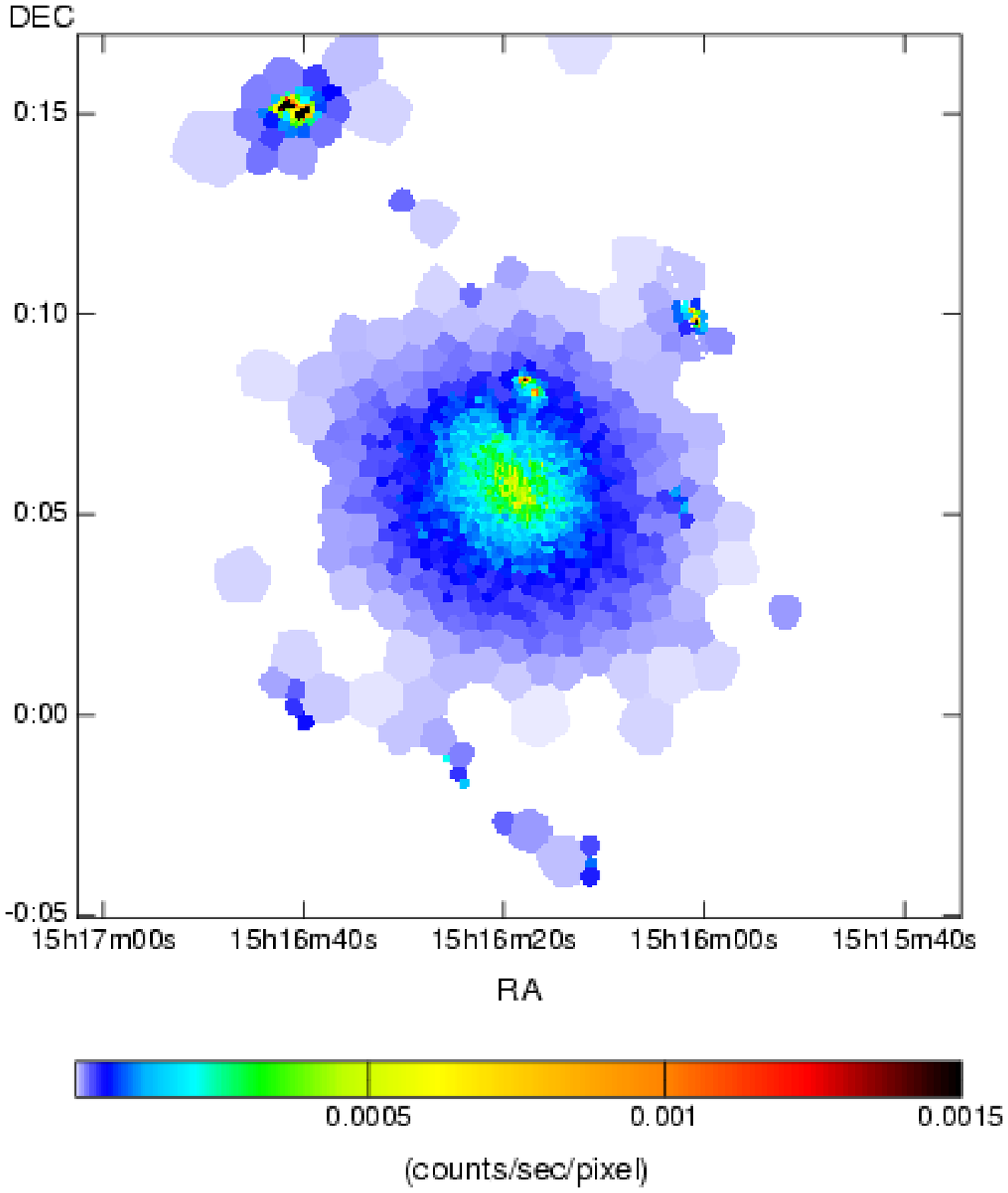}
\hfill
\includegraphics[scale=0.30,angle=0,keepaspectratio,width=0.32\textwidth]{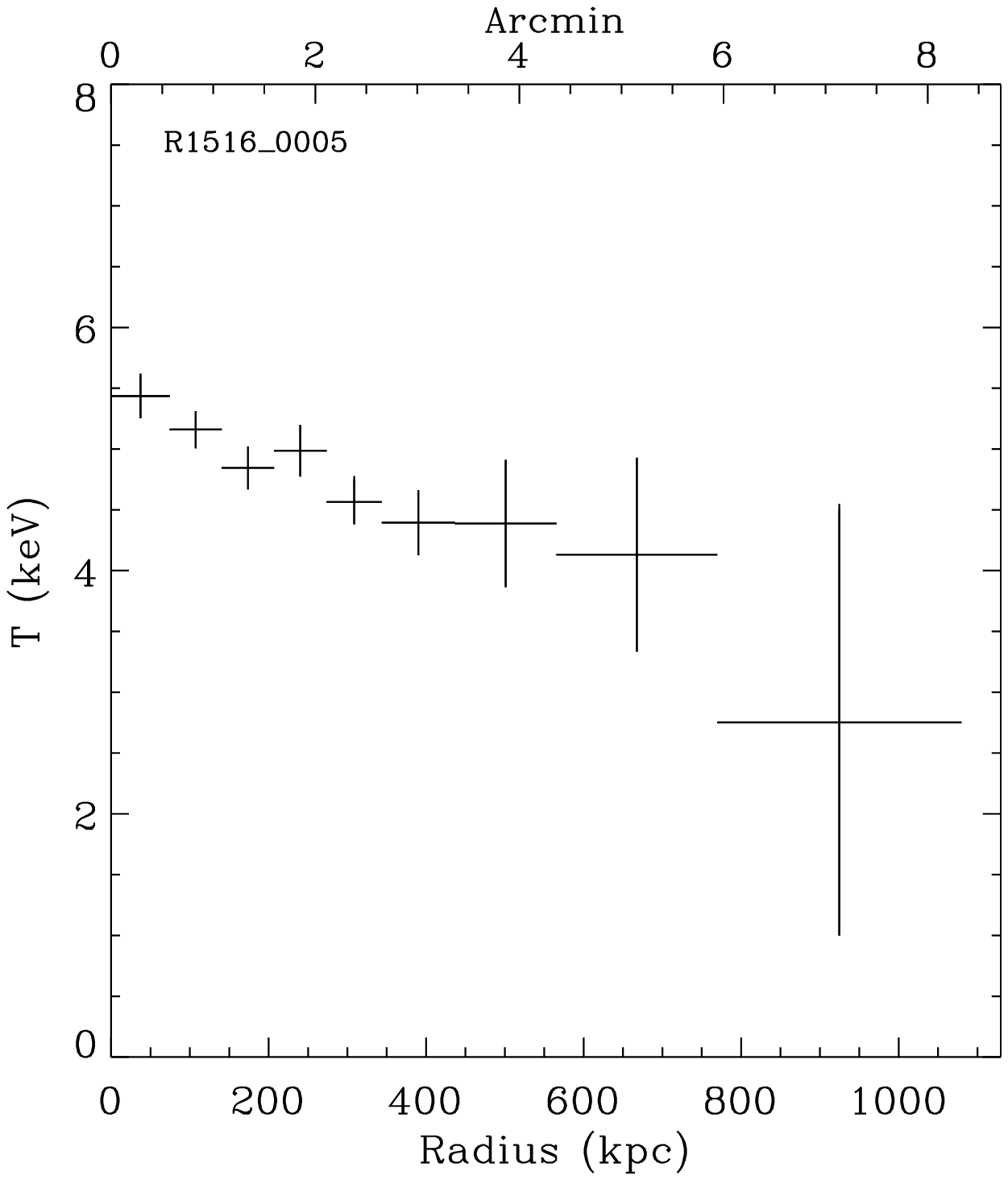}
\hfill
\includegraphics[scale=0.30,angle=0,keepaspectratio,width=0.33\textwidth]{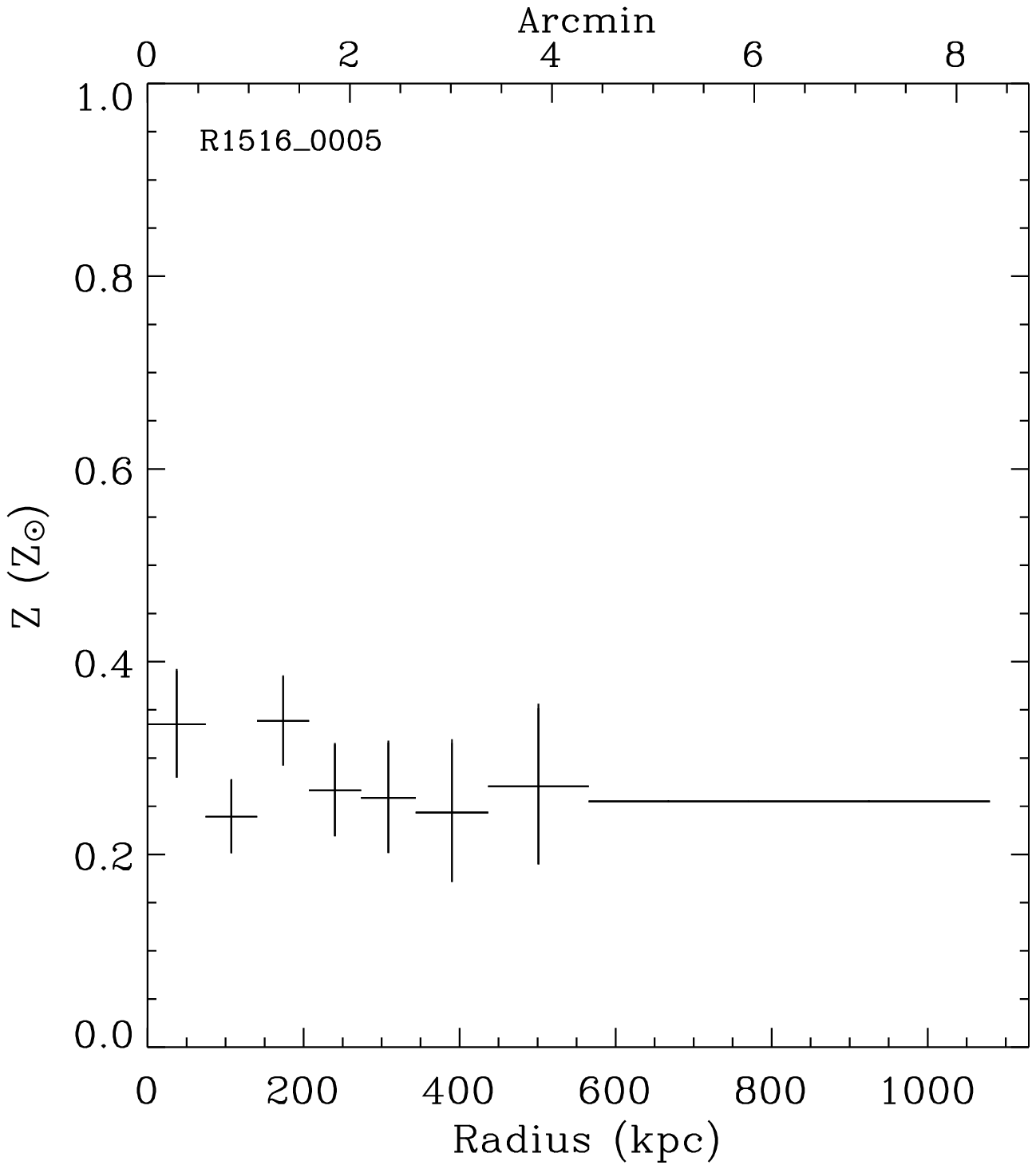}
\caption{{\footnotesize RXC\,J1516\,+0005.}}\label{fig:R15160005}
\end{centering}
\end{figure*}

The temperature profile of this cluster (Fig.~\ref{fig:R15160005})
shows no sign of cool core emission, declining linearly from the
centre to the external regions. As expected, the abundance profile is
consistent with being flat at $Z \sim 0.3\,Z_\odot$ out to 500 kpc
(the maximum radius at which we can measure abundances).

%%================
%% Figure: R1516_0005
%%
%%================

\subsection{RXC\,J1516\,+0056}

A moderate temperature ($\kT = 3.75$ keV) cluster lying at $z=0.120$,
RXC\,J1516\,+0056 is also known as A2051. The X-ray image presents
quite a lot of structure, with several possible subclumps at the
outskirts of the object. These subclumps were excluded from the annuli
used to determine the temperature profile. The background subtracted
spectrum of the external region is well fitted with a single {\sc
  MeKaL} model at 0.25 keV, with positive normalisation. 

\begin{figure*}
\begin{centering}
\includegraphics[scale=0.30,angle=0,keepaspectratio,width=0.33\textwidth]{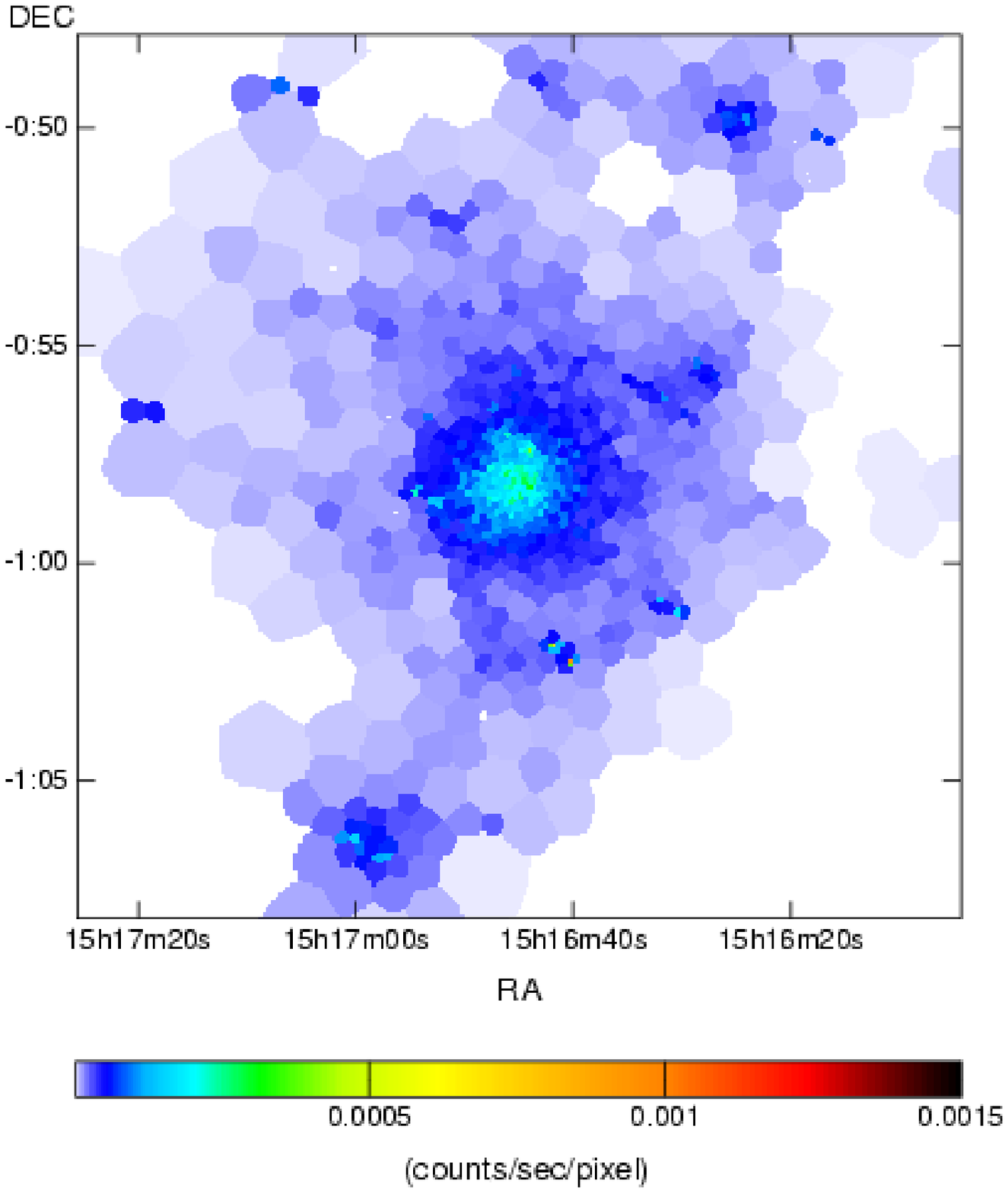}
\hfill
\includegraphics[scale=0.30,angle=0,keepaspectratio,width=0.32\textwidth]{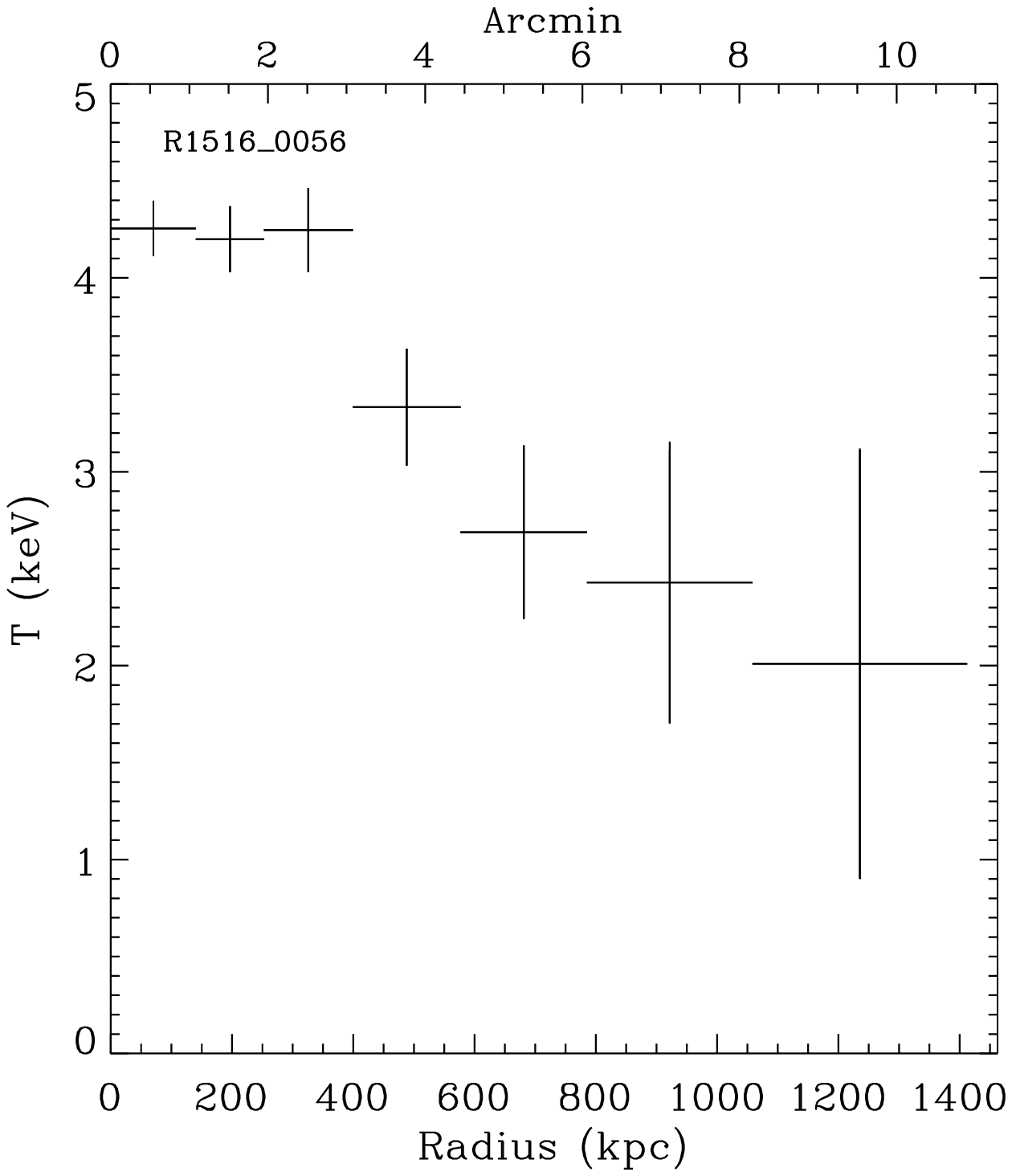}
\hfill
\includegraphics[scale=0.30,angle=0,keepaspectratio,width=0.33\textwidth]{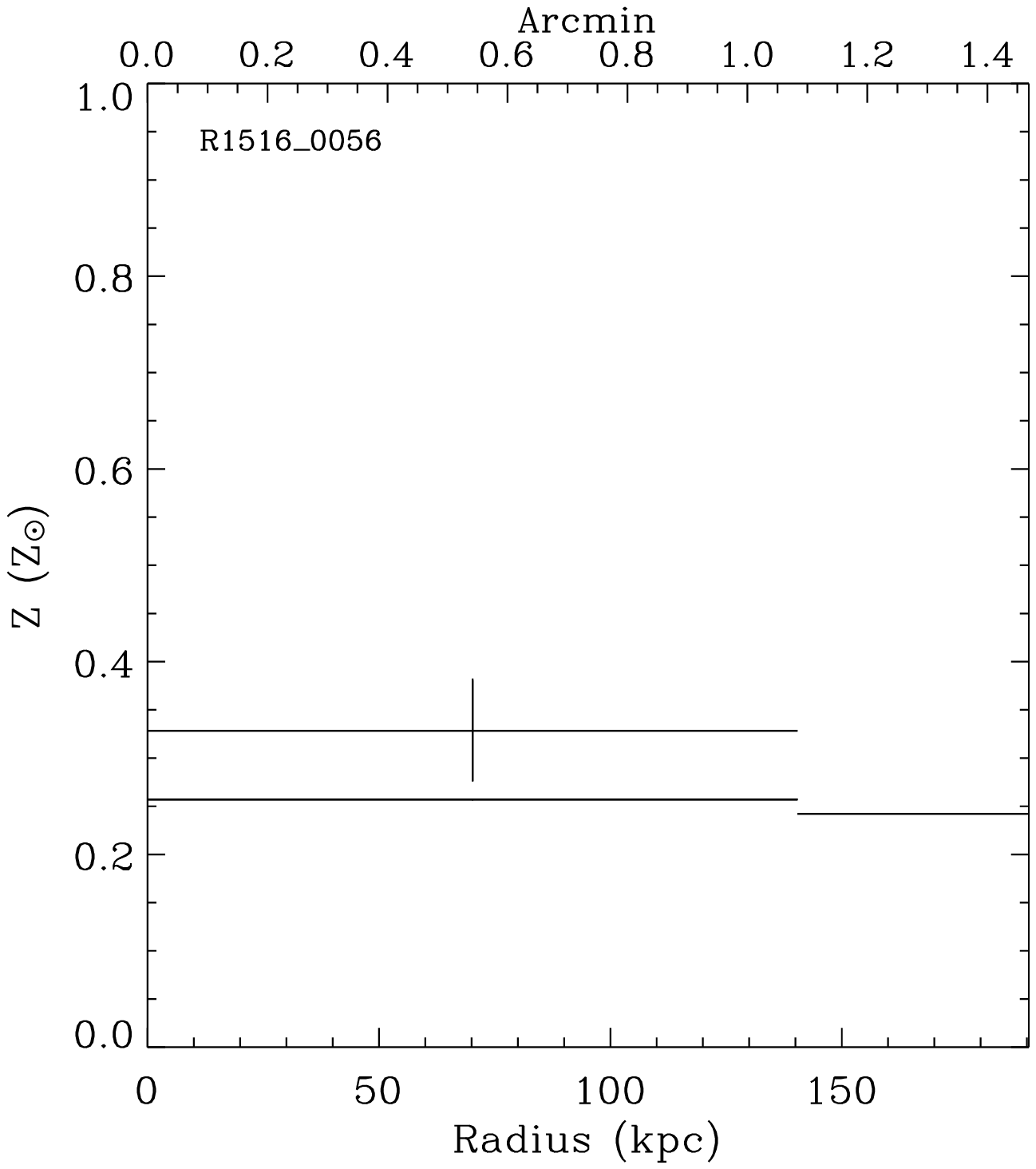}
\caption{{\footnotesize RXC\,J1516\,-0056}}\label{fig:R15160056}
\end{centering}
\end{figure*}

The temperature and abundance profiles are shown in 
Fig~\ref{fig:R15160056}. The temperature profile is flat in the inner
400 kpc, but declines by $\sim50$ per cent at the radius of maximum
detection. The abundance profile is consistent with being flat out to
the radius of maximum detection.

%%================
%% Figure: R1516_0056
%%
%%================

\subsection{RXC\,J2023\,-2056}

Also known as S868, lying at $z=0.056$, this is the lowest temperature
cluster in the present sample ($\kT = 2.7$ keV). The object has a
fairly regular appearance, but no strong evidence for a cooling
core. The background subtracted external region spectrum is well
described by a {\sc MeKal} model with positive normalisation and a
temperature of 0.23 keV. An additional power law component improves
the fit of the EPN data.

\begin{figure*}
\begin{centering}
\includegraphics[scale=0.30,angle=0,keepaspectratio,width=0.33\textwidth]{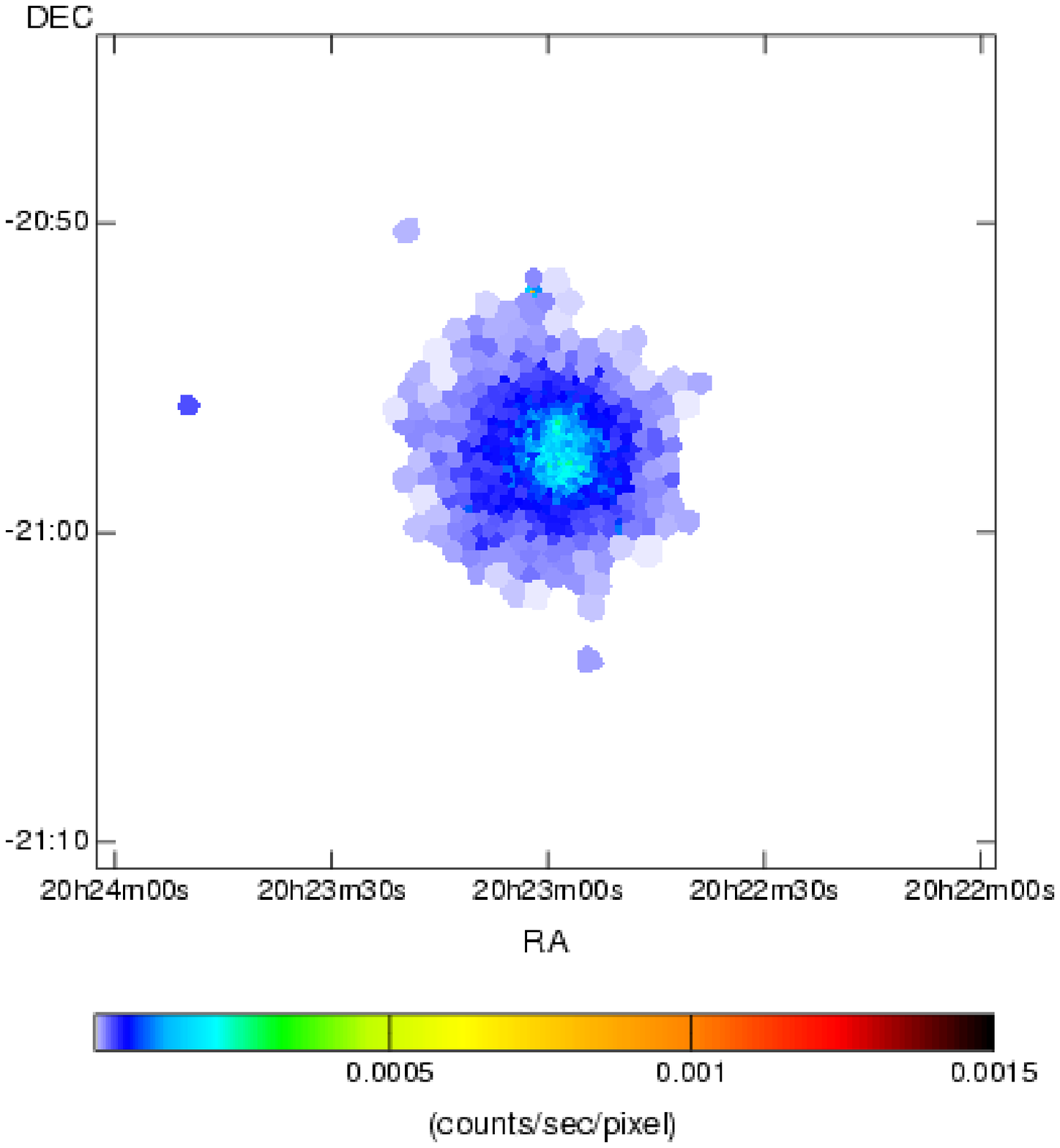}
\hfill
\includegraphics[scale=0.30,angle=0,keepaspectratio,width=0.32\textwidth]{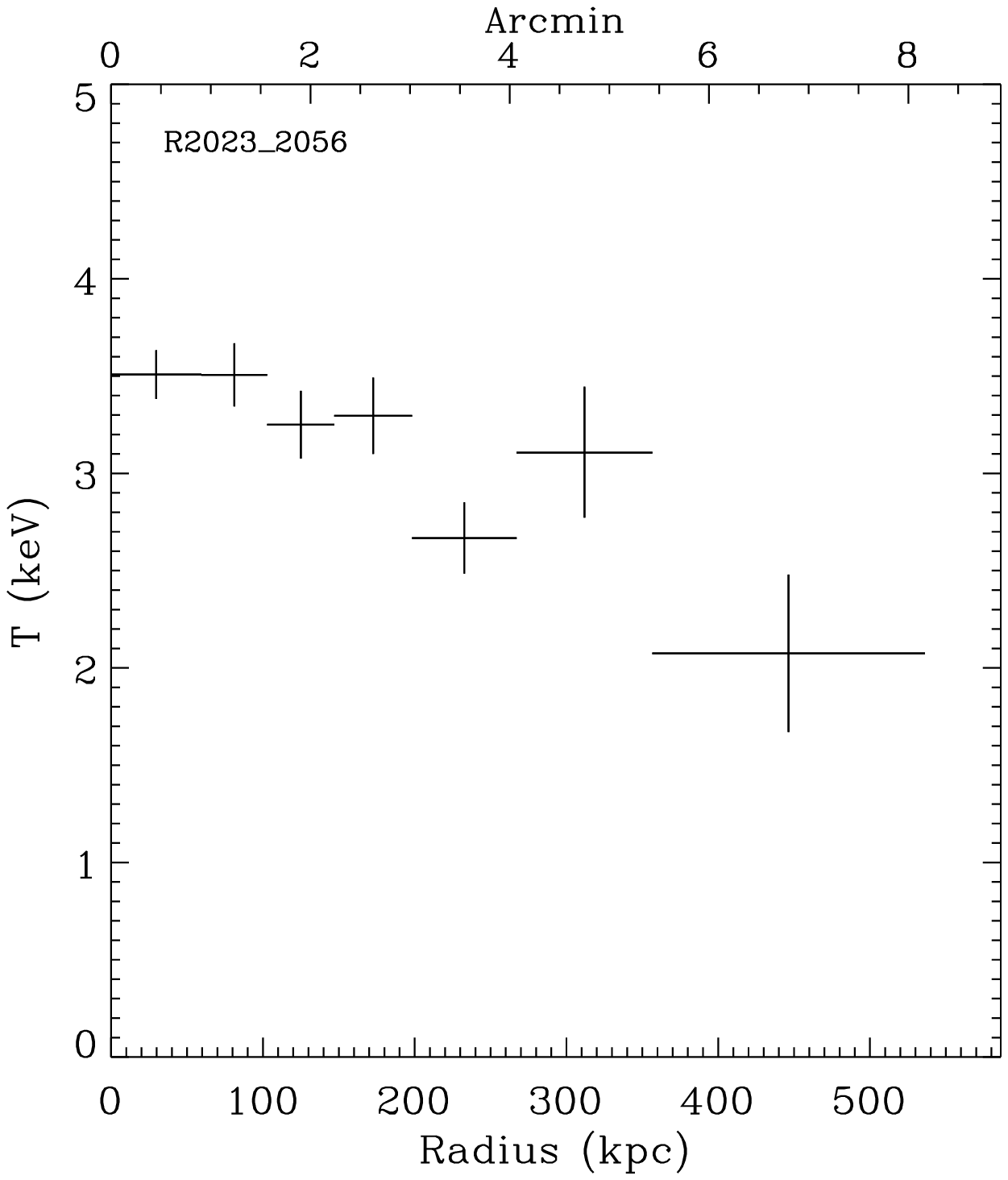}
\hfill
\includegraphics[scale=0.30,angle=0,keepaspectratio,width=0.33\textwidth]{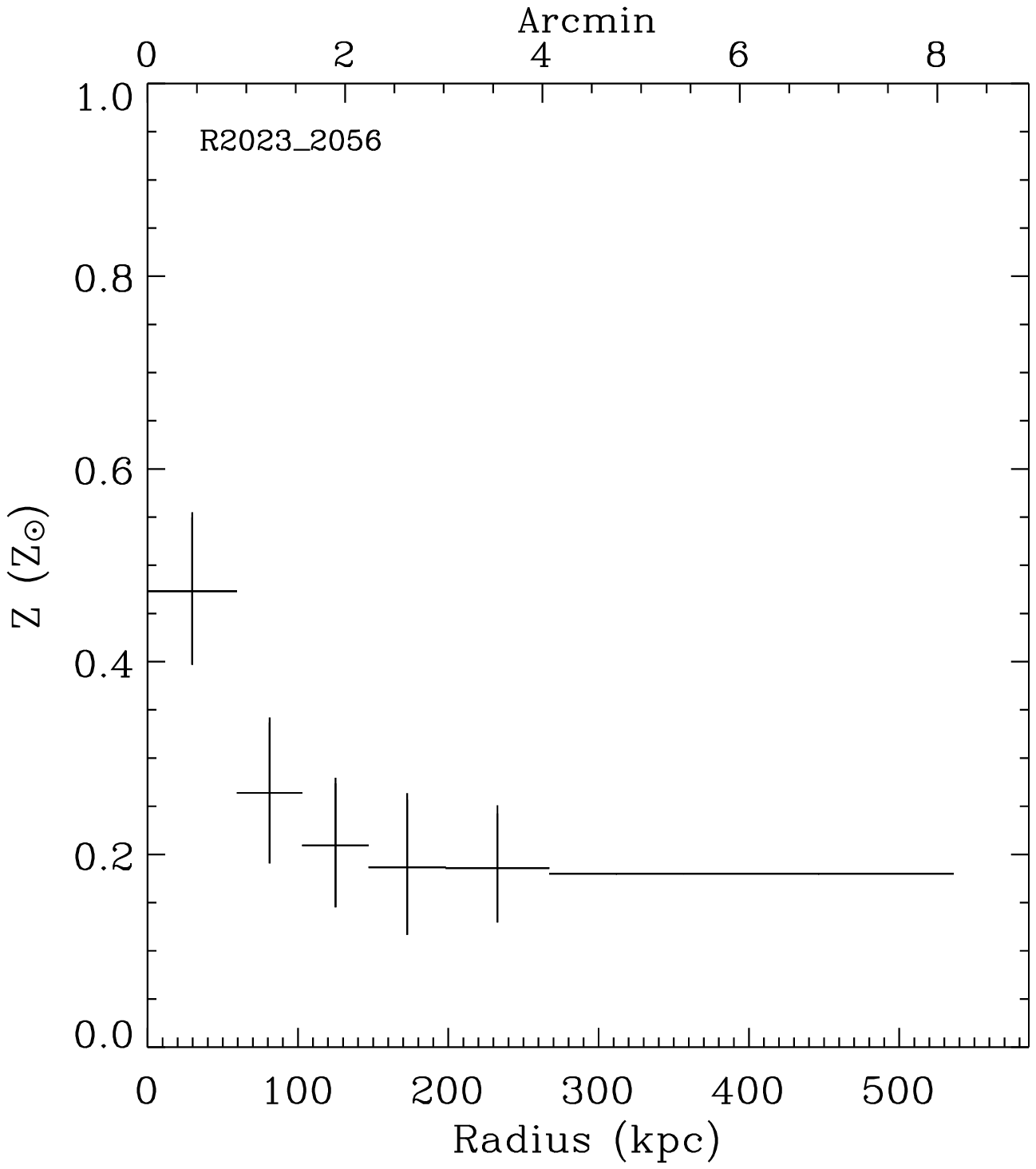}
\caption{{\footnotesize RXC\,J2023\,-2056.}}\label{fig:R2023}
\end{centering}
\end{figure*}

The temperature profile of the cluster, shown in Fig~\ref{fig:R2023}
is flat in the inner 100 kpc, 
after which there is a decline. The abundance profile declines
smoothly in power law fashion from a peak of $Z = 0.5Z_{\odot}$ in the
centre to about half that value at the outskirts.

%%================
%% Figure: R2023-2056
%%
%%================

\subsection{RXC\,J2048\,-1750}

With a temperature of $\kT = 4$ keV and lying at $z=0.085$,
RXC\,J2048\,-1750 presents a fairly disturbed appearance.  The
background subtracted spectrum of the region external to the cluster
emission can be fitted with a simple thermal model at 0.20 keV, with
positive normalisation. 

\begin{figure*}
\begin{centering}
\includegraphics[scale=0.30,angle=0,keepaspectratio,width=0.33\textwidth]{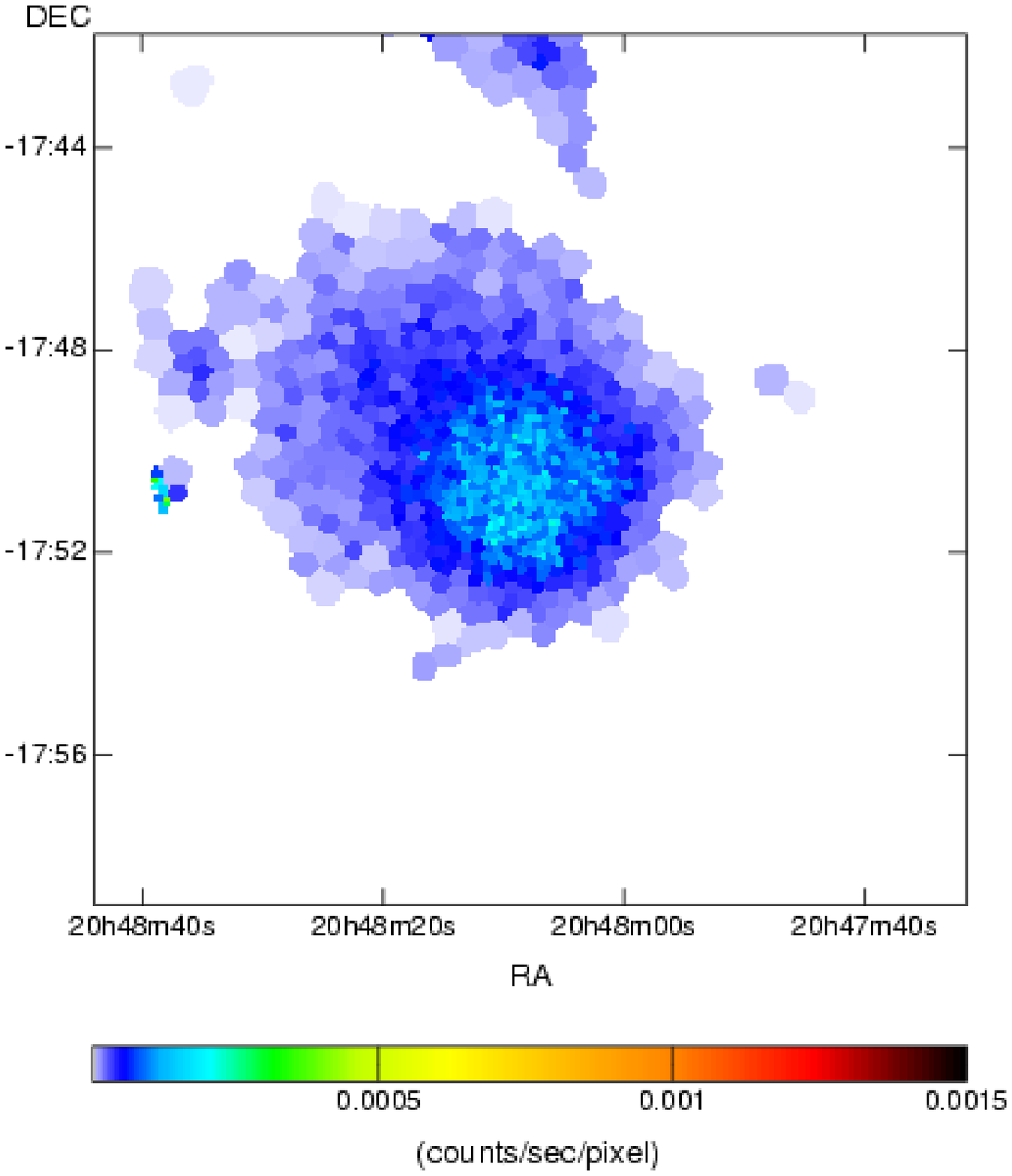}
\hfill
\includegraphics[scale=0.30,angle=0,keepaspectratio,width=0.32\textwidth]{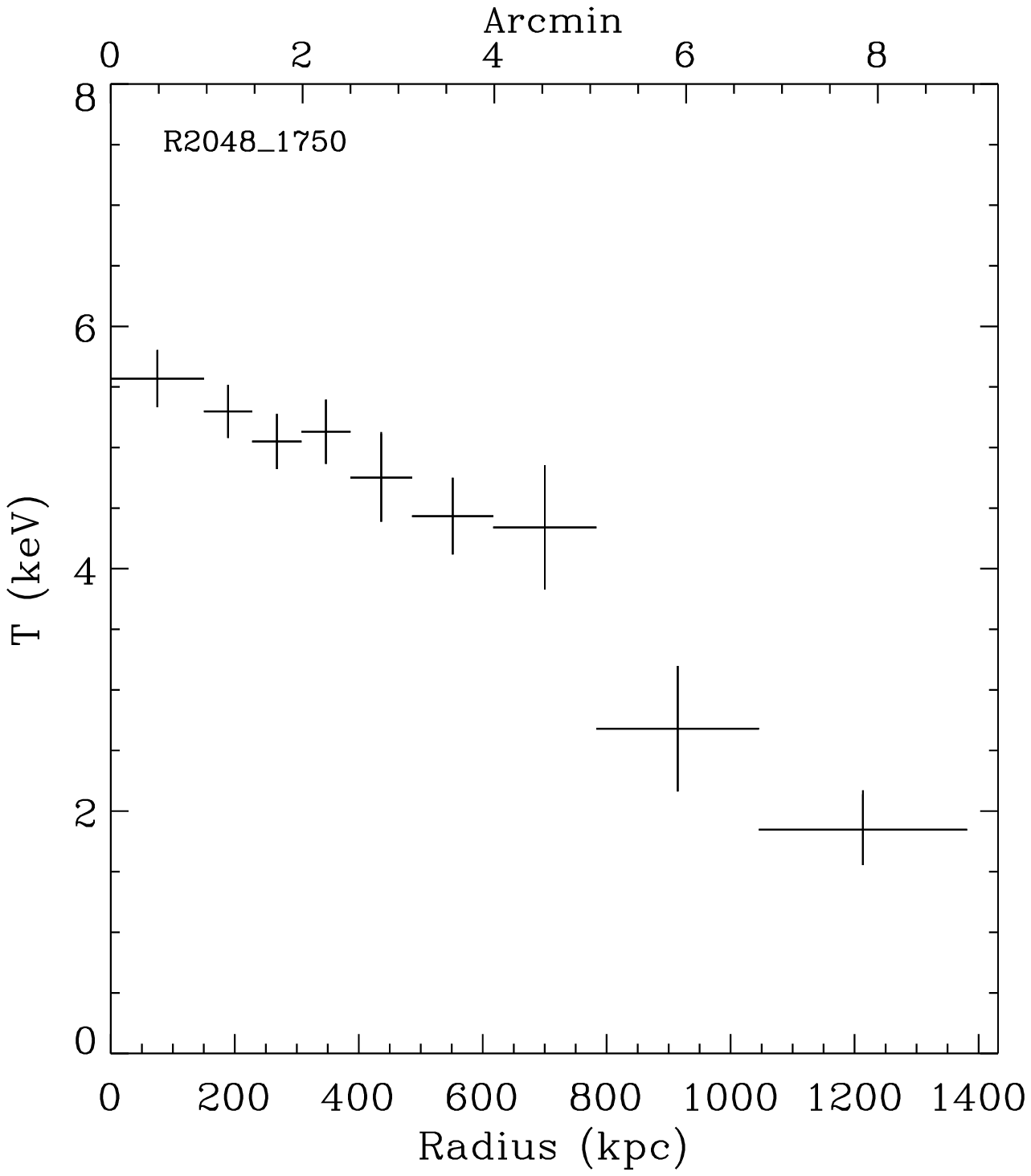}
\hfill
\includegraphics[scale=0.30,angle=0,keepaspectratio,width=0.33\textwidth]{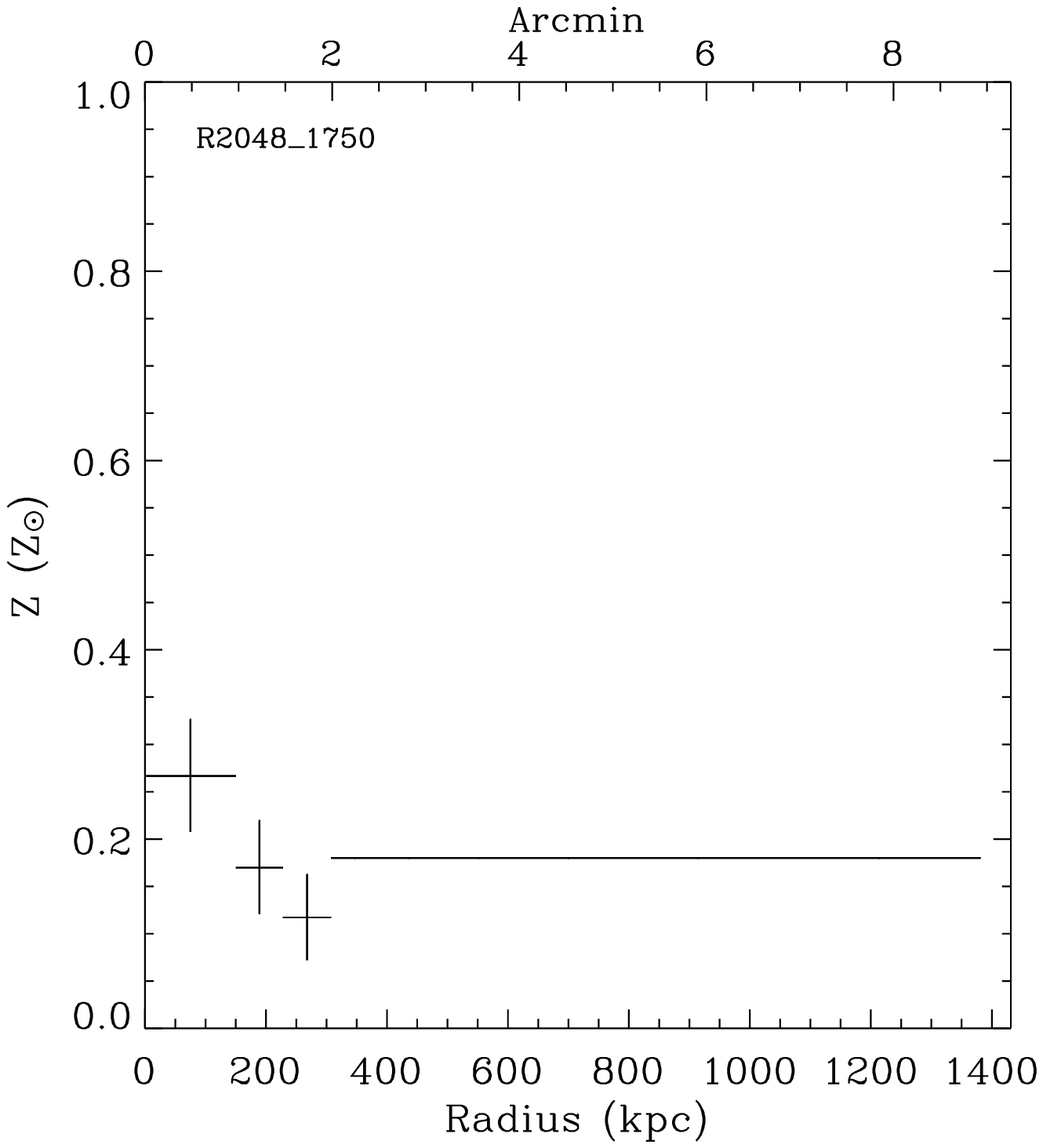}
\caption{{\footnotesize RXC\,J2048\,-1750.}}\label{fig:R2048}
\end{centering}
\end{figure*}

The temperature profile of the cluster (Fig.~\ref{fig:R2048}) declines
linearly, by more than a factor of two, from the centre to the
external regions. The abundance profile is very poorly constrained,
and we can only measure the three inner bins.

%%================
%% Figure: R2048-1750
%%
%%================

\subsection{RXC\,J2129\,-5048}

A moderate temperature ($\kT = 3.8$ keV) cluster also known as A3771,
RXC\,J2129\,-5048 lies at $z=0.08$. The X-ray image is disturbed, with
a distinct elongation in emission from the centre towards the NE. The
background subtracted spectrum of the external region can be fitted
with a thermal model at $\kT =0.33$ keV with an additional power law
improving the fit in all three cameras.

\begin{figure*}
\begin{centering}
\includegraphics[scale=0.30,angle=0,keepaspectratio,width=0.33\textwidth]{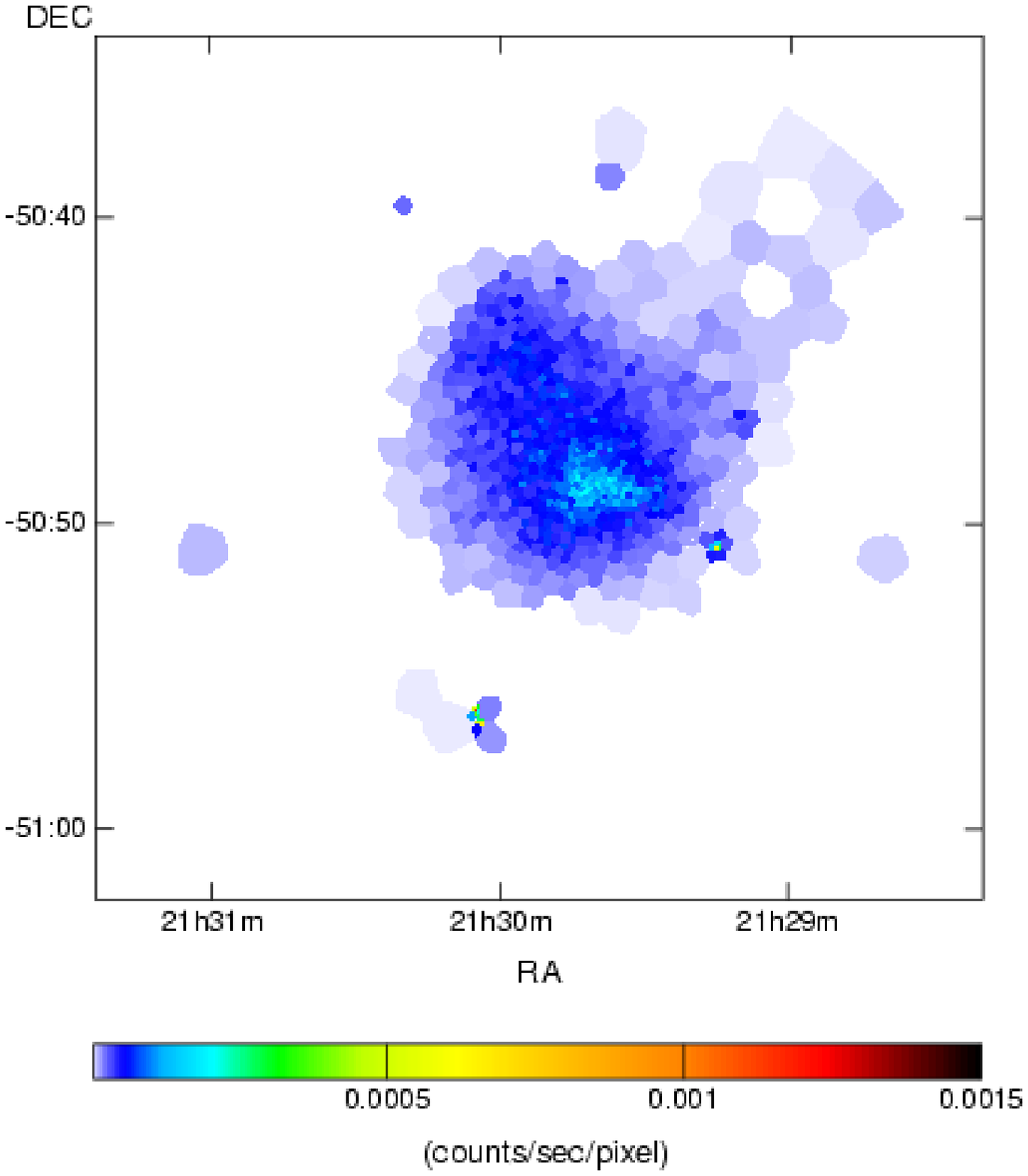}
\hfill
\includegraphics[scale=0.30,angle=0,keepaspectratio,width=0.32\textwidth]{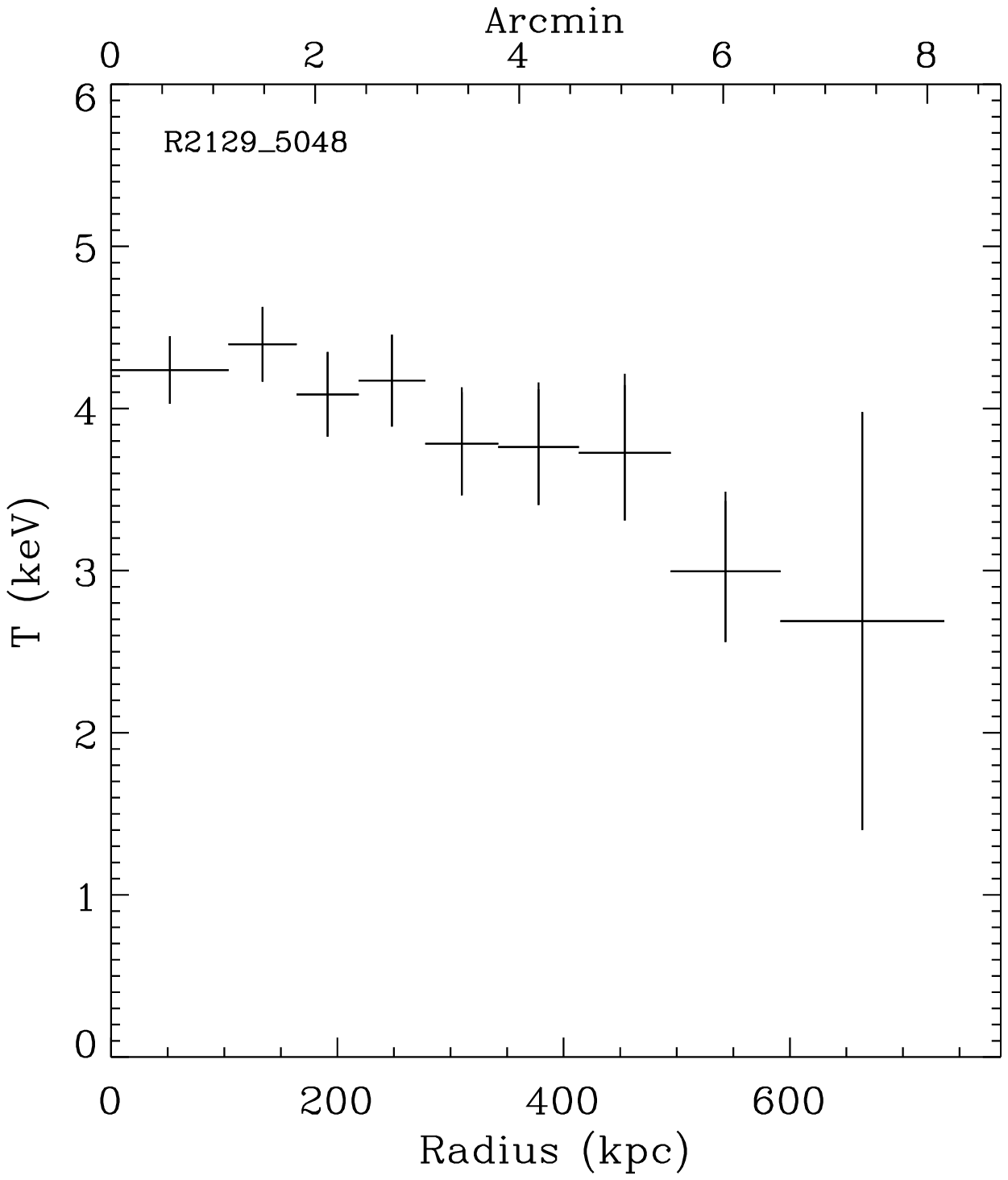}
\hfill
\includegraphics[scale=0.30,angle=0,keepaspectratio,width=0.32\textwidth]{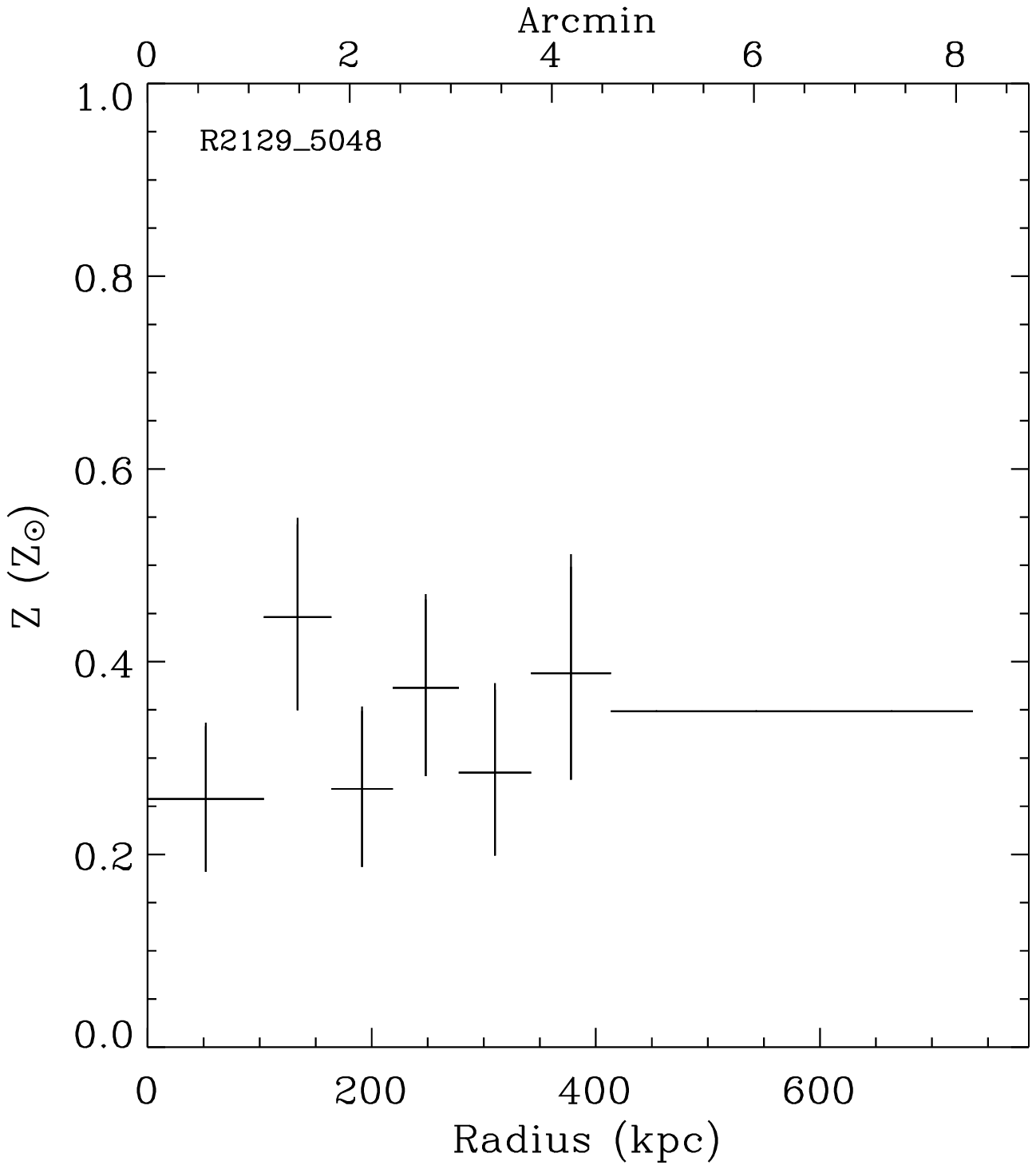}
\caption{{\footnotesize RXC\,J2129\,-5048.}}\label{fig:R2129}
\end{centering}
\end{figure*}

The temperature profile of the cluster (Fig.~\ref{fig:R2129}) is
relatively flat in the inner 200 kpc or so, but declines beyond
this. The abundance profile is relatively poorly constrained, but is
consistent with being flat at an average of $Z \sim 0.35Z_{\odot}$.

%%================
%% Figure: R2129-5048
%%
%%================

\subsection{RXC\,J2217\,-3543}

One of the more distant clusters in the sample, having an average
temperature of $\kT = 4.6$ keV and lying at $z=0.148$,
RXC\,J2217\,-3543 is also known as A3584. The X-ray image is quite
compact and symmetric, although the cluster does not present
strongly-peaked core emission. The spectrum of the background
subtracted external region presents strongly negative residuals below
1 keV and can be fitted with a thermal model at 0.65
keV, with negative normalisation.

\begin{figure*}
\begin{centering}
\includegraphics[scale=0.30,angle=0,keepaspectratio,width=0.33\textwidth]{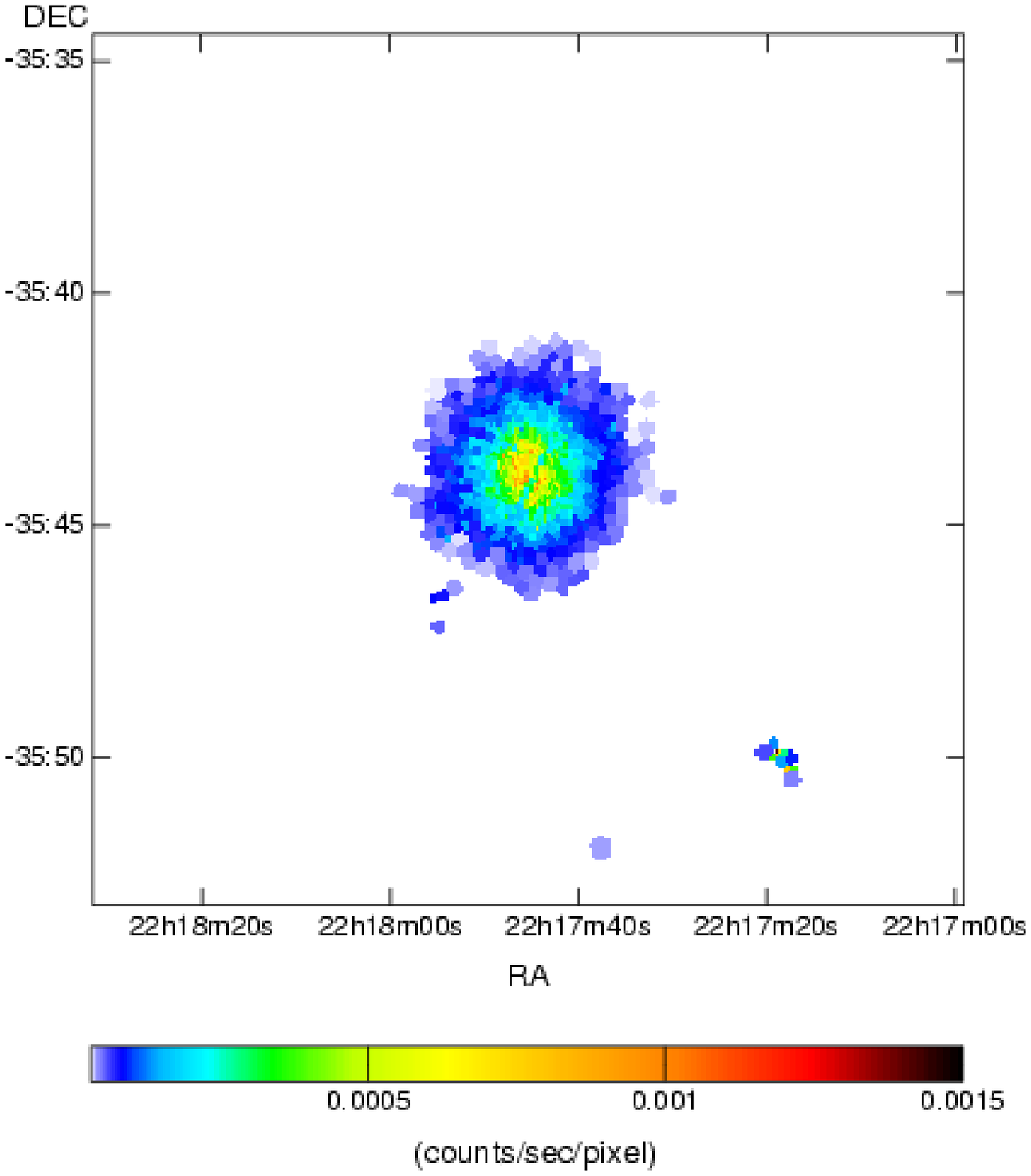}
\hfill
\includegraphics[scale=0.30,angle=0,keepaspectratio,width=0.32\textwidth]{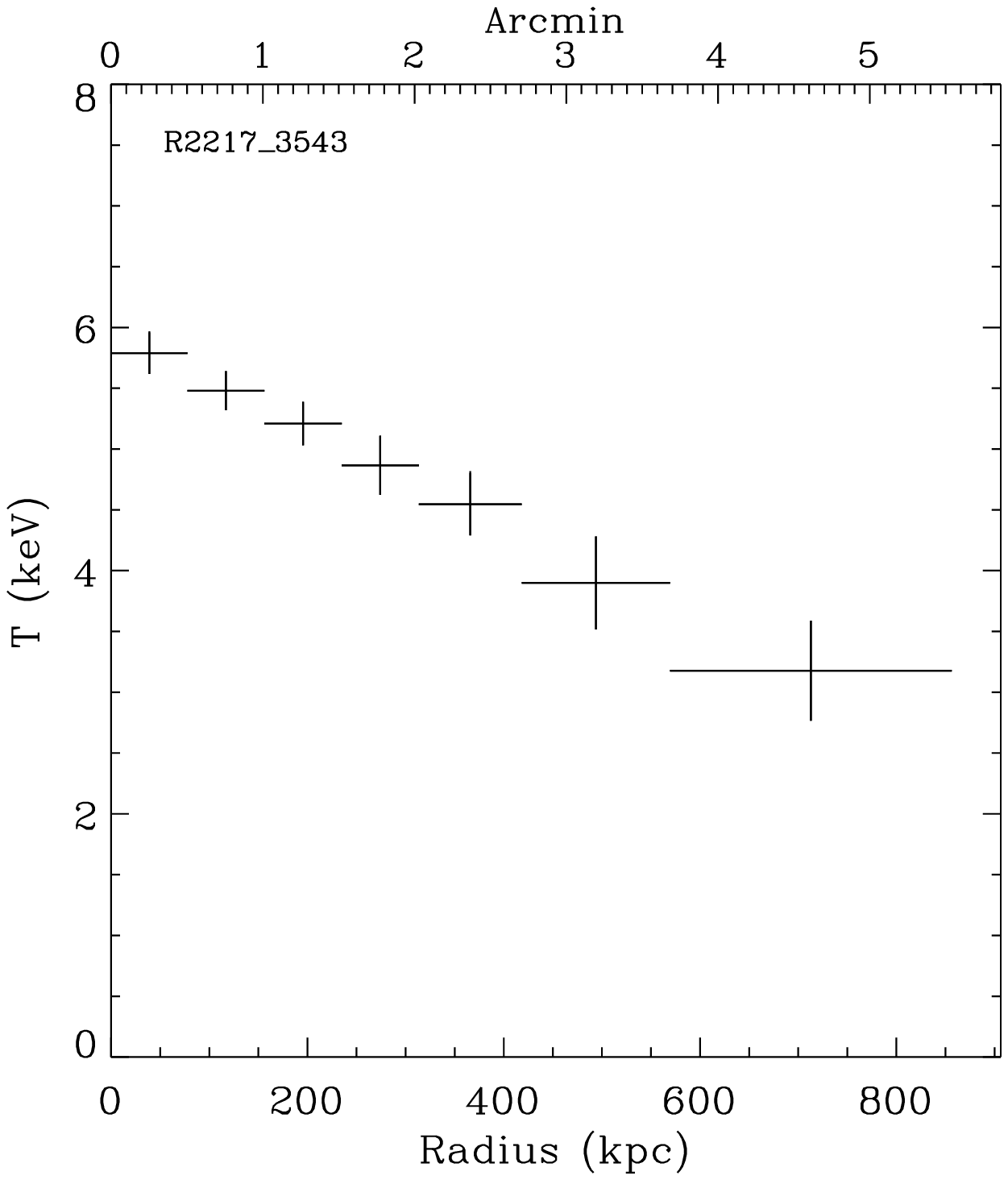}
\hfill
\includegraphics[scale=0.30,angle=0,keepaspectratio,width=0.33\textwidth]{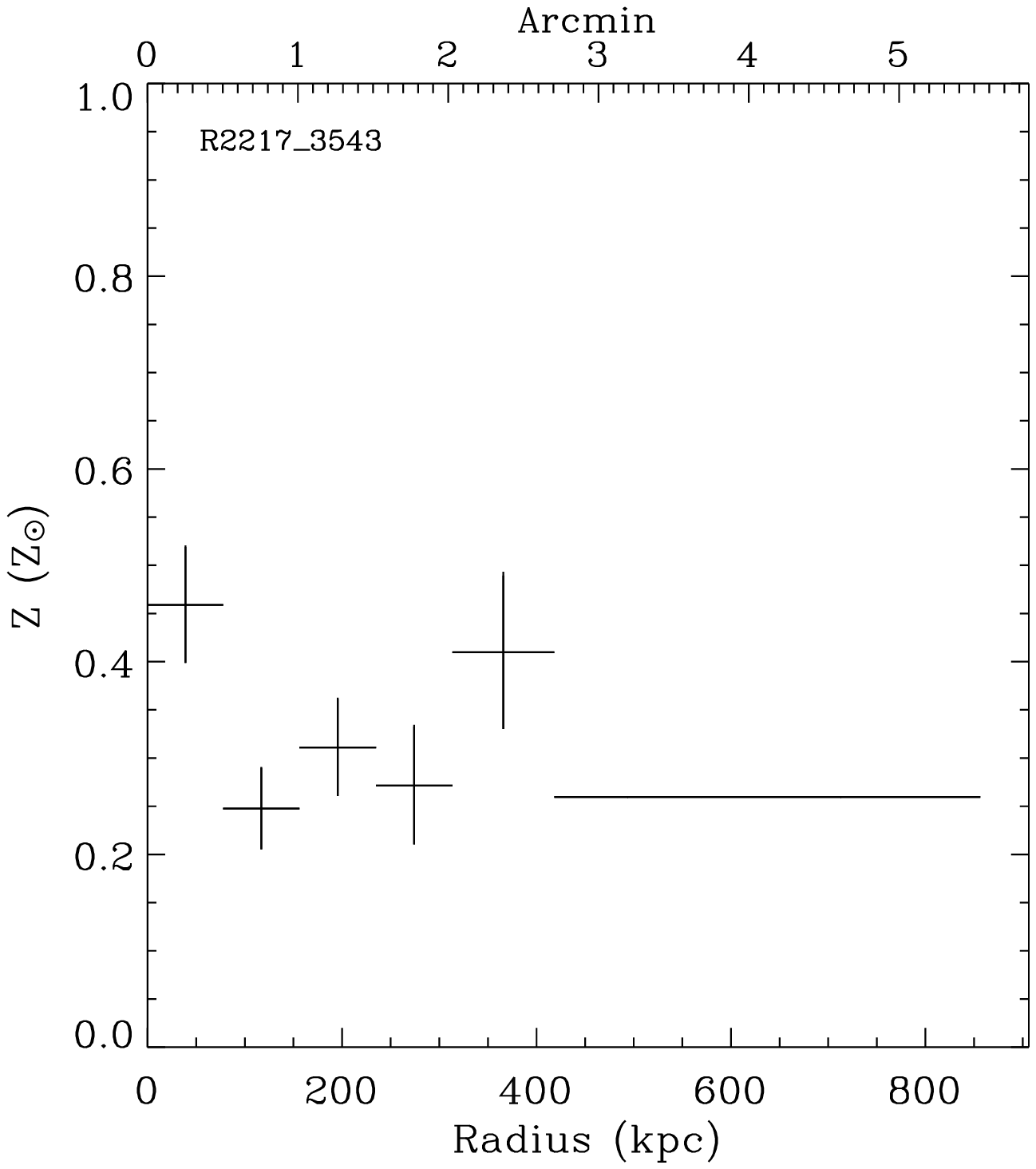}
\caption{{\footnotesize RXC\,J2217\,-3543.}}\label{fig:R2217}
\end{centering}
\end{figure*}

The temperature profile shown in Fig.~\ref{fig:R2217} declines
linearly from the peak of 5.5 keV at the centre to 3 keV at the
maximum radius at which we can measure the temperature. The abundance
profile does not show any trends with radius.

%%================
%% Figure: R2217-3543
%%
%%================

\subsection{RXC\,J2218\,-3853}

RXC\,J2218\,-3853 is also known as A3856, has an average temperature
of $\kT = 5.8$ keV and lies at $z=0.09$. The X-ray image is
elliptical, presenting an elongation in the SE-NW direction. The
background subtracted spectrum of the external region is well fitted
with a thermal model at 0.26 keV, with an additional power law
component improving the fit for all three cameras.

\begin{figure*}
\begin{centering}
\includegraphics[scale=0.30,angle=0,keepaspectratio,width=0.33\textwidth]{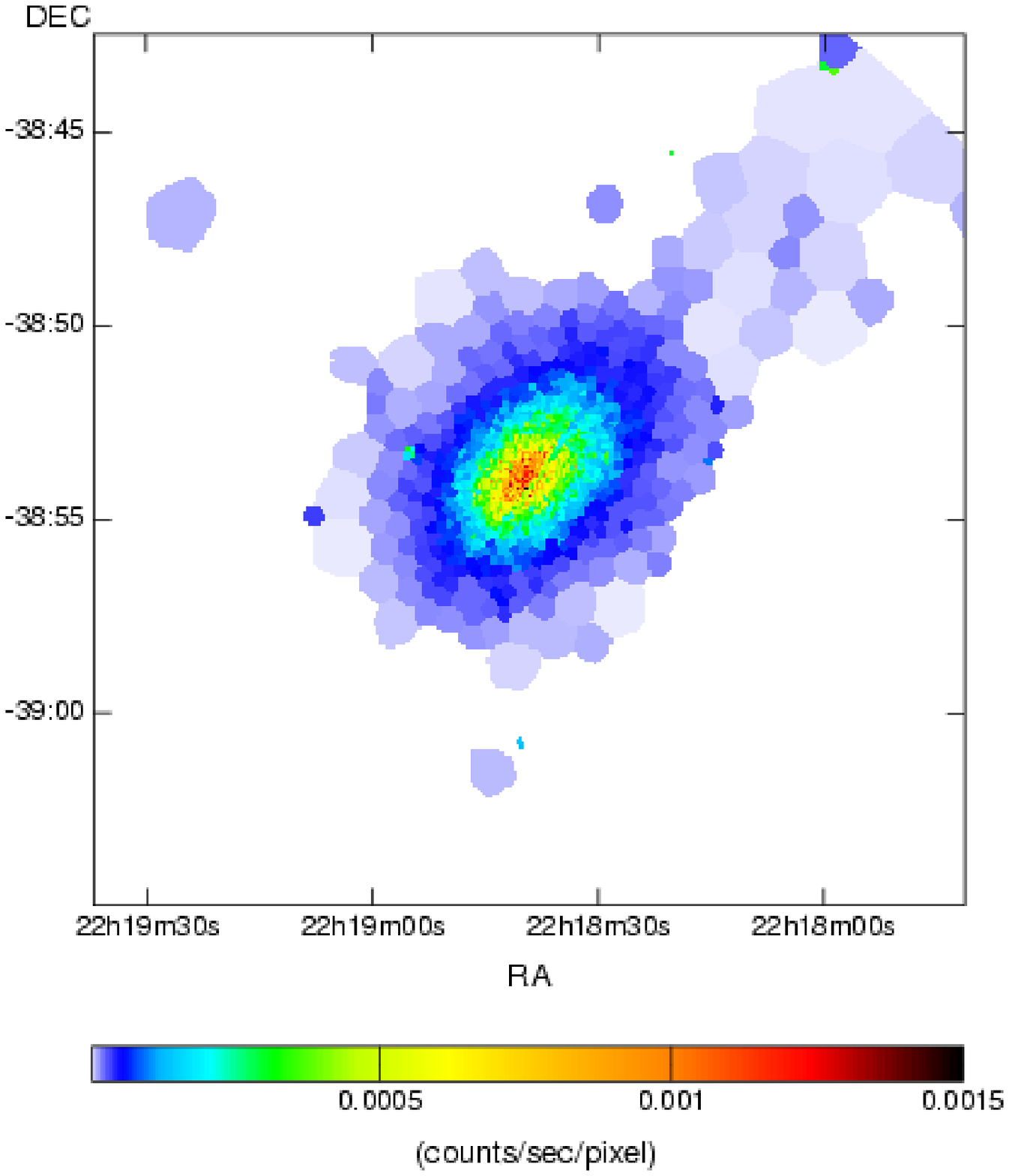}
\hfill
\includegraphics[scale=0.30,angle=0,keepaspectratio,width=0.32\textwidth]{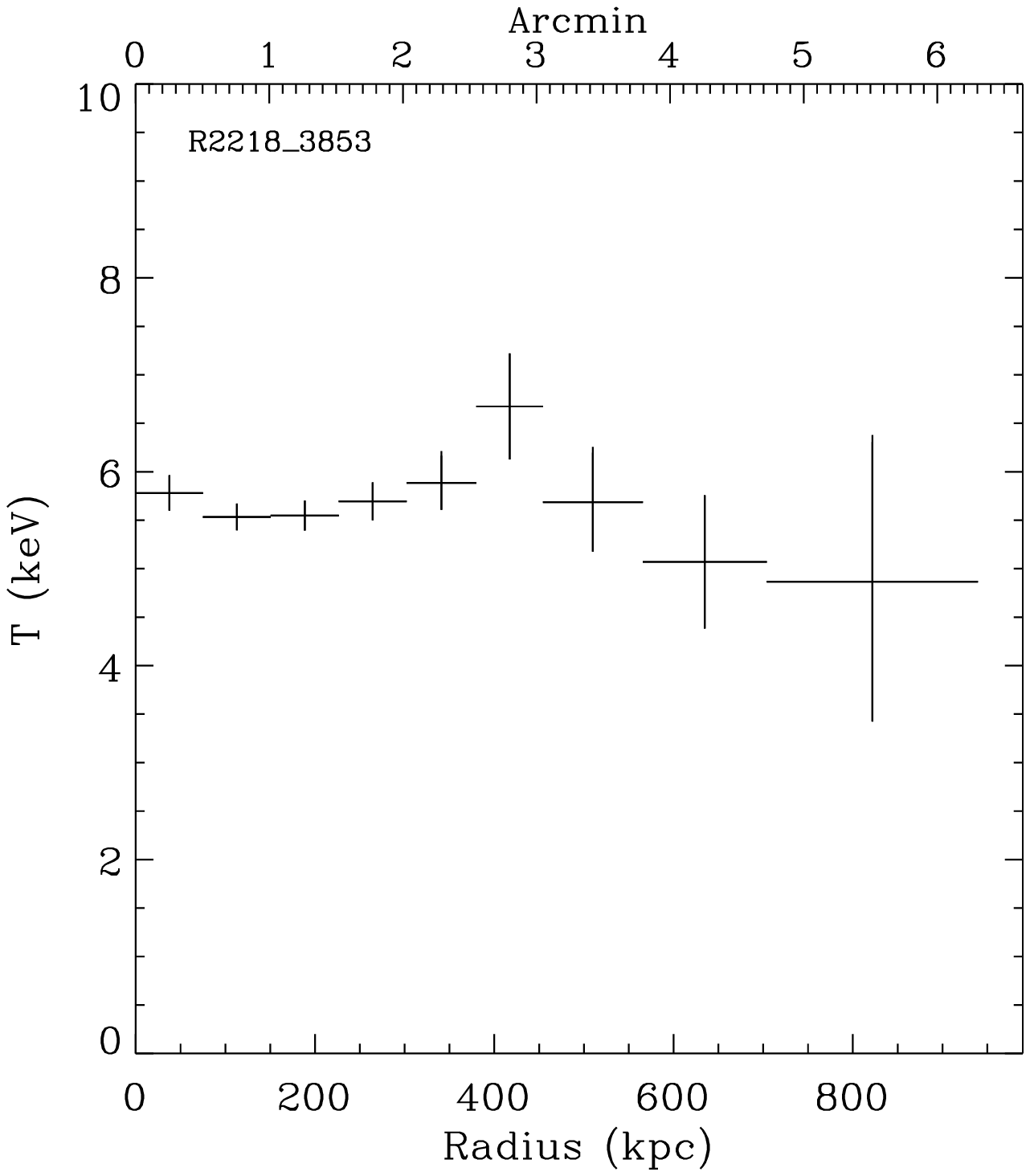}
\hfill
\includegraphics[scale=0.30,angle=0,keepaspectratio,width=0.33\textwidth]{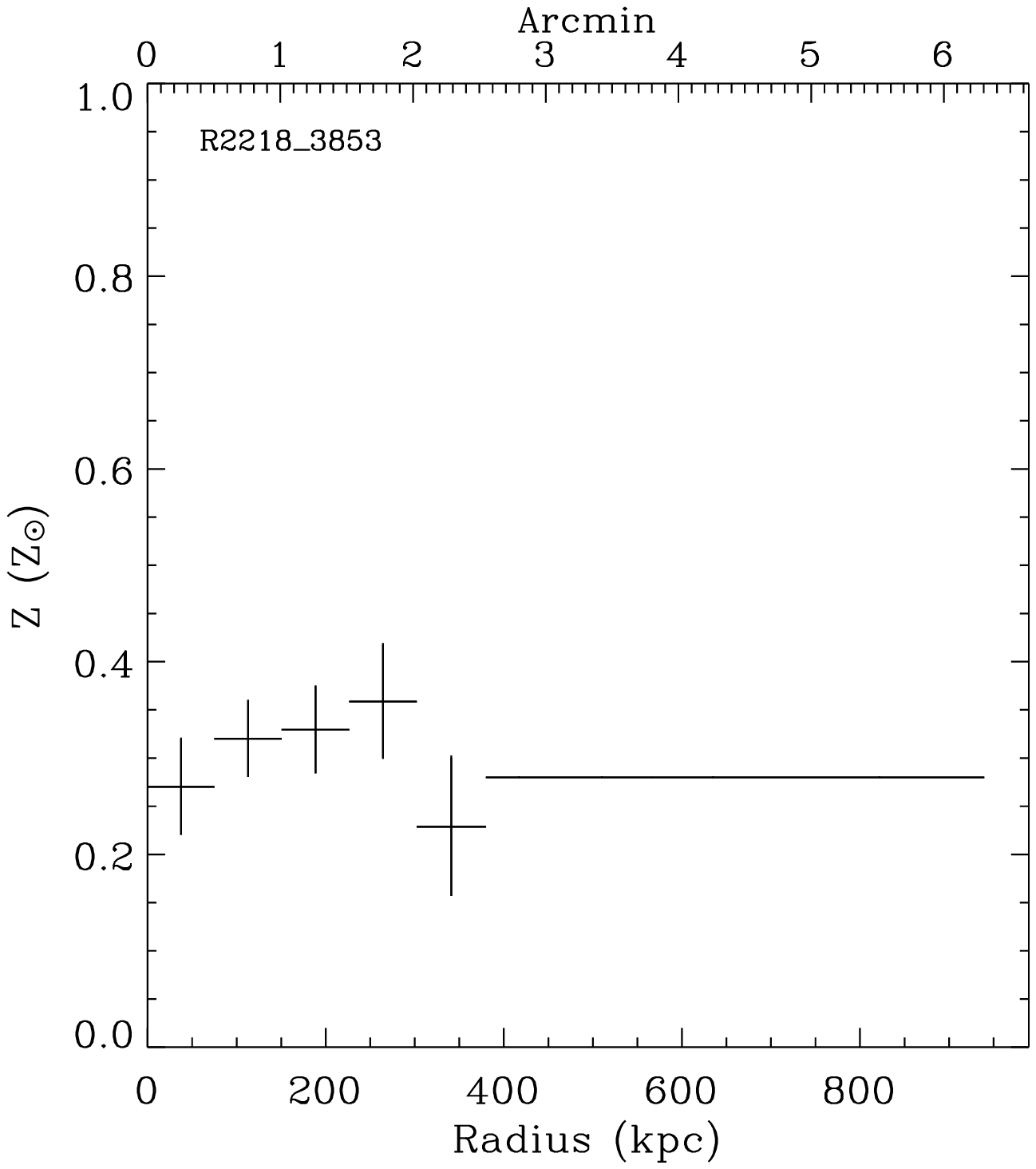}
\caption{{\footnotesize RXC\,J2218\,-3853.}}\label{fig:R2218}
\end{centering}
\end{figure*}

The temperature profile (Fig.~\ref{fig:R2218}) is flat in the inner
regions, rises to a peak at $\sim400$ kpc, and then declines (although
not significantly). The abundance profile is consistent with being
flat at an average of $Z = 0.3Z_{\odot}$ out to 400 kpc, the detection
limit. 

%%================
%% Figure: R2218-3853
%%
%%================

%%\section{Background subtraction issues}

%%We have undertaken a comparison of our blank sky background event
%%files and those of \citet*{NML}, which were produced using different
%%flare filtering bin sizes and thresholds. We extracted light curves in
%%the high energy [10-12]/[12-14] (EMOS/EPN) and medium energy [1-5] keV
%%bands in 1000s bins, and calculated the mean and $1 \sigma$ dispersion
%%of the resulting count rates. In general, our mean values were lower
%%than those found by \citep{NML}. However, the dispersion in both high
%%and medium energy count rates is comparable between the two data sets,
%%at $\sim 5-10\%$. We also accumulated spectra across the entire field
%%of view from both our blank sky background and that of
%%\citeauthor{NML}. Apart from a slightly higher EPN normalisation in
%%the \citeauthor{NML} background files,
%%there is essentially no difference between the spectra of the
%%different blank sky data sets.


\begin{thebibliography}{}

\bibitem[Allen et al.(2001)Allen, Schmidt \& Fabian]{allen01}Allen,
  S.W., Schmidt, R.W., \& Fabian, A.C.~2001, MNRAS, 328, L37

\bibitem[Arnaud et al.(2001)]{arnaud01}Arnaud, M., Neumann, D., Aghanim,
  N.  et al. 2001, A\&A, 365, L80 

%%\bibitem[Arnaud et al.(2002)]{a02}Arnaud, M., et al., 2002, A\&A, 390,
%%  27

%%\bibitem[Pratt, Arnaud \& Aghanim(2001)]{p01}Pratt, G.W., Arnaud, M.,
%%  Aghanim, N., astro-ph/0105431

%%\bibitem[Pratt \& Arnaud(2002)]{pa02}Pratt, G.W. \& Arnaud, M., 2002,
%%  A\&A, 394, 375

%%\bibitem[Arnaud et al.(2004)Arnaud, Pratt \&
%%  Pointecouteau]{arnaud04}Arnaud, M., Pratt, G.W. \& Pointecouteau,
%%  E., 2004, Mem. Soc. Astron. It, 75, 529

\bibitem[Arnaud et al.(2005)Arnaud, Pointecouteau \& Pratt]{app}
   Arnaud, M., Pointecouteau, E. \& Pratt, G.W.~2005, A\&A, 441, 893 

\bibitem[Anders \& Grevesse(1989)]{agr}Anders, E., \& Grevesse,
  N.~1989, Geochim. et Cosmochim. Acta, 53, 197

\bibitem[Andersson \& Madejski(2004)]{am}Andersson, K.E. \& Madejski,
  G.M.~2004, ApJ, 607, 190

\bibitem[B\"ohringer et al.(2004)]{reflex}B\"ohringer, H., Schuecker,
  P., Guzzo, L., et al.~2004, A\&A, 425, 367

\bibitem[Borgani et al.(2004)]{borg04}Borgani, S., Murante, G.,
    Springel, V., et al.~2004, MNRAS, 348, 1078

\bibitem[Briel \& Henry(1994)]{bh94}Briel, U.G., \& Henry, J.P.~1994,
  Nature, 372, 439

\bibitem[Buote \& Tsai(1995)]{bt}Buote, D.A., \& Tsai, J.C.~1995,
  MNRAS, 452, 522

\bibitem[Croston et al.(2006)]{croston}Croston, J.H., Arnaud, M.,
  Pratt, G.W., \& B\"ohringer, H.~2006, in The X-ray Universe 2005,
  ed. A. Wilson, ESA SP-604, p737 

\bibitem[David et al.(1995)David, Jones \& Forman]{david}David, L.P.,
  Jones, C., \& Forman, W.R.~1995, ApJ, 445, 578

\bibitem[De Grandi \& Molendi(2002)]{dm01}De Grandi, S. \& Molendi,
  S.~2001, ApJ, 551, 153

\bibitem[De Grandi \& Molendi(2002)]{dm02}De Grandi, S. \& Molendi,
  S.~2002, ApJ, 567, 163

\bibitem[Diehl \& Statler(2005)]{ds}Diehl, S., \& Statler, T.S.~2006,
  MNRAS, 368, 497

\bibitem[Evrard et al.(1996)Evrard, Metzler \& Navarro]{emn}Evrard,
  A.E., Metzler, C.A., Navarro, J.F.~1996, ApJ, 469, 494

\bibitem[Eyles et al.(1991)]{eyles}Eyles, C.J., Watt, M.P., Bertram,
  D., Church, M.J., \& Ponman, T.J.~1991, ApJ, 376, 23

\bibitem[Fabricant et al.(1980)Fabricant, Lecar \&
  Gorenstein]{fab1}Fabricant, D., Lecar, M., \& Gorenstein, P.~1980,
  ApJ, 241, 552 

\bibitem[Fabricant \& Gorenstein(1983)]{fab2}Fabricant, D., \&
  Gorenstein, P.~1983, ApJ, 267, 535

\bibitem[Finoguenov et al.(2001)]{fin01}Finoguenov, A.A., Arnaud, M.,
  \& David, L.P.~2001, ApJ, 555, 191

\bibitem[Girardi et al.(1997)]{gir}Girardi, M., Fadda, D., Escalera,
  E., Giuricin, G., Mardirossian, F., Mezzetti, M.~1997, ApJ, 490, 56

%%\bibitem[Grevesse \& Sauval(1998)]{grsa}Grevesse, N. \& Sauval, A.J.~1998,
%%  Space Sci. Rev., 85, 116

\bibitem[Henry et al.(1993)Henry, Briel \& Nulsen]{hbn}Henry, J.P.,
  Briel, U.G., \& Nulsen, P.E.J.~1993, A\&A, 271, 413

\bibitem[Henry \& Briel(1995)]{hb95}Henry, J.P. \& Briel, U.G.~1995,
  ApJ, 443, L9

\bibitem[Hughes et al.(1988)Hughes, Gorenstein \&
  Fabricant]{hughes}Hughes, J.P., Gorenstein, P., \& Fabricant,
  D.~1988, ApJ, 329, 82

\bibitem[Irwin et al.(1999)Irwin, Bregman \& Evrard]{irwin99}Irwin,
  J.A., Bregman, J.E., \& Evrard, A.E.~1999, ApJ, 519, 518

\bibitem[Irwin \& Bregman(2000)]{ib00}Irwin, J.A., Bregman, J., 2000,
  ApJ, 538, 543  

\bibitem[Isobe et al.(1986)]{ifn}Isobe, T., Feigelson, E.D. \& Nelson,
  P.I.~1986, ApJ, 306, 490 

\bibitem[Jeltema et al.(2005)]{jelt}Jeltema, T., Canizares, C.R.,
  Bautz, M.W., \& Buote, D.A.~2005, ApJ, 624, 606

\bibitem[Kaastra et al.(2004)]{kaa}Kaastra, J.S., Tamura, T.,
  Petersen, J.R., et al., 2004, A\&A, 413, 415

\bibitem[Kay et al.(2004)]{kay04}Kay, S.T., Thomas, P.A., Jenkins,
A. \& Pearce, F.R. 2004, MNRAS, 355, 1091

\bibitem[Kotov \& Vikhlinin(2006)]{kotov}Kotov, O., \& Vikhlinin, A.,
  2006, ApJ, 641, 752

\bibitem[Koyama et al.(1991)Koyama, Takano \& Tawara]{koy}Koyama, K.,
  Takano, S., \& Tawara, Y.~1991, Nature, 350, 135

\bibitem[Lewis et al.(2000)]{lew00} Lewis, G.F., Babul, A., Katz, N.,
Quinn, T., Hernquist, L., \& Weinberg, D.H.~ 2000, ApJ, 536, 623

\bibitem[Loken et al.(2002)]{lok02}Loken, C., Norman, M.L., Nelson,
  E., Burns, J., Bryan, G.L. \& Motl, P. 2002, ApJ, 579, 571

\bibitem[Markevitch et al.(1998)]{mark98} Markevitch, M., Forman, W.,
  Sarazin, C., Vikhlinin, A.~1998, ApJ, 503, 77 

%%\bibitem[Nevalainen et al.(2005)Nevalainen, Markevitch \&
%%  Lumb]{NML}Nevalainen, J., Markevitch, M. \& Lumb, D.~2005, ApJ, 629,
%%  172

\bibitem[Neumann \& Arnaud(1999)]{neuarn}Neumann, D.M., \& Arnaud, M.~1999,
  A\&A, 348, 711

\bibitem[Piffaretti et al.(2005)]{piff}Piffaretti, R., Jetzer, P.,
Kaastra, J.S., Tamura, T., 2005, A\&A, 433, 101

\bibitem[Pointecouteau et al.(2005)Pointecouteau, Arnaud \&
  Pratt]{point}Pointecouteau, E., Arnaud, M., \& Pratt, G.W.~2005,
  A\&A, 435, 1

\bibitem[Ponman et al.(2003)Ponman, Sanderson \&
  Finoguenov]{psf}Ponman, T.J., Sanderson, A.J.R., \& Finoguenov,
  A.A.~2003, MNRAS, 343, 331

\bibitem[Pratt et al.(2006)Pratt, Arnaud \& Pointecouteau]{pap}Pratt,
  G.W., Arnaud, M., \& Pointecouteau, E.~2006, A\&A, 446, 429

\bibitem[Read \& Ponman(2003)]{rp03}Read, A.M.  \& Ponman, T.J.~2003,
A\&A, 409, 395

\bibitem[Reiprich et al.(2006)]{reiprich}Reiprich, T.H., Hudson, D.S.,
  Erben, T., \& Sarazin, C.L.~2006, in Relativistic Astrophysics and
  Cosmology - Einstein's Legacy, eds. B. Aschenbach,, V. Burwitz,
  G. Hasinger, B. Leibundgut, (Berlin, Germany: ESO Astrophysics
  Symposia, Springer Verlag)
  (\href{http://arxiv.org/abs/astro-ph/0603129}{\tt 
    astro-ph/0603129}) 

%%\bibitem[Snowden et al.(1997)]{snow97} Snowden, S., Egger, R.,
%%  Freyberg, M.J., et al.~1997, ApJ, 485, 125

\bibitem[Springel(2005)]{spr05}Springel, V. 2005, MNRAS, 463, 1105

\bibitem[Tornatore et al.(2003)]{torna}Tornatore, L., Borgani, S.,
Springel, V., Matteucci, F., Menci, N. \& Murante, G. 2003, MNRAS,
342, 1025

\bibitem[Valdarnini (2003)]{valda}Valdarnini, R. 2003, MNRAS, 339, 1117

\bibitem[Vikhlinin et al.(2005)]{vikh05}Vikhlinin, A., Markevitch, M.,
  Murray, S.S., Jones, C., Forman, W., Van Speybroeck, L.~2005, ApJ,
  628, 655

\bibitem[White(2000)]{whi00}White, D.A., 2000, MNRAS, 312, 663

\bibitem[Zhang et al.(2004)]{zhang}Zhang, Y.Y., Finoguenov, A.,
  B\"ohringer, H., Ikebe, Y., Matsushita, K., Schuecker, P., 2004,
  A\&A 413, 49

\end{thebibliography}
\end{document}